\documentclass[fleqn, usenatbib]{mnras}

\usepackage[T1]{fontenc}

\DeclareRobustCommand{\VAN}[3]{#2}
\let\VANthebibliography\thebibliography
\def\thebibliography{\DeclareRobustCommand{\VAN}[3]{##3}\VANthebibliography}

\usepackage{graphicx}        
\usepackage{amsmath}         
\usepackage{amssymb}         
\usepackage{booktabs}        
\usepackage{scalerel}        
\usepackage{xcolor}          
\usepackage{upquote}         
\usepackage{subcaption}      
\usepackage{multirow}        
\usepackage{pdflscape}       
\usepackage{microtype}       
\usepackage{pgffor}          
\usepackage{crossreftools}   
\usepackage{orcidlink}       

\usepackage{enumitem}        
\setlist[enumerate]{style=standard,
                    align=left,
                    labelindent=1em,
                    labelwidth=!,
                    leftmargin=*,
                    itemindent=0em}

\definecolor{key1}{RGB}{122, 5, 137}
\definecolor{key2}{RGB}{34, 19, 163}

\usepackage{listings} 
\newcommand*{\dpi}{600} 

\newcommand*{\orcid}[1]{\;\!\orcidlink{#1}\;\!}

\newcommand*{\units}[1]{\scalebox{0.8}{(#1)}}

\newcommand*{\Sun}{\protect\scalebox{0.7}{$\odot$}}

\newcommand*{\Helio}{\protect\scalebox{0.6}{${\rm H}$}}
\newcommand*{\Heliosmall}{\protect\scalebox{0.4}{${\rm H}$}}

\newcommand*{\Range}[2]{[#1\,\text{,}\,#2]}

\newcommand*{\pd}[1]{\times\!10^{#1}}

\newcommand*{\SM}[1]{{\scaleto{\rm #1}{3.25pt}}}

\newcommand*{\var}[1]{\mbox{\footnotesize$(#1)$}}

\newcommand*{\BV}{{B\textnormal{-}\:\:\!\!\!V}}

\newcommand*{\BPRP}{{G_{\rm BP}\!-\!G_{\rm RP}}}
\newcommand*{\gr}{{g\;\!\textnormal{-}\;\!r}}
\newcommand*{\FeH}{{\rm [Fe/H]}}
\newcommand*{\AlphaFe}{{\rm [\alpha/Fe]}}

\newcommand*{\nbody}{\hbox{\textit{N}\!\!\:-body}}

\newcommand*{\GtrSim}{\smallrel\gtrsim}
\newcommand*{\LessSim}{\smallrel\lesssim}

\newcommand*{\Approx}{\smallrel\sim}

\makeatletter
\newcommand*{\smallrel}[2][.8]{%
  \mathrel{\mathpalette{\smallrel@{#1}}{#2}}%
}
\newcommand*{\smallrel@}[3]{%
  \sbox0{$#2\vcenter{}$}%
  \dimen@=\ht0 %
  \raise\dimen@\hbox{%
    \scalebox{#1}{%
      \raise-\dimen@\hbox{$#2#3\m@th$}%
    }%
  }%
}

\definecolor{mycolour}{HTML}{D62728}

\makeatletter
\newcommand{\optionaldesc}[2]{%
  \phantomsection
  #1\protected@edef\@currentlabel{#1}\label{#2}%
}
\makeatother

\title[M68 stream DESI spectroscopy]{Improved constraints on the Milky Way potential using the M68 stream and DESI spectroscopic data}

\author[C. G. Palau, W. Wang, J. Han]{Carles~G.~Palau\orcid{0000-0002-7583-534X}$^{1,2}$\thanks{E-mail: cgpalau@sjtu.edu.cn},
Wenting~Wang\orcid{0000-0002-5762-7571}$^{1,2}$\thanks{E-mail: wenting.wang@sjtu.edu.cn},
Jiaxin~Han\orcid{0000-0002-8010-6715}$^{1,2}$\thanks{E-mail: jiaxin.han@sjtu.edu.cn},
J.~Aguilar$^{3}$,
S.~Ahlen\orcid{0000-0001-6098-7247}$^{4}$,\newauthor
F.~Beutler\orcid{0000-0003-0467-5438}$^{5}$,
D.~Bianchi\orcid{0000-0001-9712-0006}$^{6,7}$,
D.~Brooks$^{8}$,
A.~Carnero Rosell\orcid{0000-0003-3044-5150}$^{9,10}$,
F.~J.~Castander\orcid{0000-0001-7316-4573}$^{11,12}$,\newauthor
T.~Claybaugh$^{3}$,
A.~Cuceu\orcid{0000-0002-2169-0595}$^{3}$,
A.~de la Macorra\orcid{0000-0002-1769-1640}$^{13}$,
J.~E.~Forero-Romero\orcid{0000-0002-2890-3725}$^{14,15}$,\newauthor
E.~Gaztañaga\orcid{0000-0001-9632-0815}$^{11,16,12}$,
Satya~{Gontcho A Gontcho}\orcid{0000-0003-3142-233X}$^{17}$,
G.~Gutierrez$^{18}$,
J.~Guy\orcid{0000-0001-9822-6793}$^{3}$,
D.~Kirkby\orcid{0000-0002-8828-5463}$^{19}$,\newauthor
A.~Kremin\orcid{0000-0001-6356-7424}$^{3}$,
M.~Landriau\orcid{0000-0003-1838-8528}$^{3}$,
L.~Le~Guillou\orcid{0000-0001-7178-8868}$^{20}$,
A.~Leauthaud\orcid{0000-0002-3677-3617}$^{21,22}$,
G.~E.~Medina\orcid{0000-0003-0105-9576}$^{23}$,\newauthor
A.~Meisner\orcid{0000-0002-1125-7384}$^{24}$,
R.~Miquel$^{25,26}$,
S.~Nadathur\orcid{0000-0001-9070-3102}$^{16}$,
H.~E.~Noriega\orcid{0000-0002-3397-3998}$^{27,13}$,
E.~Paillas\orcid{0000-0002-4637-2868}$^{28,29}$,\newauthor
N.~Palanque-Delabrouille\orcid{0000-0003-3188-784X}$^{30,3}$,
W.~J.~Percival\orcid{0000-0002-0644-5727}$^{31,32,33}$,
F.~Prada\orcid{0000-0001-7145-8674}$^{34}$,
I.~P\'erez-R\`afols\orcid{0000-0001-6979-0125}$^{35}$,\newauthor
C.~Ravoux\orcid{0000-0002-3500-6635}$^{36}$,
G.~Rossi$^{37}$,
R.~Ruggeri\orcid{0000-0002-0394-0896}$^{38}$,
E.~Sanchez\orcid{0000-0002-9646-8198}$^{39}$,
C.~Saulder\orcid{0000-0002-0408-5633}$^{40}$,
D.~Schlegel$^{3}$,\newauthor
M.~Schubnell$^{41,42}$,
M.~Siudek\orcid{0000-0002-2949-2155}$^{12,10}$,
G.~Tarl\'{e}\orcid{0000-0003-1704-0781}$^{42}$,
B.~A.~Weaver$^{24}$,
and H.~Zou\orcid{0000-0002-6684-3997}$^{43}$
}

\date{Accepted XXX. Received YYY; in original form ZZZ}

\pubyear{2026}

\begin{document}
\label{firstpage}
\pagerange{\pageref{firstpage}--\pageref{lastpage}}
\maketitle

\begin{abstract}
We present a selection of stars belonging to the stellar stream of the M68 (NGC 4590) globular cluster, also known as Fjörm. This star selection is an improvement on previous ones that used only \textit{Gaia} data, as it incorporates spectroscopic measurements from the DESI survey and photometric data from the DESI Legacy Surveys. The selection contains 96 stars, each with five phase-space parameters from \textit{Gaia}-DR3 and radial velocity from DESI, covering the entire observed section of the stream. This constitutes the largest selection of M68 stream stars with measured radial velocities to date. The observed stream is wider than expected from \nbody\ simulations, and the stars farthest from the centre of the stream appear to be correlated in radial velocity space. This suggests that these stars cannot have been stripped from the cluster in a static axisymmetric potential. By modelling a mock sample of stream stars created using an \nbody\ simulation, we found that we could reliably constrain the disc mass $M_{\rm d}$ and the dark matter halo axis ratio $q_{\rm h}$ of the Milky Way. This is because the stream flows close to and almost parallel to the disc. Using the 44 stars that are consistent with having been stripped from the cluster, combined with measurements of the Milky Way's rotation curve, we constrain the Galactic potential, obtaining $M_{\rm d} = 5.34 \pm 0.57 \pd{10}$~M$_{\Sun}$ and an oblate halo of $q_{\rm h} = 0.83^{+0.06}_{-0.05}$. Additionally, by fitting the stream track, we estimate the Heliocentric distance of M68 to be $r_{\Helio}^{\SM{M68}}=10.55\pm0.09$~kpc.
\end{abstract}

\begin{keywords}
Galaxy: kinematics and dynamics - Galaxy: structure - Galaxy: halo - Globular clusters: M68.
\end{keywords}

\defcitealias{2008gady.book.....B}{BT08}
\defcitealias{2019MNRAS.488.1535P}{PM19} 
\defcitealias{2023MNRAS.524.2124P}{PM23} 
\defcitealias{2025MNRAS.539.2718P}{P25} 
\defcitealias{2026MNRAS.545f1974P}{P26} 



\section{Introduction}\label{introduction}

The \textit{Gaia} mission \citep{2016A&A...595A...1G} provides astrometric and photometric data for about 1.5 billion stars across the entire sky. Several tens of stellar streams in the Solar neighbourhood have been discovered using this catalogue \citep{2023MNRAS.520.5225M}. These streams form when stars are stripped from their progenitor clusters by tidal forces while orbiting the Milky Way (MW) \citep[e.g.][]{2025NewAR.10001713B}. The discovery of these structures opens up many new lines of research \citep[e.g.][]{2025NewAR.10001721H}. For instance, streams can be used to study the mass loss and internal dynamics of their progenitor clusters \citep[e.g.][]{2021NatAs...5..957G, 2025ApJ...980L..18C}, as well as the structure and formation history of the Milky Way \citep[e.g.][]{2021MNRAS.501.2279V, 2025MNRAS.541..214D, 2026arXiv260507918B}. They can also be used to constrain the potential of the Galaxy, as they approximately follow the orbit of their progenitor cluster \citep[e.g.][]{1999ApJ...512L.109J, 2013MNRAS.433.1826S, 2023MNRAS.521.4936K, 2025ApJ...985L..22N}.

The third release of the \textit{Gaia} catalogue (GDR3) \citep{2023A&A...674A...1G} contains spectroscopic data and radial velocity estimates for bright stars \citep{2023A&A...674A...5K}. This data can be complemented by incorporating dedicated spectroscopic surveys dedicated to fainter stars. The Dark Energy Spectroscopic Instrument (DESI) is a multi-object spectrograph located at the Mayall 4~m telescope at the Kitt Peak National Observatory \citep{2022AJ....164..207D}. It can obtain spectra of almost 5000 objects simultaneously over a field of $\Approx 3$~deg$^2$ \citep{2016arXiv161100037D, 2024AJ....168...95M, 2024AJ....168..245P}. This instrument is currently taking measurements of about 7 million stars at Galactic latitudes $|b|>20$~deg \citep{2023AJ....166..259S}. This data focuses on thick disc and halo stars and includes member stars of stellar streams. Together with the DESI Legacy Surveys \citep{2019AJ....157..168D}, which provide star magnitudes in the optical bands, these data allow us to improve our current star selections in stellar streams.

The observed section of the stellar stream produced by the disruption of the globular cluster M68 (NGC 4590) is one of the longest and most populated streams within the DESI footprint. Discovered by \citet{2019ApJ...872..152I} and named Fjörm, this stream was associated with its progenitor cluster by \citet{2019MNRAS.488.1535P}, hereafter \citetalias{2019MNRAS.488.1535P}. Its length and velocity dispersion are comparable to those of the GD-1 \citep[e.g.][]{2025ApJ...980...71V, 2026arXiv260420958J} and Palomar 5 \citep[e.g.][]{2020ApJ...889...70B} stellar streams. The main differences are that the M68 stream is much closer to the Sun than Palomar 5, and unlike GD-1, it has a known progenitor. The most recent star selection was made by \citet{2026MNRAS.545f1974P}, hereafter \citetalias{2026MNRAS.545f1974P}, using GDR3 data and only includes radial velocity measurements for two stars obtained with the \textit{Gaia} spectrometer. This selection can be improved by using DESI photometry and spectroscopy to remove erroneously selected foreground stars and include previously discarded stream stars. Furthermore, precise spectroscopic data of the stream, such as radial velocities and metallicities, will improve our understanding of its structure, origin, and evolution.

Several mass models of the Milky Way have been introduced in recent years \citep[\S7.7 in][]{2025NewAR.10001721H}. Due to the difficulty of obtaining observational data that can constrain the Galactic potential beyond the disc, these models generally assume a spherical dark matter halo. However, the quality of observational data has improved significantly since the publication of the \textit{Gaia} catalogue and other spectroscopic surveys such as DESI. Furthermore, several methodologies have been developed to constrain the potential beyond the Galactic plane. For example, the dark halo has been constrained by assuming a distribution of halo stars \citep{2019MNRAS.485.3296W, 2021MNRAS.508.5468H} and disc stars \citep{2020MNRAS.494.6001N, 2024A&A...689A.280B} in equilibrium. Similarly, stellar streams have been used for this purpose \citep[e.g.][]{2016ApJ...833...31B, 2019MNRAS.483.4724F, 2019MNRAS.486.2995M, 2021MNRAS.502.4170R, 2024ApJ...967...89I}. Despite efforts to improve methodology and quality of observational data, the exact shape of the dark matter halo remains unclear.

The M68 stream provides a unique opportunity to determine the shape of the inner dark matter halo. This is because the stream is located close to the Galactic centre and flows almost parallel to the disc, at a distance of about 5 kpc from the symmetry plane. \citet{2023MNRAS.524.2124P}, hereafter \citetalias{2023MNRAS.524.2124P}, used this stream, together with those of NGC~3201 and Palomar~5, for this purpose. They created a model of the stream and varied parameters that characterise the phase-space position of the Sun, the clusters, and the Galactic potential, in order to determine the configuration of the model that best fits the observational data. In this paper, we build on their methodology and test its validity and accuracy using \nbody\ simulations and mock observational catalogues. Then, we use the improved star selection obtained using DESI data to further constrain the Galactic potential.

To this end, we have structured the present paper as follows: Section~\ref{selection} presents the selection methodology and the selected stars belonging to the M68 stream. Section~\ref{val_const_meth} introduces the constraining methodology and determines which parameters can be constrained and to what degree of accuracy. Section~\ref{obs_constraints} presents the constraints obtained using the observational data of the stream, and Section~\ref{conclusion} the conclusion.

\section{M68 stream star selection}\label{selection}

We use the spectroscopic data of stars within the Milky Way observed by the DESI collaboration. These stars are part of the DESI Milky Way Survey (MWS) \citep{2023ApJ...947...37C}, a programme that mainly operates during nights with bright moonlight, known as bright time. We use the internal data of the DESI MWS, which includes data collected during the first three years of observations. This survey builds upon the publicly available DESI Data Release 1 catalogue\footnote{\url{https://data.desi.lbl.gov/doc/releases/dr1/vac/mws/}} \citep{2025JCAP...07..028A, 2026AJ....171..285D, 2026OJAp....955260K}, extending it from $\Approx4$ to $\Approx12$ million stars. Its main star sample includes sources of DESI $r$-band magnitude within the range $16 < r < 19$~mag, and for specially selected targets, such as Blue Horizontal Branch (BHB) stars, it extends down to $21$~mag.

The radial velocities and atmospheric parameters contained in this catalogue are produced by the RVS pipeline (\S6.1 in \citealt{2023ApJ...947...37C}, \citealt{2023AJ....165..144G}), and are provided in the \texttt{RVTAB} table\footnote{\url{https://desi-mws-dr1-datamodel.readthedocs.io/en/latest/mwsall.html\#hdu1}} of the \texttt{mwsall} database. We select the objects from the database that fulfil the following conditions:
\begin{enumerate}
    \setlength\itemsep{0.4em}
    \item[(1)] \texttt{RR\_SPECTYPE} = \texttt{STAR}
    \item[(2)] \texttt{PRIMARY} = True
    \item[(3)] \texttt{RVS\_WARN} = 0
\end{enumerate}
The first condition ensures that the object is a star. The second condition eliminates repeated measurements of a single star by selecting the one with the highest quality. The third condition ensures that the observations are well fitted by a stellar template, and that the star radial velocity or Heliocententric line of sight velocity $v_r \in \Range{-1495}{1495}$~km~s$^{-1}$, with an uncertainty $\sigma_{v_r}<100$~km~s$^{-1}$.

\begin{figure*}
    \includegraphics[width=1.0\textwidth]{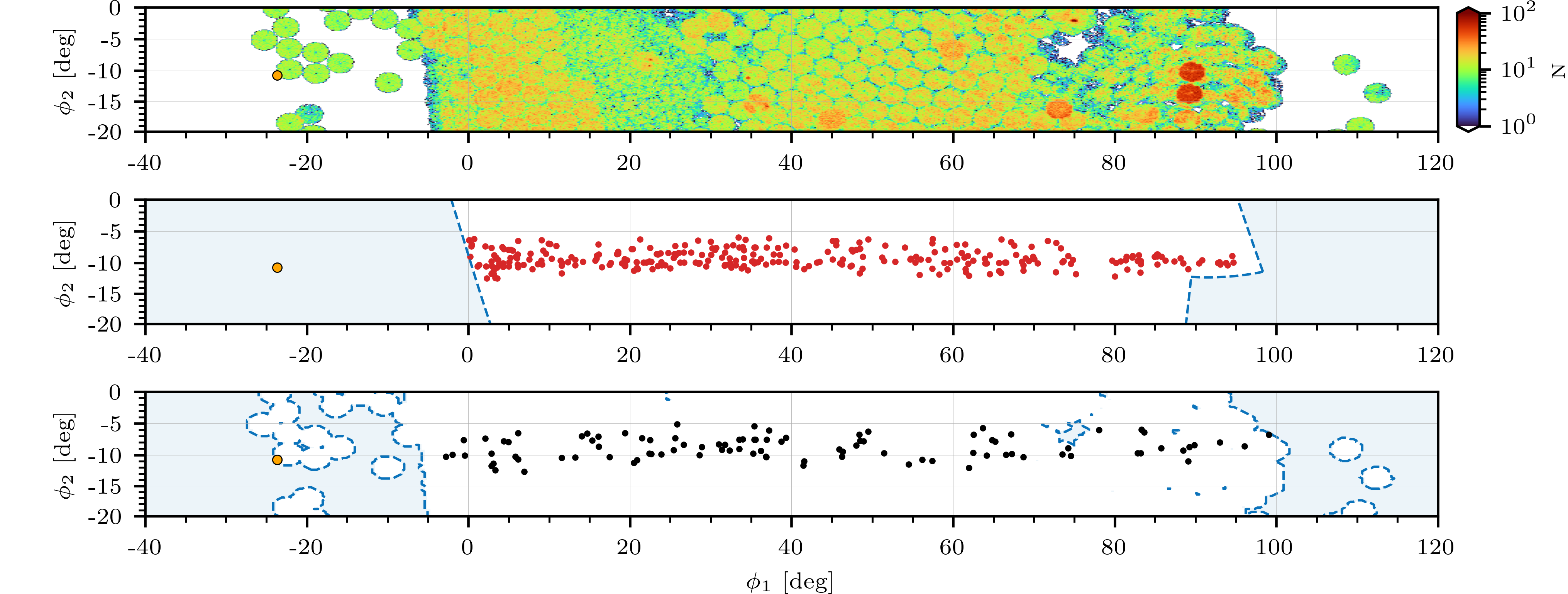}
    \caption{Sky coordinates aligned with the M68 stellar stream ($\phi_1, \phi_2$). The position of the M68 globular cluster is indicated with a large orange dot. \textit{Top:} Number of DESI stars $N$ that pass the cuts (1-3) in bins of area of $\Approx0.0207$~deg$^2$. \textit{Middle:} The red dots mark the $291$ stream star candidates selected using only GDR3 data (\S4 in \citetalias{2026MNRAS.545f1974P}). The areas with high level of star contamination are shaded in blue, and their boundaries are highlighted with dashed blue lines. \textit{Bottom:} The black dots mark the $96$ stars that pass the cuts (1-16) and constitute the DESI star selection (Section~\ref{final_sel}). The areas with no DESI data are shaded in blue, and their boundaries are highlighted with dashed blue lines.}
    \label{surface_density}
\end{figure*}

In the top panel of Figure~\ref{surface_density}, we plot the $\Approx11$ million stars that pass the above cuts in bins of area of $\Approx0.0207$~deg$^2$. The position of the M68 globular cluster is indicated with a large orange dot. The data are shown in sky coordinates aligned with the stream ($\phi_1, \phi_2$). These coordinates are obtained by rotating ($\delta, \alpha$) so that the curvature of the projected stream is minimised, and the closest extreme to the progenitor cluster is located at the origin (Eq. A5 in \citetalias{2026MNRAS.545f1974P}). The middle panel of Figure~\ref{surface_density} shows the $291$ stream member candidates selected using only GDR3 data (\S4 in \citetalias{2026MNRAS.545f1974P}). The blue shaded areas, which are bounded by blue dashed lines, show the sky zones that have been excluded due to a high level of foreground star contamination. We observe that the DESI footprint approximately covers the angular range $\phi_1\in\Range{-5}{100}$~deg, which is roughly equal to the range extended by the observable part of the stream.

The M68 stream has a low surface brightness. Consequently, it cannot be observed when projected onto the sky using a colour-magnitude filter. We have verified that incorporating radial velocities and metallicities from the DESI catalogue into the filter, does not enable us to effectively separate the stream stars from the foreground. For this reason, we restrict our data to stars whose astrometric properties have been measured by the \textit{Gaia} mission. In particular, it is the proper motions of the stars that enable us to observe the stellar stream. This is because the proper motions increase as the stream approaches the Sun and become significantly larger than those of the foreground stars. (Fig.~4 in \citetalias{2019MNRAS.488.1535P}). The proper motions are largest at the point of closest approach, located at $\phi_1\simeq 39.8$~deg, which is about $5.1$ kpc from the Sun. This point lies at the centre of the visible section of the stream. Furthermore, this section has the highest apparent surface density of the entire stream (Fig.~16 in \citetalias{2026MNRAS.545f1974P}), and it is easier to observe because it is projected onto the halo, an area with low star contamination.

The GDR3 measurements for each DESI object are provided in the \texttt{GAIA} table\footnote{\url{https://desi-mws-dr1-datamodel.readthedocs.io/en/latest/mwsall.html\#hdu5}} of the \texttt{mwsall} database. We select the DESI stars that fulfil the following \textit{Gaia} quality conditions\footnote{\url{https://gea.esac.esa.int/archive/documentation/GDR3/Gaia_archive/chap_datamodel/sec_dm_main_source_catalogue/ssec_dm_gaia_source.html}}:
\begin{enumerate}
    \setlength\itemsep{0.5em}
    \item[(4)] \texttt{bp\_rp} $\neq$ Null
    \item[(5)] \texttt{ruwe} $<1.2$
    \item[(6)] \texttt{visibility\_periods\_used} $\geqslant10$
    \item[(7)] \texttt{duplicate\_source} $=$ False,
\end{enumerate}
and reduce the amount of data by selecting stars with:
\begin{enumerate}
    \setlength\itemsep{0.5em}
    \item[(8)] \texttt{parallax} $<1/0.3$ mas
    \item[(9)] $-40 \leqslant \phi_1 \leqslant 120$ deg
    \item[(10)] $-20 \leqslant \phi_2 \leqslant 0$ deg.
\end{enumerate}
In order to remove stars with low proper motions, we define a volume in phase-space that follows the cluster orbit. This volume is defined by a bundle of orbits computed using various initial conditions for the cluster and different potentials of the Galaxy. The stars are considered to be Gaussian distributions in phase-space, and are defined by the mean values and uncertainties of the GDR3 observations. We calculate the intersection of each star with the phase-space volume, which is denoted by $P_\SM{REG}$, and select those with intersections above a certain threshold. The evaluation of the intersection does not include radial velocities. This is because we analyse them independently. This method was introduced in \citetalias{2019MNRAS.488.1535P} and is used here with the parameters listed in the App. E of \citetalias{2026MNRAS.545f1974P}.

In this case, we select the stars that satisfy the following condition:
\begin{enumerate}
    \item[(11)] $P_\SM{REG} \geqslant 5\pd{-9}$ ${\rm yr}^3 \, {\rm deg}^{-2} \, {\rm pc}^{-1} \, {\rm mas}^{-3}$.
\end{enumerate}
This low threshold is the main difference compared to the GDR3 star selection in \S4 of \citetalias{2026MNRAS.545f1974P}. This choice is justified by the use of the more complete and precise DESI spectroscopic and photometric data to separate the stream stars from the foreground. The chosen value is small enough to select stars within limits of $|\mu_{\alpha\ast}|, |\mu_{\delta}| \LessSim 10$~mas~yr$^{-1}$. This ensures that no stream stars are lost, as the observable section of the stream lies within $-4 \LessSim \mu_{\alpha\ast} \LessSim 5$ and $0 \LessSim \mu_{\delta} \LessSim 7$~mas~yr$^{-1}$ (Fig. 14 in \citetalias{2026MNRAS.545f1974P}).

\begin{figure}
\includegraphics[width=1.0\columnwidth]{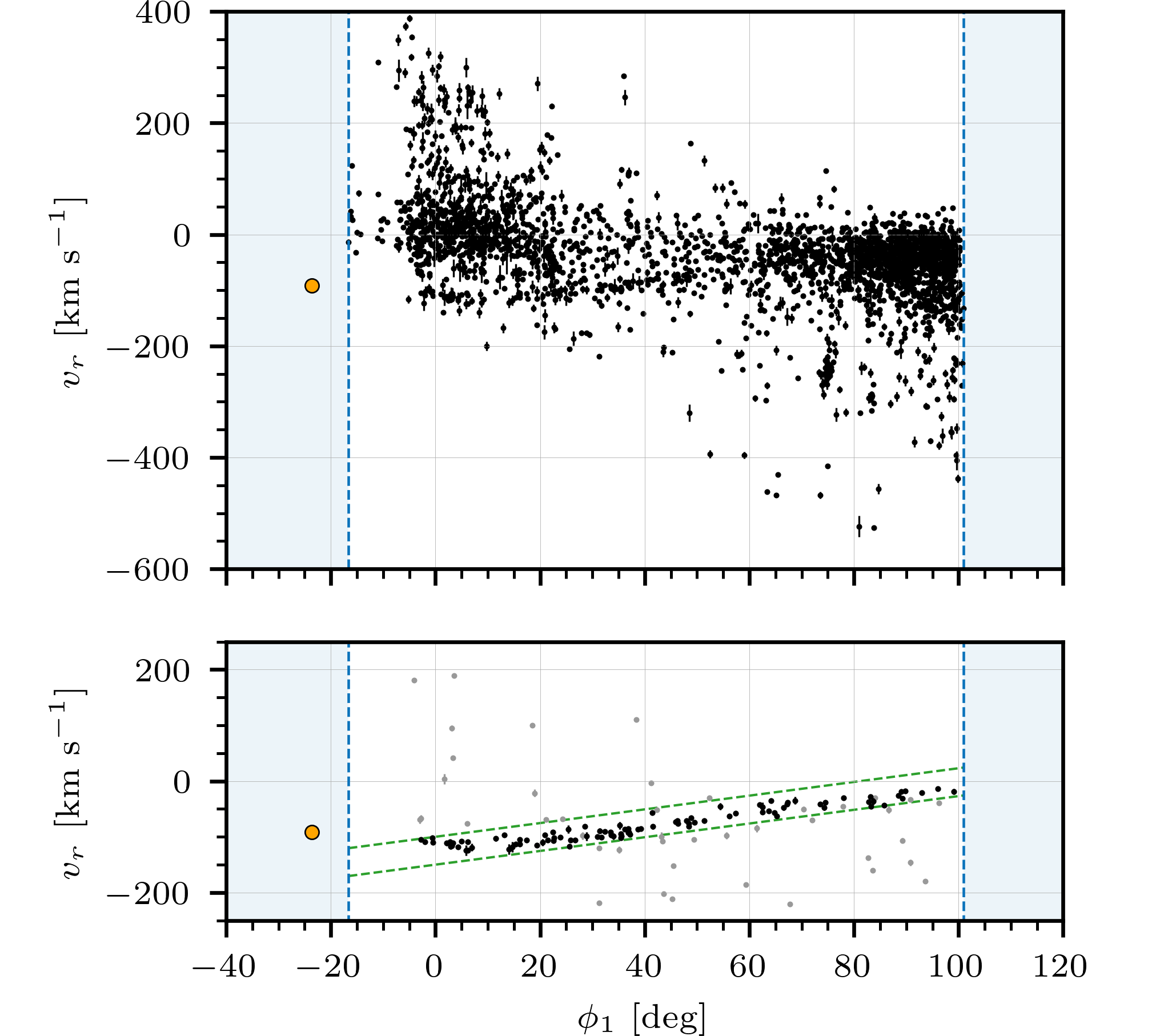}
\caption{Radial velocity $v_r$ against sky coordinate aligned with the M68 stellar stream $\phi_1$. The position of the M68 globular cluster is indicated with a large orange dot. The DESI data is only available within the approximate range $\phi_1 \in \Range{-16.7}{101}$~deg. The areas with no DESI data are shaded in blue and its boundaries are highlighted with dashed blue lines. \textit{Top:} The black dots with uncertainties mark the $2707$ stars with DESI radial velocity that pass the cuts (1-11). \textit{Bottom:} The grey dots with uncertainties mark the $140$ stars that pass the cuts (1-14). The black dots mark the $96$ stars in the DESI selection (Section~\ref{final_sel}). The green dashed lines mark the boundaries of the stream defined in the cut (13).}
\label{phi_vr}
\end{figure}

A total of $2707$ stars with DESI radial velocity pass the cuts (1-11). In the top panel of Figure~\ref{phi_vr}, we plot the radial velocities $v_r$ against the sky coordinate $\phi_1$   of these stars, including their observational uncertainties. The position of the M68 cluster is indicated with a large orange dot, and the areas with no data are shaded in blue. The stream is visible as a linear overdensity spanning from $\phi_1\in\Range{-5}{80}$~deg, with $v_r\in\Range{-120}{-10}$~km~s$^{-1}$. The $\phi_1\in\Range{-5}{25}$ and $\phi_1\in\Range{80}{100}$~deg intervals are contaminated by foreground stars because they are projected closer to the disc than the centre of the stream. Most of these stars can be filtered out based on colour, magnitude and metallicity.

\begin{figure}
\includegraphics[width=1.0\columnwidth]{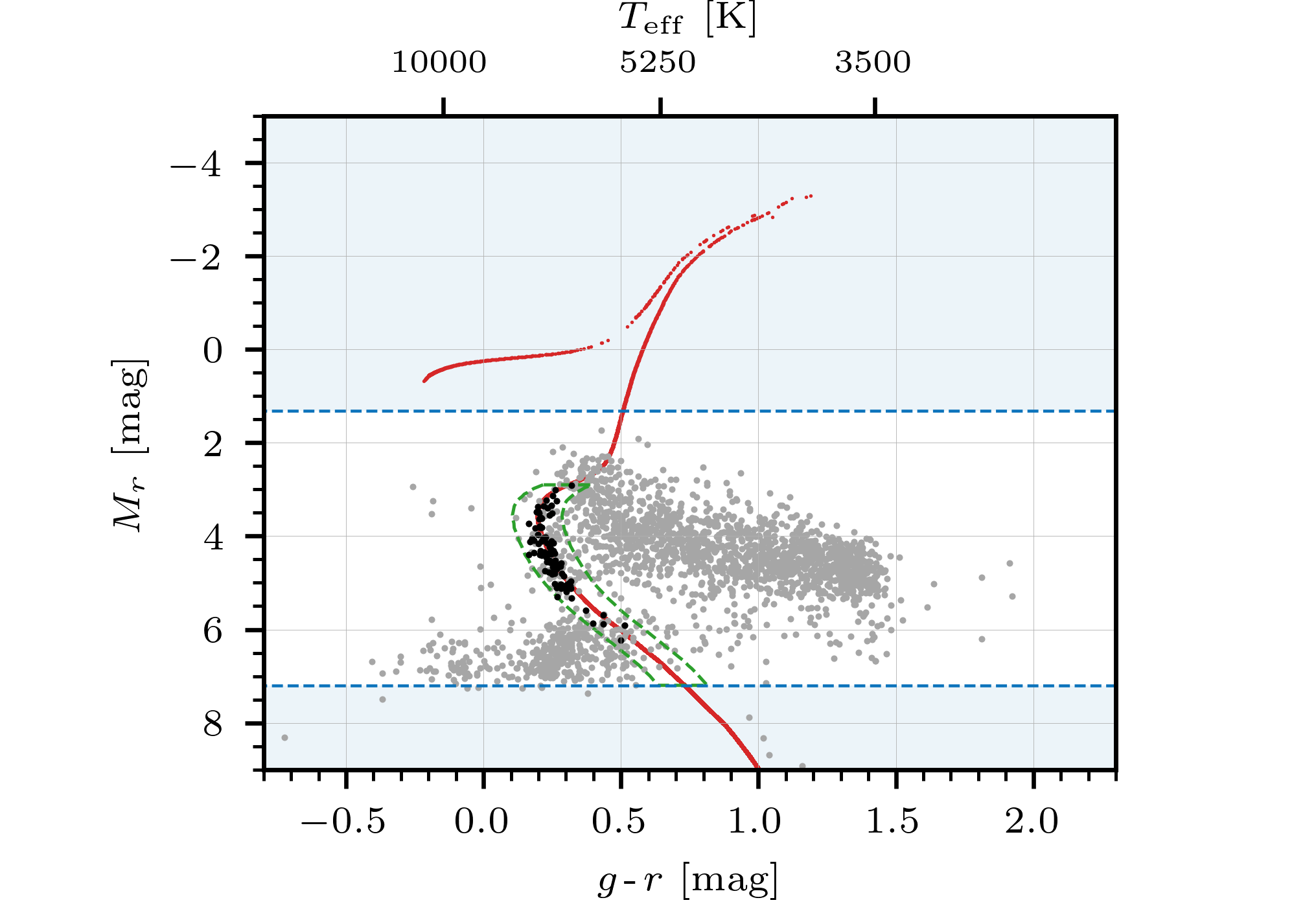}
\caption{Colour Magnitude Diagram (CMD) in DESI photometry. The absolute magnitudes $M_{\rm r}$ are computed using the estimated distances of the stars from the cluster orbit. The colour index $g\;\!\textnormal{-}\;\!r$ includes a correction for dust reddening. The $2179$ stars that pass the cuts (1-11) are shown as grey dots, and the $96$ stars in the DESI selection (Section~\ref{final_sel}) are shown in black. A synthetic stellar population of the globular cluster M68, generated by PARSEC/COLIBRI, is shown in red. The boundaries of the area containing the stars compatible with the CMD are marked with dashed green lines. The areas outside the DESI and \textit{Gaia} limiting magnitudes are shaded in blue, and their boundaries are highlighted with dashed blue lines.}
\label{cmd}
\end{figure}

In order to determine the stars compatible with the position of M68 in the Colour-Magnitude Diagram (CMD), we generate a synthetic stellar population following a theoretical isochrone from PARSEC/COLIBRI\footnote{\url{https://stev.oapd.inaf.it/PARSEC/index.html}} in \textit{Gaia}-EDR3 photometric system. We use the same properties of the cluster and free parameters of the model as in \S 2.1 of \citet{2025MNRAS.539.2718P}, hereafter \citetalias{2025MNRAS.539.2718P}. We transform the \textit{Gaia} photometry to $r$ and $\gr$ in the DESI/DECaLS Legacy Survey (AB) photometric system using polynomial approximations. Specifically, we adopt:
\begin{equation}\label{G_to_r}
 r \simeq -0.1230 + 1.0026\,(G\:\!)
\end{equation}
and
\begin{multline}
\gr \simeq -0.2352 + 0.7533\,(\BPRP) - 0.1399\,(\BPRP)^2 \\+ 0.3734\,(\BPRP)^3 - 0.1438\,(\BPRP)^4,
\end{multline}
which is valid within the limits $0<\BPRP<2$~mag. These approximations have a small systematic errors of $\Approx0.05$~mag with a median absolute deviation of $\Approx0.03$~mag. In Figure~\ref{cmd}, we show as red dots the colour index $\gr$ and the absolute magnitude in the DESI $r$-band $M_r$ of the synthetic stellar population.

The \texttt{FIBERMAP} database\footnote{\url{https://desi-mws-dr1-datamodel.readthedocs.io/en/latest/mwsall.html\#hdu3}} contains photometric measurements for $2179$ stars that pass the cuts (1-11). The flux $f$ is provided in nano-maggies, which we convert to magnitudes $m$ using the following expression:
\begin{equation}
    m = 22.5 - 2.5\log_{10}\!\var{f}.
\end{equation}
Approximately $70$ per cent of these stars are within limits of $16\LessSim r \LessSim 19$~mag, with the remainder within $19 \LessSim r \LessSim 21$~mag. The mean uncertainties for these measurements are $\sigma_{r}\simeq6$ and $\sigma_{\gr}\simeq10$~mmag. These uncertainties are significantly smaller than those of \textit{Gaia}. We therefore use DESI photometry to select stars that are compatible with the population model of M68. In order to compare the observed stars with the synthetic stellar population, we correct the magnitude $r$ and colour index $\gr$ for reddening extinction using the following colour excess:
\begin{equation}\label{red}
    E\var{\gr} = 1.049\:\!E\var{\BV} + C,
\end{equation}
and extinction correction coefficient:
\begin{equation}\label{red}
    A_r = 2.704\:\!E\var{\gr} - 0.123
\end{equation}
where $E\var{\BV}$ is obtained from the SFD Galactic dust map \citep{1998ApJ...500..525S}, the proportionality factor $1.049$ is obtained from \citet{2025OJAp....8E..83Z}\footnote{\url{https://data.desi.lbl.gov/doc/releases/dr1/vac/stellar-reddening/}}, and $A_r$ is obtained by converting $A_{\rm G}=1.98E\var{\BPRP}$ from \citet{2010A&A...523A..48J} using Eq.~\ref{G_to_r}. There is a slight discrepancy between the synthetic stellar population and the M68 stream stars observed by DESI. We attribute this discrepancy to an inaccurate reddening correction. We have therefore included a correction factor of $C=0.0167$~mag in Eq.~\ref{red} to improve the fit of the final star selection to the simulated isochrone. The absolute magnitude of a star is determined by estimating the Heliocentric distance from the closest point of the cluster orbit to the star. This method is explained in \S 5.3 of \citetalias{2025MNRAS.539.2718P}. Using a simulation of the stream, we estimated a deviation of $\Approx\pm100$~pc between the exact distance of the stars and the distance obtained from the cluster orbit. In addition, we expect a slight systematic deviation for stars at $\phi_1\GtrSim60$~deg if the real Milky Way potential is not accurately represented by the potential used to compute the orbit. However, we assume that these discrepancies are negligible given that the observed section of the stream is located between approximately $5$ and $10$~kpc from the Sun.

Figure~\ref{cmd} shows the stars that pass the cuts (1-11) as grey dots. These stars are contained within the limits $1.31 \LessSim M_r \LessSim 7.19$~mag, which are indicated by blue dashed lines. The lower limit is determined by assuming a star with a magnitude equal to the DESI limit of $r=16$~mag \citep[\S 3.1.2 in][]{2026OJAp....955260K}, located at a distance of $r_{\Helio}\simeq8.7$~kpc, which is the maximum distance at which stars in the observable section of the stream can be found. Similarly, the upper limit is determined for a star with a magnitude equal to the \textit{Gaia} limit $G\simeq20.84$~mag, located at $r_{\Helio} \simeq 5.1$~kpc, corresponding to the point of closest approach of the stream to the Sun. Stars with magnitudes $M_r$ under the upper limit are faint stars whose $r$-band magnitude significantly deviates from their expected \textit{Gaia} $G$ magnitude according to Eq.~\ref{G_to_r}.

There is little overlap between the stars that pass the cuts (1-11) and the synthetic stellar population (red dots). The only area of significant overlap is close to the turnoff point for $\gr\Approx0.4$~mag and $M_r\Approx2.5$~mag. To exclude this region, we set up the lower limit of the area containing the stars consistent with the cluster to $M_r=2.9$~mag. Figure~\ref{cmd} shows the boundaries of this area as dashed green lines. This area is approximately $\pm0.1$~mag in width with respect to the stellar population model. This width accounts for the inaccuracies in the stellar population model, observational errors, and the intrinsic dispersion of the real cluster. These boundaries define a new condition for selecting the stream stars:
\begin{enumerate}
    \item[(12)] Compatible with the CMD of M68 using DESI photometry.
\end{enumerate}

\begin{figure}
\includegraphics[width=1.0\columnwidth]{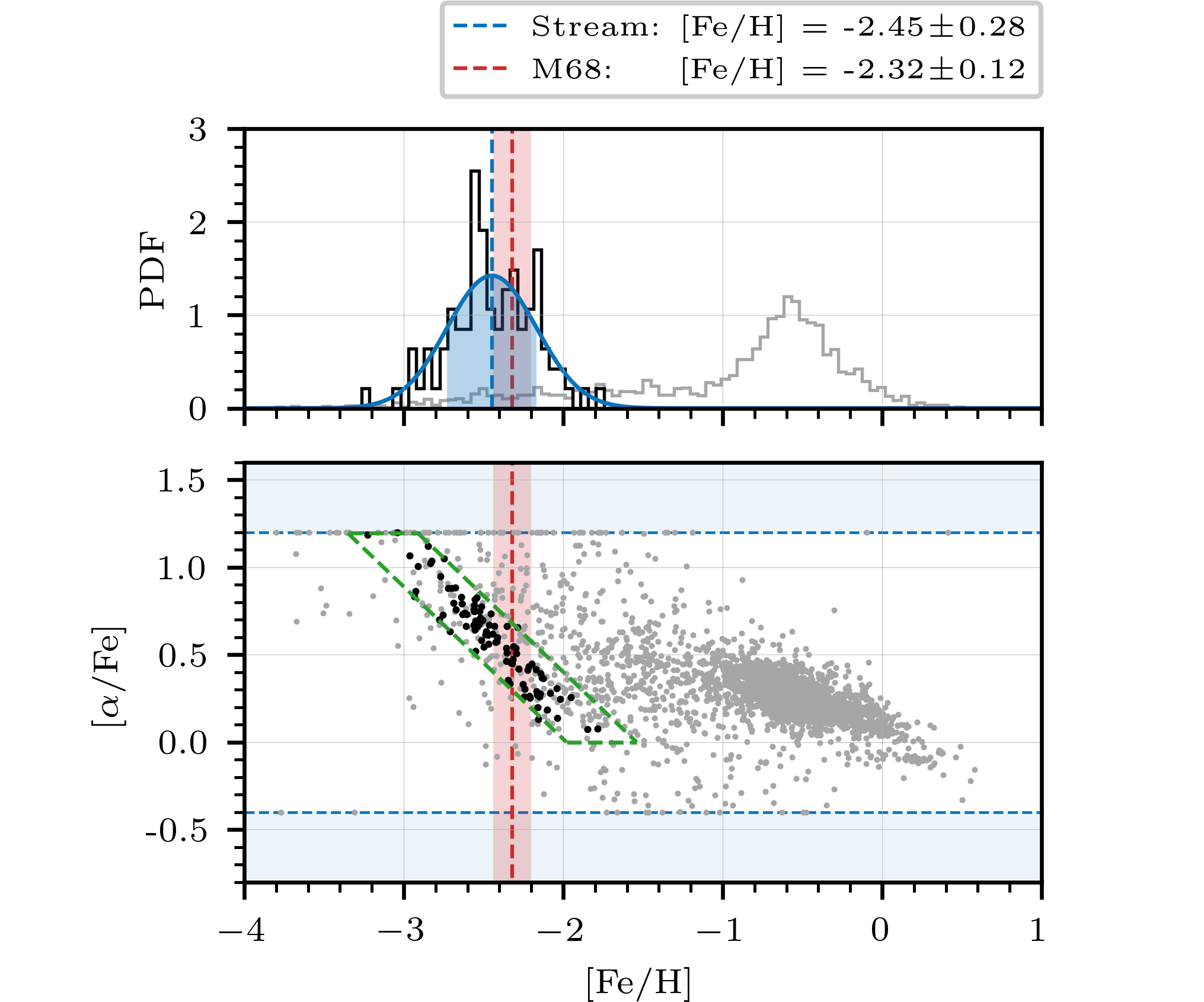}
\caption{\textit{Top:} Distributions of iron abundance $\FeH$. The grey histogram corresponds to the $2179$ stars that pass the cuts (1-11), and the black histogram corresponds to the $96$ stars in the DESI selection (Section~\ref{final_sel}). The blue line shows the best-fitting Gaussian distribution to the DESI selection. Its mean is marked with a dashed blue line, and its standard deviation with a shaded blue area. The dashed red line indicates the metallicity of M68 as measured by \citet{2023MNRAS.519..192W}, and the shaded red area indicates the error bars. \textit{Bottom:} Iron abundance $\FeH$ against alpha-elements abundance $\AlphaFe$. The stars that pass the cuts (1-11) are shown as grey dots, and the stars in the DESI selection are shown as black dots. The areas outside the DESI observational limits are shaded in blue, and their boundaries are highlighted with dashed blue lines. The boundaries of the area containing the stars compatible with the metallicity of the progenitor cluster are marked with green dashed lines.}
\label{metallicity}
\end{figure}

The \texttt{RVTAB} database includes metallicity estimates for the DESI MWS stars. In the bottom panel of Figure~\ref{metallicity}, we plot the alpha-elements abundance $\AlphaFe$ against the iron abundance $\FeH$ of the stars passing the cuts (1-11) as grey dots. These stars surround the orbit of the cluster in phase-space, and are predominantly halo stars. Their observational uncertainties are generally small, with an average of $\sigma_{\FeH}\simeq0.085$ and $\sigma_{\AlphaFe}\simeq0.031$. We therefore do not show them in the plot. Note that $\AlphaFe$ is limited to the range $\Range{-0.4}{1.2}$. This results in a non-physical accumulation of stars at the boundaries, which are indicated by the dashed blue lines. The top panel of Figure~\ref{metallicity} shows the distribution of $\FeH$ as a grey histogram. Most of the stars are concentrated in a peak centred at $\FeH\Approx-0.55$, however, there are also a large number of low-metallicity outliers. Some of these outliers have a metallicity similar to that of the M68 globular cluster, which is estimated to be $\FeH=-2.32 \pm 0.12$ \citep{2023MNRAS.519..192W}. The mean of this measurement is shown as a dashed red line in both panels of Figure~\ref{metallicity}, and the associated uncertainty is shown as a shaded red area.

As can be seen in the bottom panel of the same figure, the abundances of iron and alpha-elements are correlated. For low-metallicity stars, we consider this correlation to be non-physical and attribute it to correlated observational uncertainties. Additionally, as \citet{2026OJAp....955260K} noted, the uncertainties for low-metallicity stars provided by the DESI MWS survey are underestimated. We cannot verify the presence of large dispersions and correlations in the globular cluster M68 because it has not been observed by DESI. However, in Appendix~\ref{App0}, we demonstrate that the globular cluster M92 (NGC 6341) exhibits similar characteristics. This suggests that these characteristics are caused by observational limitations rather than being intrinsic to the stream. Consequently, we do not restrict the star selection to those compatible with the metallicity of M68. Instead, we define a much broader selection criterion. We find that the stream stars can be isolated from the foreground stars by selecting those within the green dashed lines shown in the bottom panel of Figure~\ref{metallicity}. These boundaries are defined by the following conditions:
\begin{enumerate}
\setlength\itemsep{0.5em}
    \item[(13)] $0 < \AlphaFe < 1.2$
    \item[(14)] $-1 \leqslant \dfrac{\AlphaFe -a -b\,\FeH}{\epsilon} \leqslant 1$,
\end{enumerate}
where $a\simeq-1.534$, $b\simeq-0.872$, and $\epsilon=0.22$.

\begin{figure}
\includegraphics[width=1.0\columnwidth]{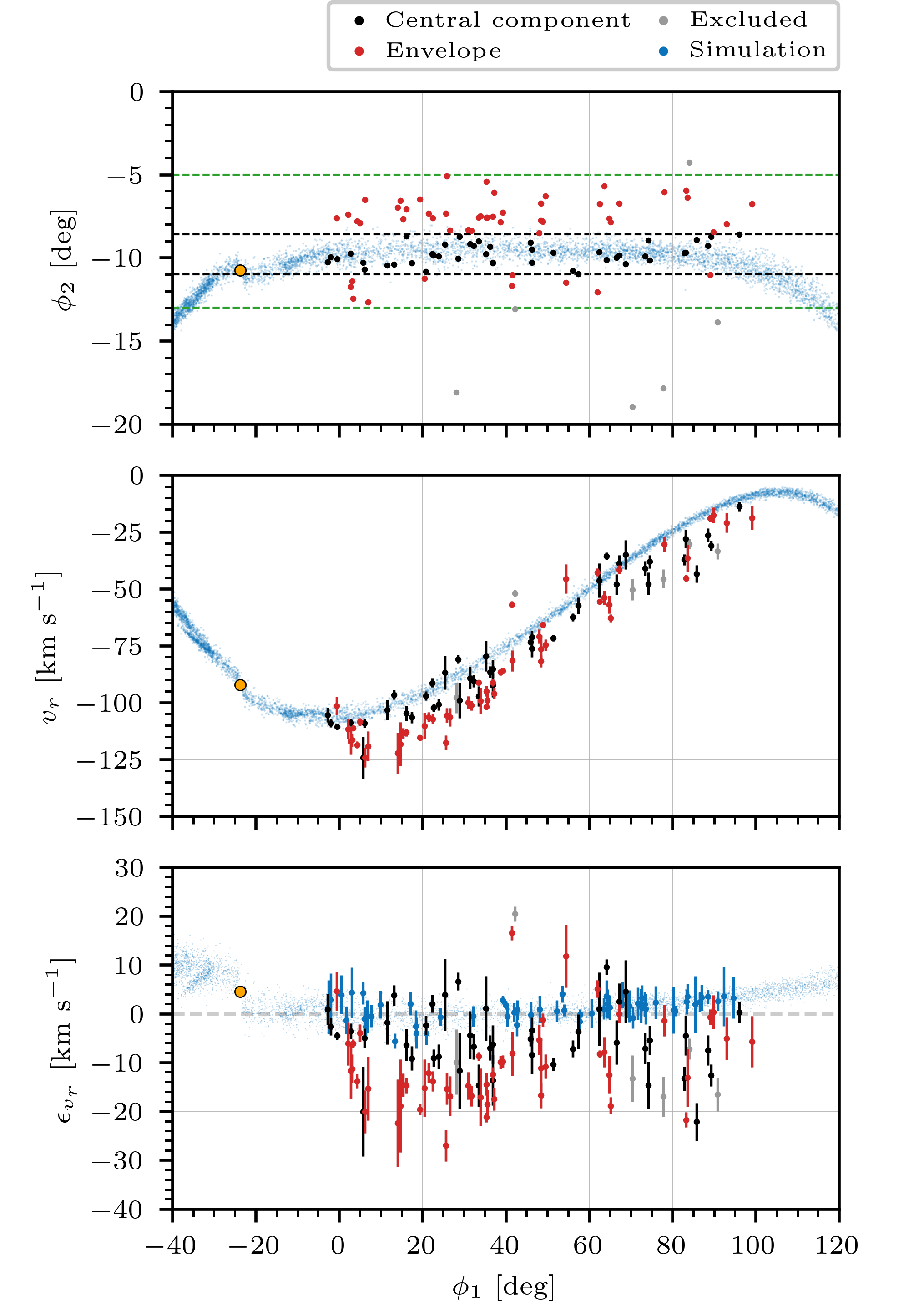}
\caption{The $44$ stars from the DESI selection are defined as the central component of the stream (black dots with error bars). The remaining $52$ stars from the DESI selection are defined as the envelope (red dots with error bars). The stars that pass the cuts (1-15) and are excluded by the cut (16) are shown as grey dots. The position of the M68 globular cluster is indicated by a large orange dot, and the \nbody\ simulation of its stream is shown as small blue dots. \textit{Top:} Sky coordinates aligned with the M68 stellar stream ($\phi_1, \phi_2$) with an aspect ratio of 4. The horizontal black dashed lines show the boundaries of the central component of the stream, and the green dashed lines show the boundaries of the envelope. \textit{Centre:} Radial velocity $v_r$ against the sky coordinate $\phi_1$. \textit{Bottom:} Difference between the radial velocity from the DESI MWS survey and the radial velocity estimated from the cluster orbit $\epsilon_{v_{r}}$ against the sky coordinate $\phi_1$. The value $\epsilon_{r_{\Heliosmall}}=0$ is highlighted in grey. The blue dots with error bars show a mock observational sample obtained from the \nbody\ simulation.}
\label{envelope}
\end{figure}

The selection based on colour, magnitude, and metallicity (12-14) eliminates $\Approx95$ per cent of the stars that pass the cuts (1-11). This yields a total of $140$ stream star candidates. These stars are shown as grey dots in the $(\phi_1, v_r)$ space in the bottom panel of Figure~\ref{phi_vr}. In the same panel, two green dashed lines mark the boundaries of the linear overdensity that constitutes the stream. We eliminate the stars outside these boundaries by including the following condition:
\begin{enumerate}
    \item[(15)] $-1 \leqslant \dfrac{v_r -a -b\,\phi_1}{\epsilon} \leqslant 1$,
\end{enumerate}
where $a=-125$~km~s$^{-1}$, $b=1.23$~km~s$^{-1}$~deg$^{-1}$, and $\epsilon=25$~km~s$^{-1}$. We finally remove six stars off-track by cutting in sky coordinate $\phi_2$:
\begin{enumerate}
    \item[(16)] $-13 \leqslant \phi_2 \leqslant -5$~deg.
\end{enumerate}
These boundaries are shown as green lines in the top panel of Figure~\ref{envelope}, and stars located beyond these limits are shown as grey dots.

\subsection{The DESI star selection}\label{final_sel}

A total of $96$ stars pass the cuts (1-16) and constitute the DESI star selection. Their $\phi$ coordinates are plotted in the bottom panel of Figure~\ref{surface_density}. The area with no DESI observations is shaded in blue, and its boundaries are marked with blue dashed lines. The length of the observable section of the stream in coordinate $\phi_1$ is limited by the DESI observational footprint. Conversely, the width of the stream in coordinate $\phi_2$ is fully covered. Within the DESI footprint boundaries, the stream surface density is limited by the completeness of the DESI catalogue, in terms of both photometry and spectroscopy.

The radial velocities of the DESI star selection are shown as black dots against the sky coordinates $\phi_1$ in the bottom panel of Figure~\ref{phi_vr}, and their colour and magnitude are also shown as black dots in Figure~\ref{cmd}. In the top panel of Figure~\ref{metallicity}, we show the distribution of $\FeH$ of the DESI star selection as a black histogram. The best-fitting Gaussian distribution is shown in blue, with its mean marked by a dashed blue line, and its standard deviation shown by a shaded blue area. The metallicity of the stream stars $\FeH=-2.45 \pm 0.27$ is consistent with that of the M68 globular cluster, which is shown in red in the same figure. In addition, our results are also consistent with the metallicity measurements obtained by the \textit{Pristine} survey for the Fjörm stream \citep[Fig.~6 in][]{2022MNRAS.516.5331M}. Its metallicity distribution lies within $\FeH\in\Range{-3.5}{-1.5}$, with a mean of $\langle\FeH\rangle=-2.33\pm0.05$.

\subsection{Additional stars}

The DESI star catalogue can be supplemented with stars for which DESI has measured the radial velocity and metallicity, but for which the DESI/DECaLS photometry is unavailable. Of the stars that pass cuts (1–11), $528$ fulfil this condition. Applying the cut (12) using \textit{Gaia} photometry to select stars that are compatible with the CMD of the M68 cluster, results in an additional $18$ stars passing cuts (1–16). We do not include these stars in the DESI selection since the GDR3 photometry is less precise than the DESI photometry. Consequently, there is a higher risk of including foreground stars in the final catalogue. We show the $\phi$ coordinates and radial velocities of these stars in Appendix~\ref{App1}.

In Appendix~\ref{App2}, we investigate the presence of radial velocity measurements for M68 stream stars in the Survey of Surveys (SoS) DR1 compilation \citep{2022A&A...659A..95T}, the \texttt{STREAMFINDER} survey \citep{2021ApJ...914..123I} and GDR3 catalogue. We check for stars included in the DESI selection and stars from the GDR3 selection (\S4 in \citetalias{2026MNRAS.545f1974P}) for which no complete DESI data is available. We found a total of $22$ stars with measured radial velocities that match our sample of stream stars. This implies that the DESI selection presented in this study is the most extensive catalogue of spectroscopically confirmed M68 stream star candidates to date. Furthermore, it is the most precise, with uncertainties in the radial velocity measurements with median and $1\sigma$ confidence intervals of $\sigma_{v_{r}}=2.77_{-1.39}^{+2.92}$~km~s$^{-1}$.

\subsection{Comparison to the GDR3 selection}\label{comparison_gdr3}

The GDR3 star selection (\S4 in \citetalias{2026MNRAS.545f1974P}) was made using data from the GDR3 catalogue only. This selection contains $291$ stars that are likely to be members of the M68 stream. Although the observed DESI stream is similar in length and width to the GDR3 selection (red dots, middle panel Figure~\ref{surface_density}), it appears sparser due to the smaller number of stars. Using DESI photometry and spectroscopy, we eliminated $8$ foreground stars from the GDR3 selection, and included $38$ that had been previously excluded due to their high proper motion observational errors. However, $225$ stars in the GDR3 star selection do not have complete DESI data, implying that their stream membership cannot be confirmed spectroscopically.

\subsection{Comparison to an \nbody\ simulation}\label{stream_width}

In Figure~\ref{envelope}, we compare the DESI stream stars to an \nbody\ simulation of the M68 stream. The simulation has been done using the collisional code PeTar \citep{2020MNRAS.497..536W}, following the methodology described in \S2.6 of \citetalias{2025MNRAS.539.2718P}, but assuming an accretion time for the cluster of $T=3.04$~Gyr. In the top panel, we plot the sky coordinates of the final selection as big coloured dots, and the simulated stars as small blue dots. We also show the position of the cluster as a large orange dot. We plot the $\phi$ coordinates with an aspect ratio of four, which implies that the stream appears wider in relation to its length. As it can be seen, the stream is much wider than the expected from the simulation, given the size of the progenitor cluster and the tidal forces caused by the Milky Way's potential model. In addition, as shown in \S4.1 of \citetalias{2026MNRAS.545f1974P}, approximately half of the stream stars appear to be projected over the cluster orbit. This implies that these stars cannot originate from M68 in an axisymmetric static potential, such as those considered in our study. These characteristics are robust to changes in the potential since the cluster's position is measured with high precision, and its orbit does not change significantly along the observable section of the stream because it is close to the cluster. This motivates the division of the stream into two groups, which are defined by the following boundaries:
\begin{equation}\label{main_comp_limits}
-11 \leqslant \phi_2 \leqslant -8.6 \,\, {\rm deg.}
\end{equation}
These boundaries are indicated as dashed black lines in the top panel of Figure~\ref{envelope}. The $44$ stars from the DESI selection within the boundaries are considered to be compatible with being stripped from the cluster, and defined as central component of the stream. These stars are shown as black dots. The $52$ stars outside the boundaries are defined as envelope, and are shown as red dots. A fraction of approximately $54$ per cent of the selected stream stars belong to the envelope.

In the central panel of Figure~\ref{envelope}, we plot the radial velocity with error bars against the sky coordinate $\phi_1$. The stars in the envelope (red) are found to have systematically lower radial velocities than the stars in the central component (black) within the approximate range $\phi_1\in\Range{0}{50}$~deg. Similarly, the stars in the envelope also have slightly larger proper motions than the stars in the central component (Fig.~14 in \citetalias{2026MNRAS.545f1974P}). There is no significant difference in metallicity between the two groups, being $\FeH=-2.47\pm0.29$ for the stars in the central component and $\FeH=-2.43\pm0.27$ for the stars in the envelope. There is also no difference in colour and magnitude between these stars. Therefore, the differences are only observed in phase-space.

In the bottom panel of Figure~\ref{envelope}, we plot the difference between the radial velocity measured by DESI and the radial velocity estimated from the cluster orbit $\epsilon_{v_{r}}$. The method of estimating phase-space coordinates from the cluster orbit is introduced in \S5.3 of \citetalias{2025MNRAS.539.2718P}, and includes a correction factor to subtract the offset between the orbit and the stream. This explains why the cluster appears with a $\epsilon_{v_{r}}=4.5$~km~s$^{-1}$. In this panel, we have included a mock observed star selection generated from the \nbody\ simulation using the method described in Section~\ref{observational} as blue dots with error bars. The distribution of mock observed stream stars has a radial velocity dispersion comparable to the central component of the stream within the range $\phi_1\in\Range{0}{50}$~deg. However, the central component is systematically wider than the simulation along the entire stream. This may indicate that some envelope stars have been erroneously classified as members of the central component. Furthermore, we observe that the observational uncertainties alone cannot account for the existence of the envelope, which deviates approximately by $\epsilon_{v_{r}}=-15$~km~s$^{-1}$ from the \nbody\ simulation within the range $\phi_1\in\Range{0}{50}$~deg.

According to simulations, perturbations such as a rotating Galactic bar or a tilting Galactic disc, as well as interactions with the Large Magellanic Cloud or dark matter subhalos can cause a stellar stream to widen (see references in \S4.1 of \citetalias{2026MNRAS.545f1974P}). In principle, these perturbations should tend to mix the stars in phase-space. However, the separation observed in radial velocity space between the central component and the envelope of the M68 stream contradicts this possibility. This separation is consistent with the hypothesis that the envelope stars are the remnants of a stellar stream that was formed by the M68 cluster due to tidal forces from its original host galaxy, before the cluster was accreted by the Milky Way. In this case, the remnants of the previous stream could have mixed with a new stream caused by the tidal forces of the Milky Way. Alternatively, if the M68 stream is located close to a strong resonance in frequency space, the stream may divide into groups that can differ after a few periods of evolution \citep{2021MNRAS.501.1791Y, 2023ApJ...954..215Y}. The observational evidence is inconclusive, so all of these possibilities remain open.

\section{Validation of the MW potential constraining methodology}\label{val_const_meth}

In this section, we build upon the fitting methodology introduced by \citetalias{2023MNRAS.524.2124P} to constrain the potential of the Milky Way using stellar streams. We focus on the observable section of the M68 stellar stream, and validate the fitting methodology using mock star samples generated from an \nbody\ simulation. We then determine which potential parameters characterising the simulation can be recovered, and estimate the expected statistical, observational, and systematic errors resulting from the application of this methodology.

\subsection{Observed stream section}\label{obs_section}

\begin{figure}
\includegraphics[width=1.0\columnwidth]{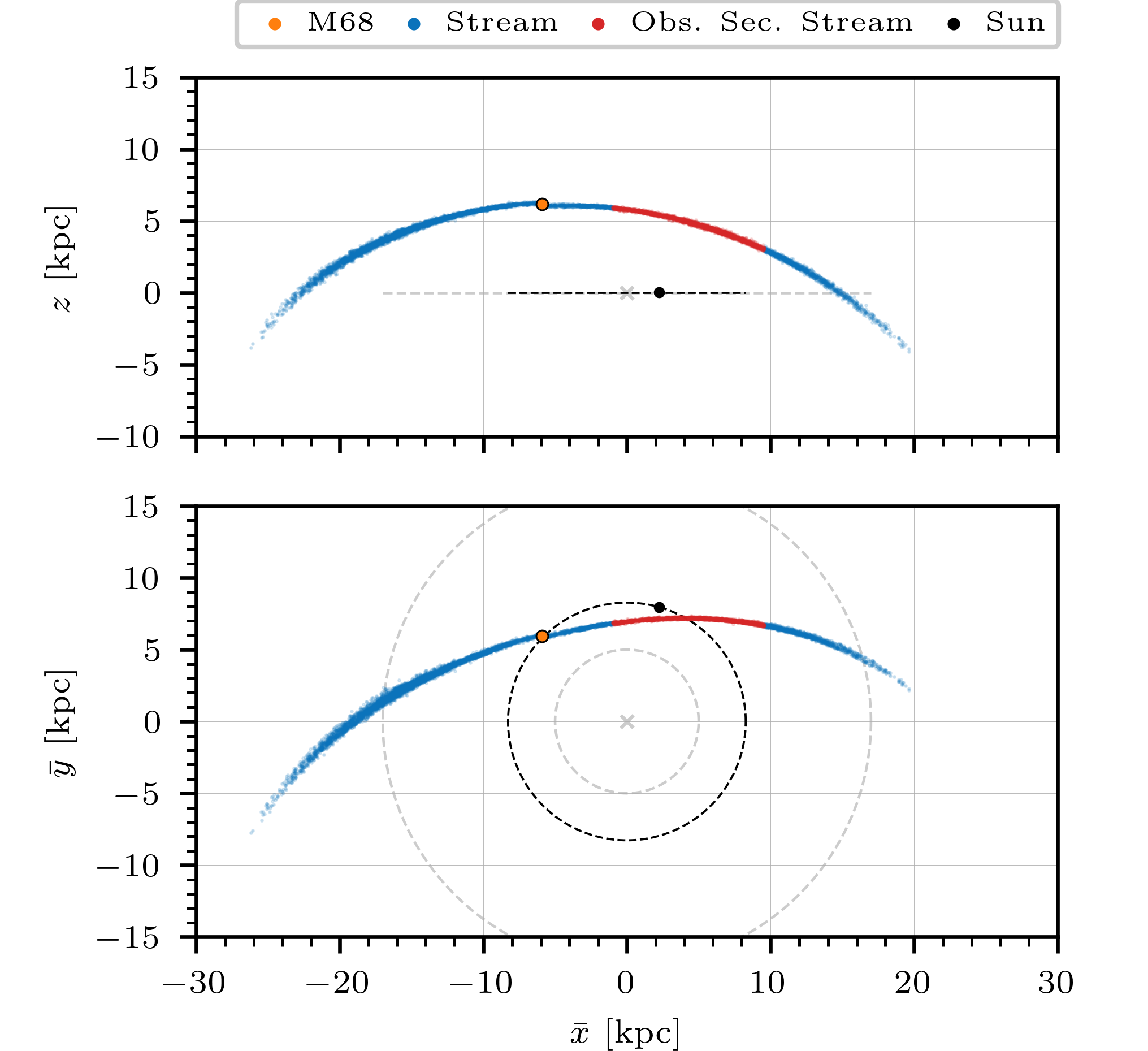}
\caption{\nbody\ simulation of the M68 stream in Galactocentric cartesian coordinates (blue dots). The axes are rotated around the $z$-axis so that the observable section of the stream (red dots) is aligned with the horizontal axis. The position of the M68 cluster is indicated by a large orange dot. The position of the Sun is marked by a black dot, and its circular orbit by a black dashed line. The Galactic centre is marked by a grey cross, and the Galactic cylindrical distances $R=5$ and $R=17$~kpc are indicated by dashed grey circles. \textit{Top:} Perpendicular direction to the Galactic plane. \textit{Bottom:} Galactic plane.}
\label{stream_xyz}
\end{figure}

The observed stream stars in the DESI selection, which we use in Section~\ref{obs_constraints} to constrain the potential, are contained within a section of the leading arm of the stream. Assuming that the stream is $3.04$~Gyr old \citepalias{2026MNRAS.545f1974P}, this section is approximately one quarter of the total length of the stream. Figure~\ref{stream_xyz} shows an \nbody\ simulation of the stream in Galactocentric cartesian coordinates. The axes have been rotated around the $z$-axis so that the observable section of the stream aligns with the horizontal axis. This section is highlighted in red, while the rest of the stream is shown in blue. The top panel shows that the observed section of the stream flows almost parallel to the disc, at a distance of approximately $3$ to $6$ kpc. The bottom panel shows that this section covers a cylindrical Galactocentric radius ranging from approximately $R\in\Range{6.8}{11.7}$~kpc. Figure~\ref{stream_xyz} also shows the position of the M68 cluster as a large orange dot. Its location, $z=6.1$~kpc and $R=8.4$~kpc, is similar to that of the observed stream stars. Consequently, the system formed by the cluster and the observed section of the stream can be used to constrain the potential of the disc and the inner dark halo.

\subsection{Fitting methodology}\label{methodology}

We assess the accuracy of the fitting methodology by applying it to recover the parameters that characterise the Galactic potential used in the \nbody\ simulation. This model is axisymmetric and static, and consists of a spherical bulge with an exponential cut-off, a Miyamoto-Nagai disc, and a spherical NFW halo. The values of the parameters characterising each component are given in Table~2 of \citetalias{2025MNRAS.539.2718P}. As the cluster and the observable section of the stream cover an approximate range of $R\in\Range{7}{12}$ kpc, they cannot constrain the inner Galaxy ($R\LessSim5$~kpc) and the outer dark halo ($R\GtrSim17$~kpc). Therefore, we only consider the following free parameters: the scale density $\rho_{0\rm h}$, the inner slope $\gamma_{\rm h}$, the scale length $a_{\rm h}$ and the density axis ratio $q_{\rm h}$ of the dark halo, and the disc mass $M_{\rm d}$. The other disc parameters are fixed to their exact values, as are the phase-space positions of the Sun (Table~C1 in \citetalias{2025MNRAS.539.2718P}) and the M68 cluster (Table~3 in \citetalias{2025MNRAS.539.2718P}).

Following \citetalias{2023MNRAS.524.2124P}, we determine the best-fitting values for the free-parameters of the model by maximising the posterior distribution obtained using the Bayes' theorem. We assume that the prior distributions are uniformly distributed within the validity range of each parameter. The likelihood function evaluates the error-convolved stream model at the position of the observed stream stars (App.~C, \citetalias{2023MNRAS.524.2124P}). Each star is modelled as a Gaussian distribution in phase-space, with its location given by its mean position and its covariance matrix given by the observational uncertainties of each coordinate. The model of the stream is constructed from the distribution of simulated stream stars in ICRS coordinates, but using parallaxes instead of Heliocentric distances. The probability density function (PDF) of the stream is estimated from a distribution of simulated stream stars using a kernel density estimate (KDE) method based on Gaussian kernels. In Appendix~\ref{App3}, we describe the method used to determine the smoothing parameter or bandwidth. The convolution between a star and the stream is evaluated by computing the inner product of the PDF representing the star with the PDF of the stream. In Appendix~\ref{App4}, we show that this computation is equivalent to evaluating multivariate Gaussian distributions, and it is therefore computationally efficient. However, to avoid unnecessary evaluations, we restrict the stream model to a region that only covers the observed section of the stream.

To determine the distribution of simulated stars for each parameter configuration, we use the method introduced in App.~D of \citetalias{2023MNRAS.524.2124P}. This method is much faster than \nbody\ simulations or particle-spray methods \citep[e.g.][]{2015MNRAS.452..301F} because the stream is not simulated for each parameter configuration. In this method, the orbit of the progenitor cluster and an \nbody\ simulation of the stellar stream are computed in a fiducial potential. The positions of the stars relative to the cluster orbit are then stored. For a new parameter configuration, a new cluster orbit is computed, and the stream stars are reallocated to their stored positions relative to the new orbit. This method is therefore based on the assumption that the stream closely follows the cluster orbit, and that its properties do not differ significantly from those obtained using the fiducial model for small variations in the free parameters. In this study, we use the potential from the \nbody\ simulation (see Section~\ref{stream_width}) as the fiducial potential. This implies that the systematic errors determined in this section may be underestimated, since the real potential differs from the one used to create the stream model.

\begin{figure*}
\includegraphics[width=1.0\textwidth]{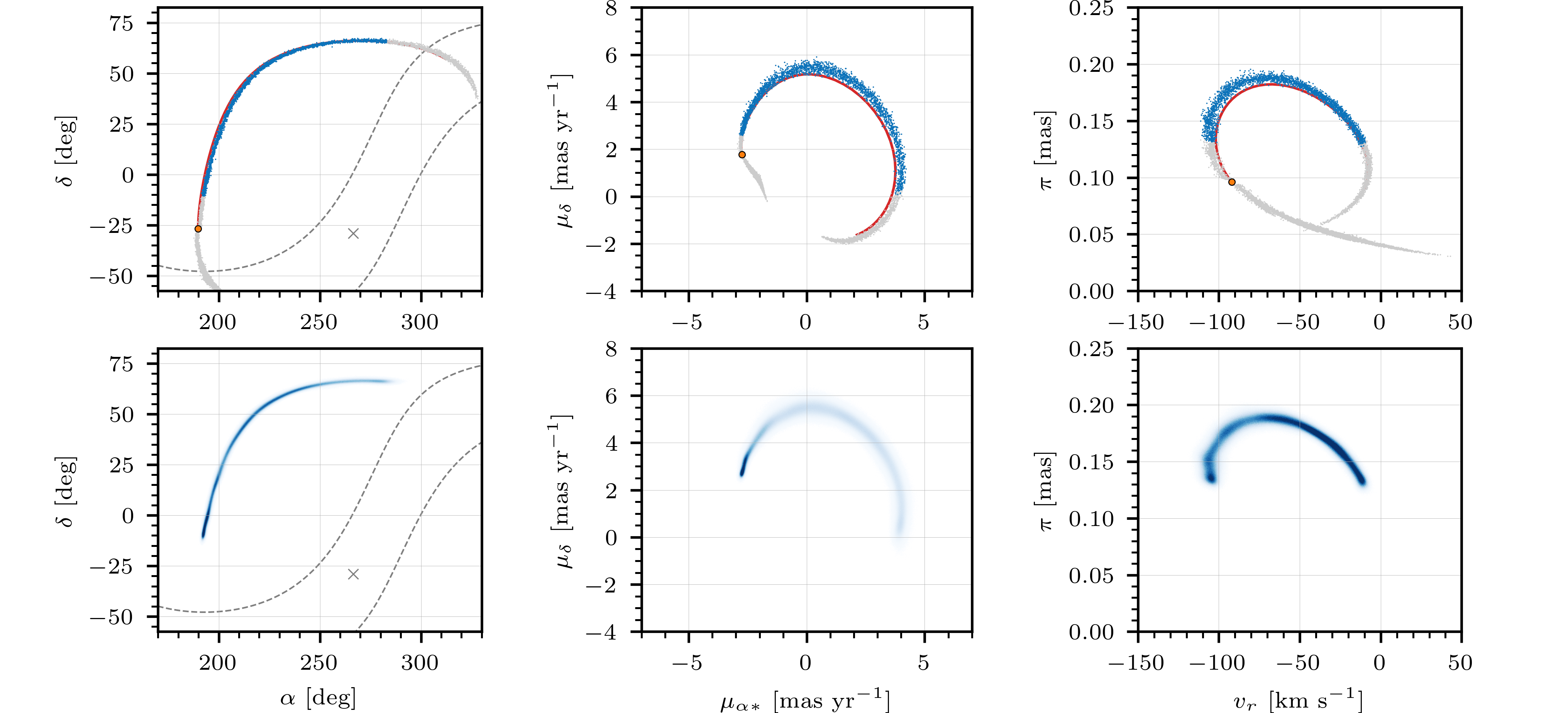}
\caption{\textit{Top:} Grey dots show the \nbody\ simulation of the stream, computed using the fiducial potential, for different pairs of observed ICRS coordinates. Blue dots indicate the positions of the stars used to construct the stream model. The position of the M68 cluster is indicated by a large orange dot, and its orbit $60$~Myr forward in time is shown as a solid red line. \textit{Bottom:} PDF of the stream model obtained using the KDE method from the blue stars depicted in the top panels. Dark blue indicates high density and light blue indicates low density. \textit{Left:} Grey dashed lines indicate a Galactic latitude $b=\pm15$~deg, and the grey cross indicates the Galactic centre.}
\label{stream_model}
\end{figure*}

In the top panels of Figure~\ref{stream_model}, we plot the simulated stream stars obtained using the fiducial potential as grey dots. The stars used to construct the stream density model are highlighted in blue and are defined within the boundaries $-7\leqslant\,\phi_1\,\leqslant95$~deg. These boundaries define a section that extends slightly beyond the observable section of the stream. This ensures that the model covers all observed stars for each parameter configuration. In the same figure, the position of the cluster is indicated by a large orange dot, and its orbit $60$~Myr forward in time is shown as a solid red line. The bottom panels show the density distribution obtained using the KDE method. This figure shows a model made of $N=2887$ stars (see Eq.~C4 of \citetalias{2023MNRAS.524.2124P} for the definition of this parameter). This example demonstrates that this method produces a smooth density model which accurately represents the distribution of stream stars across the entire defined range. In practice, we reduce the number of stars used to construct the model to $N=400$, since computation time scales linearly with the number of stars in the model. We have verified that this quantity is sufficient to generate a smooth model without any significant irregularities that could bias the results.

\subsection{Determination of the expected statistical uncertainties}\label{statistical}

We estimate the variability of the parameters that maximise the posterior distribution by optimising several samples of simulated stream stars. Each sample contains randomly selected stars from the stream. The variability is proportional to the size of the sample. Here, we study samples of $44$ stars, as this is the number of real, observed stream stars used to constrain the potential of the Milky Way (Section~\ref{star_selection}). These stars are located in the observable section of the stream, and we use their exact phase-space positions obtained from the \nbody\ simulation. For a given free parameter $\theta$, we use the Nelder-Mead simplex algorithm implemented in \texttt{scipy} \citep{2020SciPy-NMeth} to determine the values that maximise the posterior distribution. We then normalise these results by dividing them by the exact values used in the \nbody\ simulation. This means that the reference values are equal to one. We denote the normalised values that maximise the posterior distribution as $\hat{\theta}$.

\begin{figure}
\includegraphics[width=1.0\columnwidth]{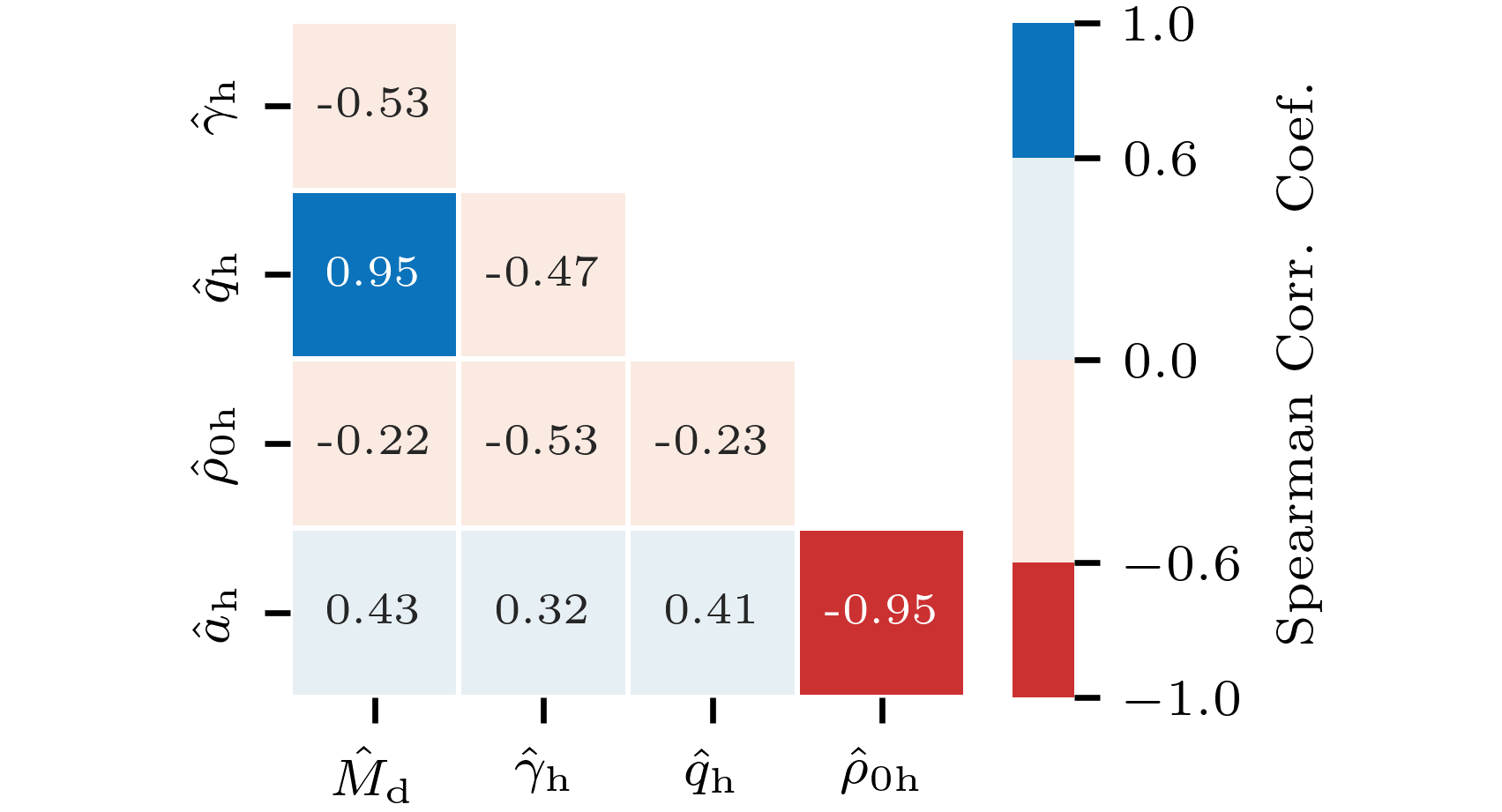}
\caption{Spearman correlation coefficient between the parameters that maximise the posterior distribution when optimising $1.5\pd{4}$ simulated streams of $44$ stars. The simulated stars are located within the observable section of the stream. We use their exact phase-space positions, which are obtained from an \nbody\ simulation.}
\label{corner_corr_stat}
\end{figure}

Figure~\ref{corner_corr_stat} shows the Spearman correlation coefficient between the best-fitting parameters obtained by optimising $1.5\pd{4}$ samples of simulated stream stars. This coefficient indicates how well a monotonic function can describe the relationship between two variables. The strongest positive correlation is found between the disc mass $M_{\rm d}$ and the dark halo axis ratio $q_{\rm h}$. This correlation arises because the stream is close by and flows almost parallel to the disc (see the top panel of Figure~\ref{stream_xyz}). Consequently, a more massive disc requires less dark matter at the position of the stream, and therefore a prolate halo. Conversely, a less massive disc requires more dark matter, which is achieved by an oblate dark halo.

\begin{figure}
\hspace{-0.05cm}\includegraphics[width=1.0\columnwidth]{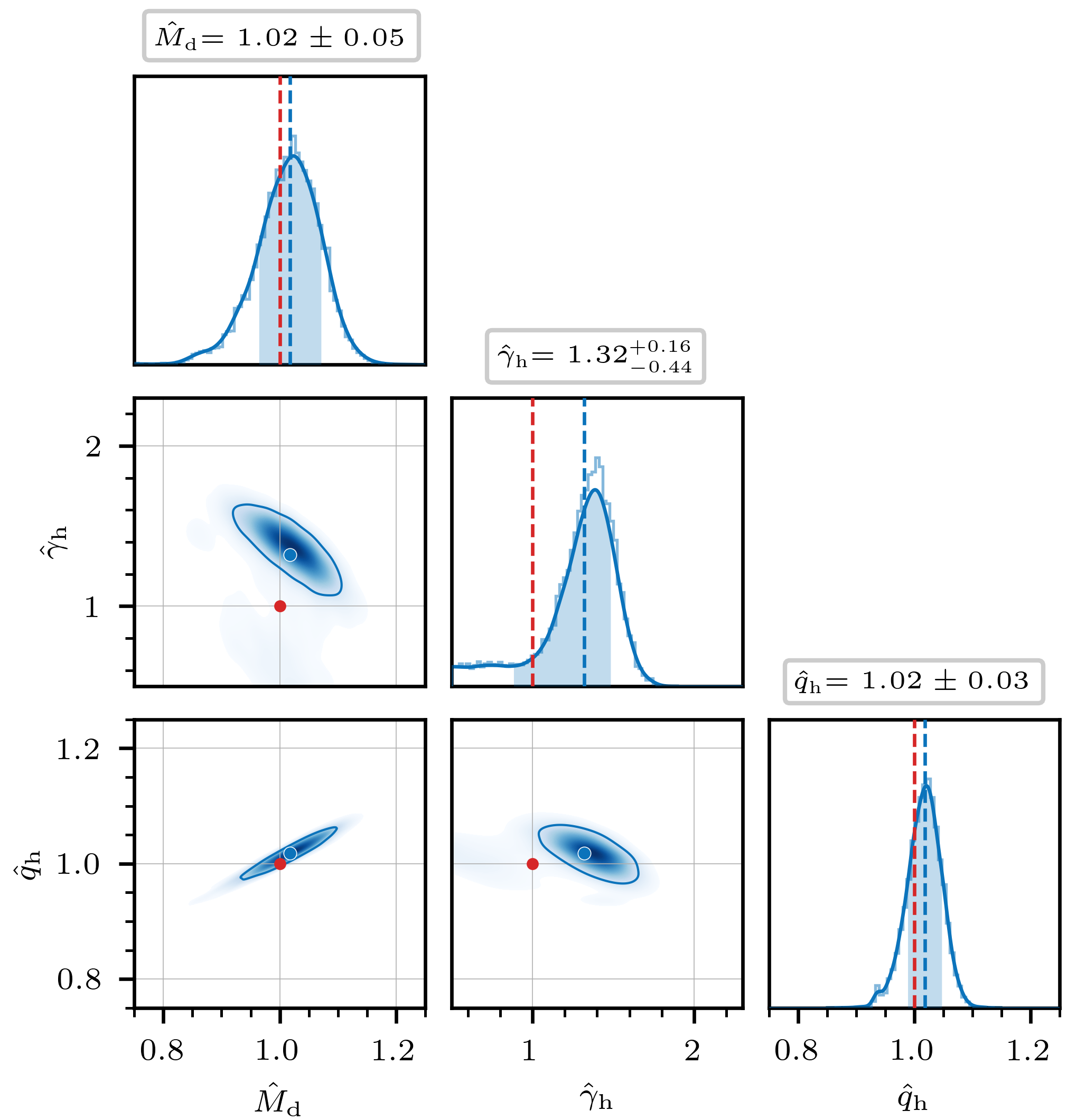}
\caption{Marginal distributions of the best-fitting parameters of the Milky Way model, obtained by optimising $1.5\pd{4}$ simulated streams of $44$ stars. The distributions are estimated using a KDE method. The parameters are divided by the values used in the \nbody\ simulation so that the reference value is one. \textit{Diagonal panels:} The solid blue line shows the estimated distribution, and the light blue line the histogram of the results. The vertical dashed blue line marks the median, and the confidence interval with areas of $1\sigma$ around the median is indicated by a shaded blue area. The numerical values are shown in the legends. The vertical red line shows the reference value. \textit{Inferior panels:} The solid blue line marks the limit containing 68 per cent of the distribution. The large blue dot marks the median, and the large red dot marks the reference value.}
\label{corner_M_q_gamma}
\end{figure}

Figure~\ref{corner_M_q_gamma} shows the marginalised distributions of the best-fitting values for $M_{\rm d}$, $\gamma_{\rm h}$, and $q_{\rm h}$ in blue. The reference configuration is indicated by a vertical red dashed line in the histograms and a red dot in the bidimensional plots. The vertical blue line and the blue dots show the median of the distribution. We quantify the variability of the distribution using the confidence interval with areas of $1\sigma\simeq0.683$ around the median. The numerical values for the disc mass and the dark halo axis ratio are $\hat{M}_{\rm d} = 1.02\pm 0.05$ and $\hat{q}_{\rm h}=1.02\pm 0.03$. These parameters are approximately unbiased, and the statistical uncertainties are smaller than $5$ per cent. We consider these discrepancies to be small, and that these two parameters can be consistently and independently recovered with samples of 44 stars in the observable section of the stream. Conversely, the inner slope of the dark matter density $\hat{\gamma}_{\rm h}=1.32_{-0.44}^{+0.16}$, presents a systematic bias of $\Approx 30$ per cent, which is comparable to the average of the uncertainties. Therefore, we conclude that this parameter cannot be reliably recovered from this stream.

\begin{figure}
\includegraphics[width=1.0\columnwidth]{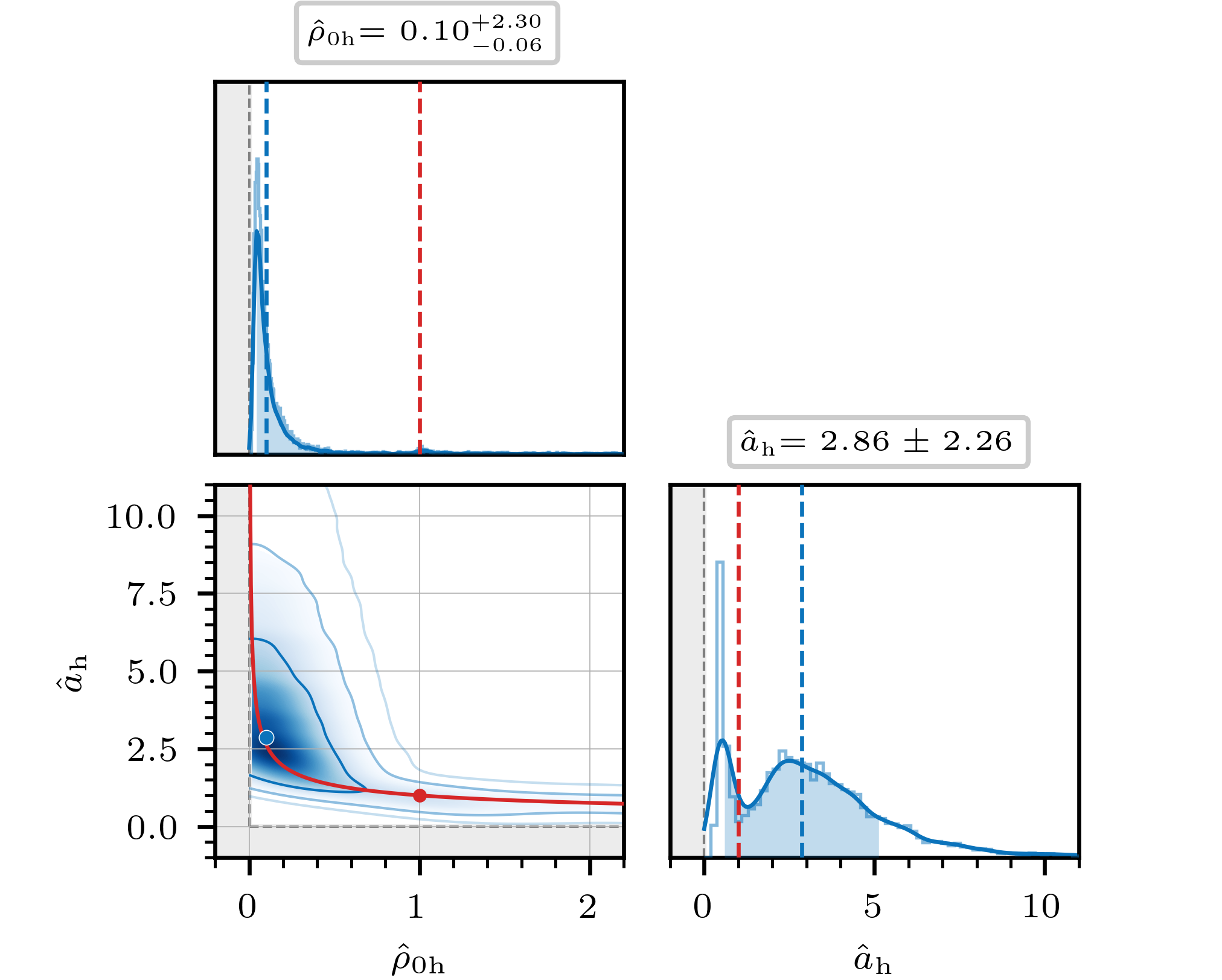}
\caption{As Figure~\ref{corner_M_q_gamma}. Areas outside the valid parameter range are shaded in grey, and their boundaries are highlighted with dashed grey lines. \textit{Bottom left:} The solid blue lines show the limits containing 68, 95, and 99 per cent of the distribution. The red solid line shows the configurations with the same dark halo mass $M_{\rm h}^{200}$ as the reference potential.}
\label{corner_h}
\end{figure}

The only other significant correlation is between the scale density $\rho_{0{\rm h}}$ and the scale length $a_{\rm h}$ of the dark halo. As these parameters are degenerate, the stream cannot constrain them independently. The bottom left panel of Figure~\ref{corner_h} shows the marginalised distribution of the best-fitting configurations for these two parameters in blue. The solid red line indicates the configurations with the same dark halo mass $M_{\rm h}^{200}$, which is defined as the enclosed mass within $r_{200} \simeq 197.28$~kpc\footnote{Galactocentric spherical radius $r_{200}$ such that the average density is equal to $\Delta_{\rm c}=200$ times the critical density of the Universe $\rho_{\rm c} \equiv 3H_0^2 / 8\pi G$, where the Hubble constant $H_0 = 71$~km~s$^{-1}$~Mpc$^{-1}$ and the Universal Gravitational constant
$G = 4.4987\pd{-12}$~kpc$^3$~M$_{\Sun}^{-1}$~Myr$^{-2}$.}, as the reference. As can be seen, the distribution of the best-fitting configurations follows the line of equal mass. However, the median of the distribution (blue dot) is significantly biased towards smaller values of $\rho_{0{\rm h}}$ and larger values of $a_{\rm h}$. This systematic deviation is explained by the fact that all configurations along the equal-mass line produce a similar fit to the stream. The location of the distribution's peak is then determined by minor factors, such as inaccuracies in the stream model. This is consistent with the results obtained by \citetalias{2025MNRAS.539.2718P} (Fig.~7) using a method based on angle-action coordinates. This suggests that the inability to constrain the values of both parameters is a characteristic of the stream and its location, rather than being a consequence of the method used to constrain the potential of the Milky Way.

\begin{figure}
\hspace{0.5cm}\includegraphics[width=0.8\columnwidth]{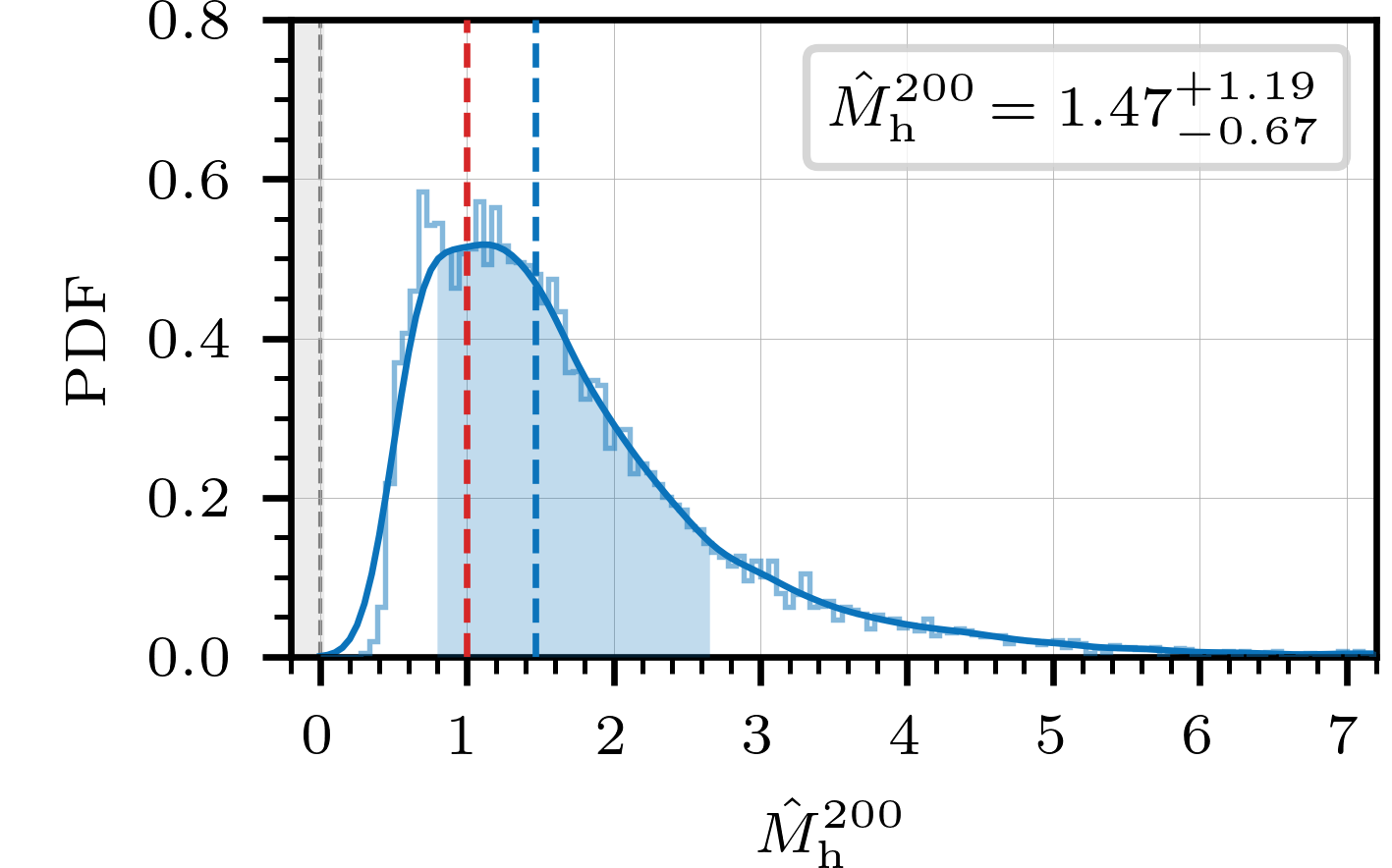}
\caption{Distribution of best-fitting dark halo masses $\hat{M}_{\rm h}^{200}$, obtained by optimising $1.5\pd{4}$ simulated streams of $44$ stars. The solid blue line shows the estimated distribution using a KDE, and the light blue line shows the histogram of the results. The vertical dashed blue line indicates the median, and the blue shaded area shows the $1\sigma$ confidence interval. The numerical values are shown in the legend. The vertical red line indicates the normalised mass of the reference dark halo. Areas outside the valid parameter range are shaded in grey, and the boundary is highlighted with a dashed grey line.}
\label{mass_hist}
\end{figure}

Figure~\ref{mass_hist} shows the distribution of dark halo masses $\hat{M}_{\rm h}^{200}$ obtained for the best-fitting configurations. The distribution peaks at the reference value, which is marked by a dashed red line. However, the distribution has a long tail that biases the median and the dispersion towards larger masses. We estimate the median and the $1\sigma$ confidence interval for $\hat{M}_{\rm h}^{200} \simeq 1.47_{-0.67}^{+1.19}$. Therefore, we consider that samples of $44$ stars in the observable section of the stream cannot reliably constrain this property of the dark halo.

\subsection{Determination of the expected observational uncertainties}\label{observational}

In order to test how the best-fitting results vary due to observational uncertainties, we study the simulated stream that exhibits the least systematic deviation with respect to the reference potential. This stream is taken from the sample of $1.5\pd{4}$ simulations studied in Section~\ref{statistical}. Using the method described in \S5 of \citetalias{2025MNRAS.539.2718P}, we generate mock observational uncertainties for this simulation. First, we generate realistic \textit{Gaia} photometry, which we then use to estimate the GDR3 selection function and the associated observational uncertainties. Similarly, we determine the DESI radial velocity uncertainty as a function of the colour and magnitude by interpolating from the Year-3 internal DESI MWS observational results introduced in Section~\ref{selection}. This procedure is similar to that described in \S7.4.2 of \citet{2023ApJ...947...37C}. In this case, we do not consider the DESI selection function due to a lack of resources at this early stage of the DESI MWS's development. Consequently, we treat all stars that pass the GDR3 selection function as if their spectra had been measured by DESI. The mock observations are generated by sampling Gaussian distributions assuming that the mean is the exact phase-space location of the stars and the covariance matrix is given by the mock observational uncertainties. Finally, we maximise the posterior distribution for $1.5\pd{4}$ samples of mock observed stream stars, using the same method described in Section~\ref{statistical}.

The main consequence of taking the GDR3 observational limitations into account is the loss of distance information. This is because the average parallax uncertainty of the DESI selection of stream stars is approximately $\pm 0.15$~mas, which corresponds to about $\pm 6.6$~kpc. The effects of the uncertainties on the other coordinates resemble those caused by statistical randomness, but resulting in greater dispersion in the best-fitting configurations. The correlations between the best-fitting parameters are also similar to those obtained in Section~\ref{statistical} and shown in Figure~\ref{corner_corr_stat}.

\begin{figure}
\includegraphics[width=1.0\columnwidth]{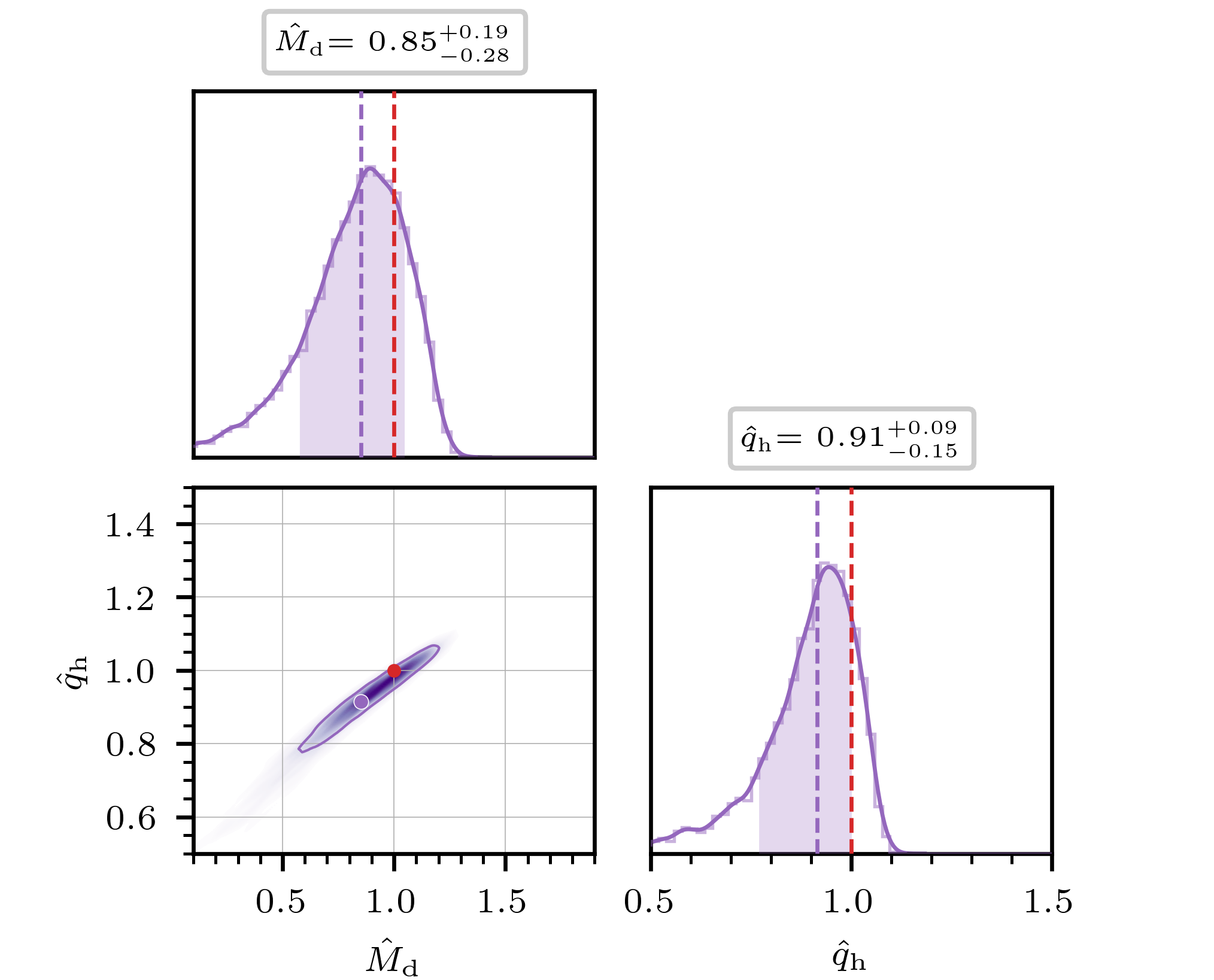}
\caption{As Figure~\ref{corner_M_q_gamma} but optimising $1.5\pd{4}$ samples of mock observations of stream stars. The mock samples are obtained from an \nbody\ simulation of the stream, and incorporate the GDR3 selection function and observational uncertainties, as well as the DESI observational uncertainties for the radial velocities.}
\label{corner_Md_q}
\end{figure}

We focus on the correlation between the disc mass $M_{\rm d}$ and the dark halo axis ratio $q_{\rm h}$. Figure~\ref{corner_Md_q} shows the marginalised distributions of the best-fitting values of these parameters. The median of the distributions is indicated by vertical purple dashed lines, the $1\sigma$ confidence interval by shaded purple areas and the reference configuration by red dashed lines. In this case, the observational uncertainties are of approximately $24$ per cent for $M_{\rm d}$ and $12$ per cent for $q_{\rm h}$. Both parameters are skewed towards smaller values than the reference configuration, and present a systematic deviation of the median from the reference values of approximately $15$ per cent for $M_{\rm d}$ and $9$ per cent for $q_{\rm h}$. Nevertheless, we consider these deviations to be small, and expect to reliably recover the values of these two parameters when using the real observed DESI star sample.

\subsection{Determination of the expected total uncertainties}\label{total}

Table~\ref{res_stat_obs} lists the statistical uncertainties $\sigma_{\rm stat}$ and observational uncertainties $\sigma_{\rm obs}$ for all free parameters of the model. These are determined by the $1\sigma$ confidence interval around the median of their marginalised distributions. For asymmetric distributions, the average of the upper and lower limits of the confidence interval is taken. We determined $\sigma_{\rm obs}$ using a particular mock sample of stream stars. Therefore, we assume that these results are independent of the sample and calculate the total uncertainty $\sigma_{\rm tot}$ as follows:
\begin{equation}\label{sigma_tot}
 \sigma_{\rm tot} = \sqrt{\sigma_{\rm stat}^2 + \sigma_{\rm obs}^2}.
\end{equation}
The parameters with the smallest uncertainties are $q_{\rm h}$ with $\sigma_{\rm tot}=12$ per cent, and $M_{\rm d}$ with $\sigma_{\rm tot}=24$ per cent. We therefore focus on these parameters when constraining the potential of the Milky Way using the DESI star selection. For these two parameters, $\sigma_{\rm obs} > \sigma_{\rm stat}$. This implies that observational errors dominate the total uncertainty. These estimates are obtained using the exact potential and stream employed in the \nbody\ simulation. In practice, however, the potential is unknown and the formation and evolution history of the stream may be more complex than assumed in the simulation. Furthermore, other sources of uncertainty should be considered, such as perturbations caused by the Large Magellanic Cloud (LMC) or other Milky Way satellites. We therefore consider these to be lower bounds on the true total uncertainties.

\begin{table}
\caption[]{Sample-to-sample variability for $44$ stream stars with exact positions $\sigma_{\rm stat}$, variability caused by observational uncertainties $\sigma_{\rm obs}$, and total uncertainty $\sigma_{\rm tot}$, as defined by Eq.~\ref{sigma_tot}, estimated for the values of the free parameters of the model $\hat{\theta}$ that maximise the posterior distribution. These uncertainties correspond to the $1\sigma$ confidence interval and are assumed to be symmetric with respect to the median of the distributions. The parameters are normalised by the values used in the \nbody\ simulation.}
\begin{center}
\begin{tabular}{lrrr}
\toprule
{\boldmath$\hat{\theta}$} & {\boldmath$\sigma_{\rm stat}$} & {\boldmath$\sigma_{\rm obs}$} & {\boldmath$\sigma_{\rm tot}$}\\
\midrule
$\hat{M}_{\rm d}$ & $0.05$ & $0.24$ & $0.24$ \\[4.5pt]
$\hat{\gamma}_{\rm h}$ & $0.30$ & $0.58$ & $0.65$ \\[4.5pt]
$\hat{q}_{\rm h}$ & $0.03$ & $0.12$ & $0.12$ \\[4.5pt]
$\hat{\rho}_{0 \rm h}$ & $1.18$ & $1.62$ & $2.00$ \\[4.5pt]
$\hat{a}_{\rm h}$ & $2.26$ & $1.25$ & $2.58$ \\
\bottomrule
\end{tabular}
\end{center}
\label{res_stat_obs}
\end{table}

\section{Observational constraints on the MW potential}\label{obs_constraints}

In this section, we use the observational data from the DESI star selection to constrain the potential of the Milky Way. We also include additional free parameters, such as the positions of the Sun and the progenitor cluster. First, we test the constraining power of the observed stream alone. Then, we combine this with observations of the Milky Way's rotation curve (RC) to create a model of the Galactic potential that is valid within the range defined by the cluster and the observable section of the stream. Our focus is on the mass of the disc and the axis ratio of the dark matter halo, since we determined in Section~\ref{val_const_meth} that these parameters can be recovered with small uncertainties and small systematic deviations from the true values.

\subsection{Observational star selection}\label{star_selection}

We restrict the DESI selection to the $44$ stars located in the central component of the stream, as defined in Section~\ref{stream_width}. The stars in the envelope are inconsistent with having been stripped from the progenitor cluster in the static, smooth potential models of the Milky Way that we define in Section~\ref{potential}. In such models, the stream stars always appear to be projected below the cluster orbit (Fig.~13 in \citetalias{2026MNRAS.545f1974P}), which roughly follows the separation boundary between the central component of the stream and the upper envelope. This is independent of the potential because the position of the cluster is measured with high precision and its projected sky coordinates do not change significantly along the stream close to the cluster. To avoid biasing the results of the fit, especially in radial velocity space, where the envelope stars are systematically separated from the central component (bottom panel Fig.~\ref{envelope}), we exclude the envelope stars from the fit. We emphasise that the separation of the components defined by Eq.~\ref{main_comp_limits} is independent of the potential, and that their exact values are only approximate. Minor variations in the boundaries can result in small variations in the number of stars in the main component of the stream used to constrain the potential. We consider the variability in the final results caused by different boundary choices to be comparable to the statistical variability quantified in Section~\ref{statistical}.

For each star, we use the parallax, sky coordinates, and proper motions from the GDR3 catalogue, and the radial velocities from the DESI MWS survey. The DESI collaboration also provides with Heliocentric distances for the stream stars, estimated from their spectra. These are included in the associated value-added catalogues\footnote{\url{https://data.desi.lbl.gov/doc/releases/dr1/\#value-added-catalogs}}. In Appendix~\ref{App5}, we analyse these distance estimates. However, we have decided not to use them, as they differ systematically from the \nbody\ simulation of the stream, which we consider to be more accurate.

\subsection{Milky Way potential model}\label{potential}

We adopt a static, axisymmetric potential model of the Milky Way, composed of three components. The bulge and the dark halo are modelled using a spheroidal density profile with an exponential cut-off, as described by the following equation:
\begin{equation}
 \rho\var{s} \equiv \rho_0 \Big( \:\!\frac{s}{a}\:\! \Big)^{-\gamma} \Big(1 +  \frac{s}{a}\;\! \Big)^{\gamma-\beta} \exp\!\!\:\bigg[\!\!\, - \Big(\frac{s}{r_{\rm cut}}\Big)^{\xi}\;\! \bigg],
\end{equation}
where:
\begin{equation}
 s^2\equiv R^2 + \frac{z^2}{q^2}.
\end{equation}
As a consequence that the stream does not cover the Galactocentric radii $R\LessSim5$~kpc, and we are not modelling the Galactic bar, we assume that the bulge model does not significantly impact the fit of the observed stream. We therefore fix its parameters to the values listed in Table~\ref{MW_table}, which are taken from \citet{2017MNRAS.465...76M} and correspond to a bulge of mass $M_{\rm b} \simeq 8.66\pd{9}$~M$_{\Sun}$.

The scale density $\rho_{0\rm h}$, the inner slope $\gamma_{\rm h}$, the scale length $a_{\rm h}$ and the density axis ratio $q_{\rm h}$ of the dark halo are taken as free parameters of the model. The priors for these parameters are assumed to be uniform across their entire valid range. Since the stream cannot constrain the potential for $R\GtrSim12$~kpc, and $a_{\rm h}\Approx15$~kpc (Section~\ref{results_str_rc}), we set the outer slope of the dark halo to be $\beta_{\rm h}=3$. This value corresponds to an NFW density profile \citep{1996ApJ...462..563N}. The other fixed parameters of the dark halo are specified in Table~\ref{MW_table}.

\begin{table}
\caption[]{Parameters of the bulge and dark halo density profiles. Parameters with no value specified are taken as free parameters.}
\begin{center}
\begin{tabular}{p{1.7em}lrrr}
\toprule
\multicolumn{2}{l}{\textbf{Spheroid}}&\textbf{Bulge}&\textbf{Dark halo}\\
\midrule
$\rho_{0}$&\units{$10^{10}\,{\rm M}_{\Sun}\,{\rm kpc}^{-3}$}&$9.6$&$-$\\
$\beta$&-&$1.8$&$3$\\
$\gamma$&-&$0$&$-$\\
$a$&\units{pc}&$75$&$-$\\
$q$&-&$0.5$&$-$\\
$\xi$&-&$2$&$2$\\
$r_{\rm cut}$&\units{kpc}&$2.1$&$\infty$\\
\bottomrule
\end{tabular}
\end{center}
\label{MW_table}
\end{table}

We neglect the internal structure of the disc, such as the thin, thick and gaseous components, because with the stream, we can only constrain the total mass of the disc. Therefore, we model the disc using a single exponential of density profile:
\begin{equation}
 \rho_{\rm d}\var{R,z} \equiv \frac{\varSigma_{\rm d}}{2\:\!h_{\rm d}} \exp\!\!\:\bigg( \!\!-\frac{R}{a_{\rm d}} - \frac{|z|}{h_{\rm d}} \,\bigg),
\end{equation}
with a total mass of:
\begin{equation}\label{mass_disc_exp}
 M_{\rm d} = 2 \:\! \pi \:\! a_{\rm d}^2 \:\! \varSigma_{\rm d}.
\end{equation}
We treat the surface density $\varSigma_{\rm d}$, scale length $a_{\rm d}$, and scale height $h_{\rm d}$ as free parameters of the model. We consider $\varSigma_{\rm d}$ to be poorly constrained by observations, and assume a uniform prior for this parameter. On the other hand, the spatial distribution of mass in the disc can be constrained using photometric measurements. We use the estimates of \citet{2016ARA&A..54..529B} as priors for $a_{\rm d}$ and $h_{\rm d}$, assuming that they follow Gaussian distributions, and specify their mean and standard deviation in the second column of Table~\ref{table_results}. For the priors, we use the measurements corresponding to the thin disc because, as the most massive component, it makes the greatest contribution to the potential.

\begin{table*}
\caption{Median and $1\sigma$ confidence interval of the marginalised posterior distributions. If a distribution cannot be fully covered by the MCMC method, the result is considered to be infinity. The second column shows the mean and standard deviation of the Gaussian priors, as well as the lower and upper limits of the uniform priors. The subsequent columns present the results obtained using only the DESI selection of stream stars (M68 stream), using only the rotation curve of the Milky Way (RC), and the combination of the M68 stream and the rotation curve (M68 Str. + RC). The final column shows the best-fitting configuration, corresponding to the maximum value of the posterior distribution when the stream and rotation curve are used as observational constraints.}
\begin{subtable}[b]{1.0\textwidth}

\begin{center}
\begin{tabular}{llrrrrr}
\toprule
\multicolumn{2}{l}{\textbf{Parameter}}&\textbf{Prior}&\textbf{M68 Stream}&\textbf{RC}&\textbf{M68 Str. + RC}&\textbf{Best-fitting}\\
\midrule
$R_{\Sun}$ & \units{kpc} & $8.275 \pm 0.034$ & $8.284 \pm 0.033$ & $8.276 \pm 0.035$ & $8.274 \pm 0.034$ & $8.274$\\[4.5pt]
$U_{\Sun}$ & \units{km s$^{-1}$} & $11.1 \pm 1.25$ & $11.96 \pm 0.92$ & $11.12 \pm 1.27$ & $10.37 \pm 1.03$ & $10.58$\\[4.5pt]
$V_{\Sun}$ & \units{km s$^{-1}$} & $12.24 \pm 2.05$ & $12.16 \pm 2.01$ & $12.15 \pm 2.02$ & $7.58 \pm 1.84$ & $7.46$\\[4.5pt]
$W_{\Sun}$ & \units{km s$^{-1}$} & $7.25 \pm 0.62$ & $7.71 \pm 0.61$ & $7.30 \pm 0.62$ & $7.59 \pm 0.61$ & $7.55$\\
\midrule
$r_{\Helio}^{\SM{M68}}$ & \units{kpc} & $\Range{0}{\infty}$ & $10.48 \pm 0.10$ & $-$ & $10.55 \pm 0.09$ & $10.57$\\[4.5pt]
$v_r^{\SM{M68}}$ & \units{km s$^{-1}$} & $-92.07 \pm 1.51$ & $-88.02 \pm 0.78$ & $-$ & $-88.59 \pm 0.69$ & $-88.53$\\[4.5pt]
$\mu_\delta^{\SM{M68}}$ & \units{mas yr$^{-1}$} & $1.779 \pm 0.024$ & $1.755 \pm 0.022$ & $-$ & $1.773 \pm 0.021$ & $1.775$\\[4.5pt]
$\mu_{\alpha*}^{\SM{M68}}$ & \units{mas yr$^{-1}$} & $-2.739 \pm 0.024$ & $-2.681 \pm 0.020$ & $-$ & $-2.685 \pm 0.020$ & $-2.682$\\
\midrule
$\varSigma_{\rm d}$ & \units{$10^{9} \,$M$_{\Sun}$ kpc$^{-2}$} & $\Range{0}{\infty}$ & $9.52_{-1.41}^{+1.79}$ & $1.37_{-0.70}^{+0.16}$ & $1.04 \pm 0.20$ & $1.01$\\[4.5pt]
$a_{\rm d}$ & \units{kpc} & $2.6 \pm 0.5$ & $3.36 \pm 0.41$ & $2.93_{-0.31}^{+0.46}$ & $2.85_{-0.22}^{+0.32}$ & $2.85$\\[4.5pt]
$h_{\rm d}$ & \units{kpc} & $0.3 \pm 0.05$ & $0.30 \pm 0.05$ & $0.30 \pm 0.05$ & $0.31 \pm 0.05$ & $0.30$\\
\midrule
$\rho_{0\rm h}$ & \units{$10^{7} \,$M$_{\Sun}$ kpc$^{-3}$} & $\Range{0}{8}$ & $\infty$ & $\infty$ & $2.30_{-1.91}^{+3.55}$ & $1.31$\\[4.5pt]
$\gamma_{\rm h}$ & - & $\Range{-\infty}{\infty}$ & $-1.16_{-0.42}^{+0.31}$ & $-2.80_{-5.29}^{+2.74}$ & $0.90 \pm 0.45$ & $1.06$\\[4.5pt]
$a_{\rm h}$ & \units{kpc} & $\Range{1}{50}$ & $\infty$ & $2.01_{-0.84}^{+3.15}$ & $11.89_{-3.53}^{+11.75}$ & $14.83$\\[4.5pt]
$q_{\rm h}$ & - & $\Range{0}{\infty}$ & $1.01_{-0.07}^{+0.09}$ & $\infty$ & $0.83_{-0.05}^{+0.06}$ & $0.83$\\
\bottomrule
\end{tabular}
\end{center}

\caption{Free parameters of the model: Position of the Sun and M68 globular cluster, disc, and dark halo.}
\label{table_results_a}

\end{subtable}
\phantom{newline}
\begin{subtable}[b]{1.0\textwidth}

\begin{center}
\begin{tabular}{llrrrrr}
\toprule
\multicolumn{2}{l}{\textbf{Parameter}}&\textbf{Prior}&\textbf{M68 Stream}&\textbf{RC}&\textbf{M68 Str. + RC}&\textbf{Best-fitting}\\
\midrule
$|K_z|$ & \units{M$_{\Sun}$ pc$^{-2}$} & $-$ & $2.43 \pm 0.26$ & $2.13_{-0.86}^{+1.21}$ & $2.10 \pm 0.15$ & $2.07$ \\[4.5pt]
$\varSigma_{\Sun}$ & \units{km$^{2}$ s$^{-2}$ pc$^{-1}$} & $-$ & $108.29 \pm 12.16$ & $78.21_{-33.22}^{+45.89}$ & $77.37 \pm 6.13$ & $76.18$ \\[4.5pt]
$M_{\rm d}$ & \units{$10^{10} \,$M$_{\Sun}$} & $-$ & $6.77 \pm 0.78$ & $6.69 \pm 3.39$ & $5.34 \pm 0.57$ & $5.14$ \\[4.5pt]
$M_{\rm bd}$ & \units{$10^{10} \,$M$_{\Sun}$} & $-$ & $7.64 \pm 0.78$ & $7.55 \pm 3.39$ & $6.20 \pm 0.57$ & $6.00$ \\[4.5pt]
$M_{\rm h}^{r \leqslant 50}$ & \units{$10^{11} \,$M$_{\Sun}$} & $-$ & $\infty$ & $\infty$ & $3.19 \pm 0.37$ & $3.23$ \\[4.5pt]
$M_{\rm h}^{200}$ & \units{$10^{11} \,$M$_{\Sun}$} & $-$ & $\infty$ & $\infty$ & $7.37_{-1.51}^{+2.13}$ & $7.59$ \\
\bottomrule
\end{tabular}
\end{center}

\caption{Derived properties of the model: Total vertical gravitational force $|K_z| \equiv |K_z\var{R_{\Sun}, 0, 1.1\, {\rm kpc}}|$, total density at the Sun's position integrated within $z\in\pm1.1$~kpc or surface density $\varSigma_{\Sun}$, disc mass, baryonic mass $M_{\rm bd} \equiv M_{\rm b} + M_{\rm d}$, dark matter halo mass enclosed within $r \leqslant 50$~kpc and within $r_{200}$.}
\label{table_results_b}

\end{subtable}
\label{table_results}
\end{table*}

\subsection{Additional free parameters}\label{add_parameters}

We use the phase-space position of the Sun to convert from ICRS to Galactocentric reference frame. In the model, we consider the Galactocentric distance of the Sun $R_{\Sun}$ to be a free parameter. We take the measurement of \citet{2021A&A...647A..59G} as a Gaussian prior. We consider the impact of the Sun's vertical separation from the Galactic plane on the fit to be negligible, and fix this variable at $z_{\Sun}=20.8$~pc, which is the mean of the estimate of \citet{2019MNRAS.482.1417B}. We also treat the velocity of the Sun $v_{\Sun}=(U_{\Sun}, V_{\Sun}, W_{\Sun})$ as free parameters, where the components of the velocity vector are given with respect to the Local Standard of Rest, and $U$ points to the Galactic centre, $V$ is positive along the direction of the Sun's rotation (clockwise when viewed from the North Galactic Pole), and $W$ is positive towards the North Galactic Pole. We use the measurements of \citet{2010MNRAS.403.1829S} as Gaussian priors. We list the mean and standard deviation of the priors in Table~\ref{table_results}.

The phase-space position of the M68 globular cluster is the initial condition of integration of the cluster orbit. First, we assume that the observational uncertainties in the sky coordinates of the cluster are negligible, and set their values to $(\delta^{\SM{M68}}, \alpha^{\SM{M68}}) = (-26.744, 189.867)$~deg, corresponding to the mean of the measurements of \citet{2021MNRAS.505.5978V}. We treat the Heliocentric distance of M68 $r_{\rm h}^{\SM{M68}}$ as a free parameter. In this case, we assume a uniform prior because we can constrain this parameter using the stream (Section~\ref{results_str_rc}). We also treat the radial velocity $v_r^{\SM{M68}}$ and proper motions $(\mu_\delta^{\SM{M68}}, \mu_{\alpha*}^{\SM{M68}})$ as free parameters in the model. We assume Gaussian priors, taking the mean and standard deviation from the measurements of \citet{2023MNRAS.519..192W} and \citet{2021MNRAS.505.5978V} respectively, and list their numerical values in the second column of Table~\ref{table_results}.

\subsection{Constraints using the M68 stream}\label{results_str}

We generate random samples that follow the posterior distribution by using our implementation of the Metropolis-Hastings algorithm, which is a Markov Chain Monte Carlo (MCMC) method. We run $72$ chains of $10^6$ steps, and discard the first $5\pd{5}$ steps from each chain to eliminate possible biases caused by the initial conditions. For each free parameter, the third column of Table~\ref{table_results} (M68 Stream) lists the median and $1\sigma$ confidence interval of the marginalised distributions.

In order to fit the DESI star selection, there is no need for significant deviations in the position and velocity of the Sun or the M68 globular cluster relative to the priors. The only exceptions are the radial velocity $v_r^{\SM{M68}}$ and the proper motion $\mu_{\alpha *}^{\SM{M68}}$, where slightly larger values are preferred. We consider these discrepancies to be small, and caused by forcing the model to obtain a marginally better fit to the data. The stream can constrain the disc mass to $M_{\rm d} = 6.77 \pm 0.78 \pd{10}$~M$_{\Sun}$ and the dark matter axis ratio to $q_{\rm h} = 1.01_{-0.07}^{+0.09}$. These two parameters are strongly correlated, with a Spearman correlation coefficient of $\rho=0.83$, as obtained in Sections~\ref{statistical} and \ref{observational} by studying mock stream star samples. However, there is no correlation between these two parameters and the halo parameters. This implies that, at the position of the stream, the mass of the disc and the shape of the halo are independent on the radial distribution of the dark matter.

\begin{figure}
\includegraphics[width=1.0\columnwidth]{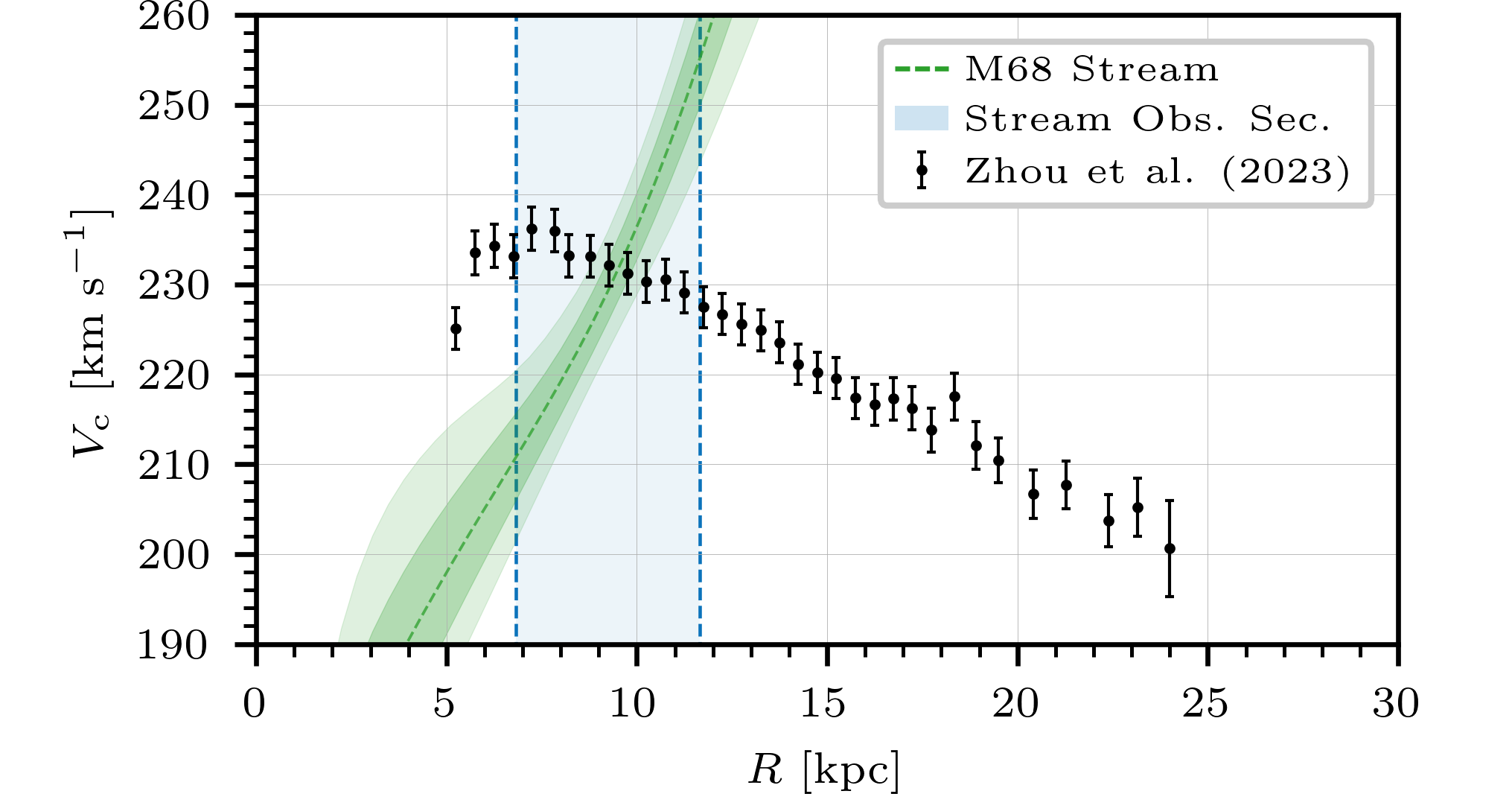}
\caption{Milky Way's rotation curve $V_{\rm c}$ against the Galactocentric cylindrical radius $R$. The dots with error bars indicate the measurements of \citet{2023ApJ...946...73Z}. The radial range covered by the observed section of the M68 stream is shaded blue and its boundaries are highlighted with dashed blue lines. The dashed green line shows the median of the rotation curve obtained by constraining the potential model with the M68 stream. The dark and light green shaded areas show the $1\sigma$ and $2\sigma$ confidence intervals, respectively.}
\label{rot_curve_no_rc}
\end{figure}

Conversely, the stream cannot constrain the scale density $\rho_{0 \rm h}$ and the scale length $a_{\rm h}$ of the dark halo. Their distributions extend beyond a reasonable parameter range, so we list in Table~\ref{table_results} their median as infinity. Additionally, this stream requires a dark matter density inner slope $\gamma_{\rm h} < 0$, which for a $a_{\rm h}> 5$~kpc, it implies an increasing rotation curve. In Figure~\ref{rot_curve_no_rc}, we show the rotation curve of the model as a function of the Galactocentric radius. For reference, we plot the observational measurements of the Milky Way's rotation curve (see Section~\ref{results_rc}) as black dots with error bars. In the middle of the range constrained by the stream, which is indicated by a blue shaded area, the rotation curve is approximately $232$~km~s$^{-1}$, which is consistent with observations. However, due to the positive slope, the discrepancies are significant at the limits of this area. We conclude that, due to the flexibility of the model and the large number of free parameters, it can produce unrealistic configurations in order to achieve marginally better fits to the data. Therefore, we consider it necessary to include additional constraints in order to obtain a model that is consistent with observations across the entire range covered by the stream.

\subsection{Constraints using the MW rotation curve}\label{results_rc}

\begin{figure}
\includegraphics[width=1.0\columnwidth]{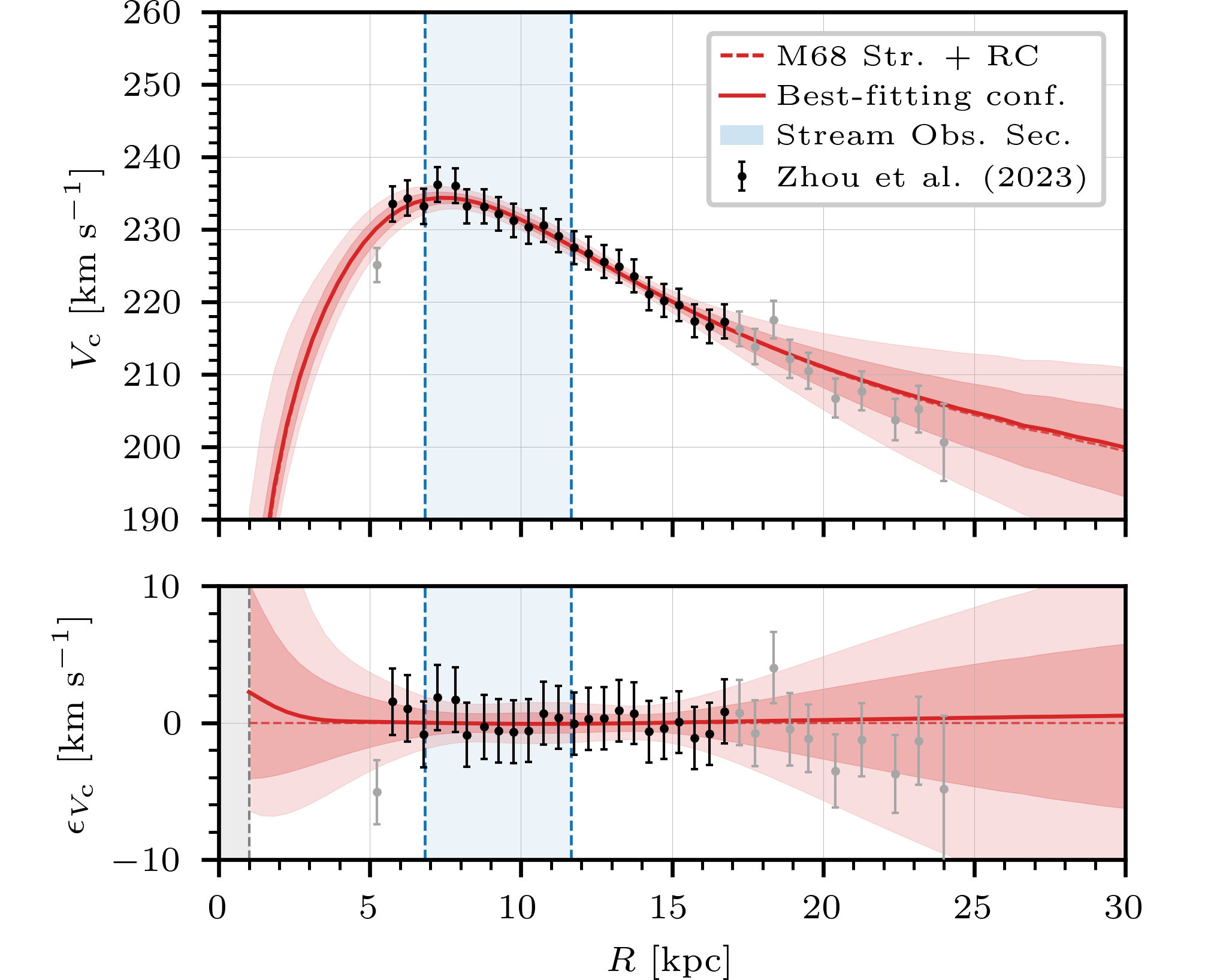}
\caption{Milky Way's rotation curve $V_{\rm c}$ against the Galactocentric cylindrical radius $R$, as in Figure~\ref{rot_curve_no_rc}. The black dots with error bars show the rotation curve measurements from \citet{2023ApJ...946...73Z} used in the likelihood function, and the grey dots show the discarded ones. \textit{Top:} The dashed red line shows the median of the rotation curve obtained by constraining the potential model with the M68 stream and the rotation curve. The dark and light red shaded areas show the $1\sigma$ and $2\sigma$ confidence intervals, respectively. The red solid line shows the rotation curve for the best-fitting configuration. \textit{Bottom:} Error between the observational measurements and the model $\epsilon_{V_{\rm c}}$.}
\label{rot_curve}
\end{figure}

In order to break the degeneracies among the dark halo parameters, we incorporate observational estimates of the Galactic rotation curve into the likelihood function. Each measurement is considered to follow a Gaussian distribution characterised by its mean and standard deviation. The likelihood function is then multiplied by the product of these distributions, as described in \S4 of \citetalias{2023MNRAS.524.2124P}. The rotation curve measurements are taken from \citet{2023ApJ...946...73Z}, who obtained them using the axisymmetric Jeans equation with a large sample of luminous red giant branch stars. These measurements range within $5 \LessSim R \LessSim 25$~kpc, with a median uncertainty of $0.55$~km~s$^{-1}$, and a systematic uncertainty of $\Approx 1$ per cent of the mean value. For $R\GtrSim17$~kpc, we consider that the rotation curve can possibly be influenced by the Galactic warp or by perturbations caused by the LMC. Therefore, to avoid possible biases caused by an improper modelling of the outer Galaxy, we only use the central part of the rotation curve, corresponding to the range $5.5 \leqslant R \leqslant 17$~kpc. We also neglect the measurement at $R=5.24$~kpc because it is abnormally low. Figure~\ref{rot_curve} shows the mean of the measurements with their total uncertainty as dots with error bars. The measurements included in the likelihood function are shown in black and the discarded measurements are shown in grey. The included observations of the rotation curve cover the entire radial range of the stream ($6.82 \LessSim R \LessSim 11.66$~kpc). This range is highlighted by the shaded blue area in Figure~\ref{rot_curve}.

As in Section~\ref{results_str}, we generate random samples from the posterior distribution using a MCMC method with $72$ chains of $10^7$ steps, discarding the first $2\pd{6}$ steps. The fourth column of Table~\ref{table_results} lists the median and the $1\sigma$ confidence interval of the marginalised posterior distributions obtained by fitting the potential model to the Milky Way's rotation curve. In this case, the M68 stream is excluded. The rotation curve does not significantly constrain either the position of the Sun or the mass of the disc, being $M_{\rm d} = 6.69 \pm 3.39\pd{10}$~M$_{\Sun}$. The four free parameters that characterise the dark halo are highly degenerate. This implies that the rotation curve cannot constrain these parameters simultaneously. The distributions of $\rho_{0 \rm h}$ and $q_{\rm h}$ cannot be fully covered by random samples generated using the MCMC method. We therefore assume that the distributions extend to infinity. These results ensure that the rotation curve does not introduce any bias, and that the posterior distributions for $M_{\rm d}$ and $q_{\rm h}$ obtained in the following section are determined only by the stream.

\subsection{Constraints using the M68 stream and the MW rotation curve}\label{results_str_rc}

Combining the stream stars from the DESI selection with the rotation curve of the Milky Way allow us to break the degeneracies between the halo parameters and obtain a realistic model of the inner density profile of the Galaxy. Similarly to Sections~\ref{results_str} and \ref{results_rc}, we generate samples from the posterior distribution using a MCMC method. We run $72$ chains of $10^6$ steps, and discard the firsts $10^5$ steps. The top panel of Figure~\ref{rot_curve} shows the rotation curve against the Galactocentric cylindrical radius of this model. The dashed red line indicates the median value estimated from the random sample. The dark and light red shaded areas show the $1\sigma$ and $2\sigma$ confidence intervals, respectively. The bottom panel shows the error between the observational measurements and the model. We observe that the uncertainties of the model are smaller than the observational uncertainties and that there is no significant systematic deviation. Therefore, we consider this model to be in very good agreement with the rotation curve measured by \citet{2023ApJ...946...73Z}.

The fifth column of Table~\ref{table_results} (M68 Str. + RC) lists the median and the $1\sigma$ confidence interval of all model's free parameters. Firstly, it can be seen that the stream and rotation curve are compatible with the Sun's position as defined by the priors. The only significant discrepancy is in the Sun's velocity in the direction of the disc rotation, which is estimated to be $V_{\Sun}=7.58\pm1.84$~km~s$^{-1}$, approximately $5$~km~$s^{-1}$ slower than the estimates of \citet{2010MNRAS.403.1829S}. There is no significant correlation (Spearman) between the Sun's position and velocity and the other free parameters of the model that could explain this result. Conversely, the literature contains significant variability in the measurement of $V_{\Sun}$, with values ranging from 4 to 22~km~s$^{-1}$ \citep{2026PhR..1150....1P}. Our result is compatible with recent measurements \citep[e.g.][]{2019MNRAS.484.3544R}, so we conclude that this discrepancy is a consequence of the arbitrarily chosen prior.

Our model allows for variations in the initial phase-space position of the M68 globular cluster. We observe deviations towards larger values compared to the priors for the radial velocity $v_r^{\SM{M68}}$ and the proper motion $ \mu_{\alpha*}^{\SM{M68}}$. These deviations are also obtained without the rotation curve (Section~\ref{results_str}), so they are caused by the stream. We consider these deviations to be minor and most likely caused by forcing the model to achieve a slightly better fit to the particular random sample of stream stars in the DESI selection. In addition, there is no correlation between the position of M68 and the other free parameters of the model. The only exception is the correlation between $\mu_\delta$ and $q_{\rm h}$, with a Spearman correlation coefficient of $\rho=-0.65$. As the proper motion $\mu_\delta$ is fully compatible with the prior, we conclude that this parameter does not affect the estimate of $q_{\rm h}$.

We have imposed a uniform prior on the Heliocentric distance of the cluster $r_{\Helio}^{\SM{M68}}$. We therefore estimate $r_{\Helio}^{\SM{M68}}$ using only the stellar stream, the rotation curve, and the potential model of the Galaxy. This give us an estimate of $r_{\Helio}^{\SM{M68}}=10.55\pm0.09$~kpc. This result is fully compatible with the average of 24 independent measurements $\langle r_{\Helio}^{\SM{M68}}\rangle=10.4\pm0.1$~kpc, compiled by \citet{2021MNRAS.505.5957B}. This methodology provides a novel approach to estimating the distances of globular clusters and can be applied to other clusters with known stellar streams.

\begin{figure}
\includegraphics[width=1.0\columnwidth]{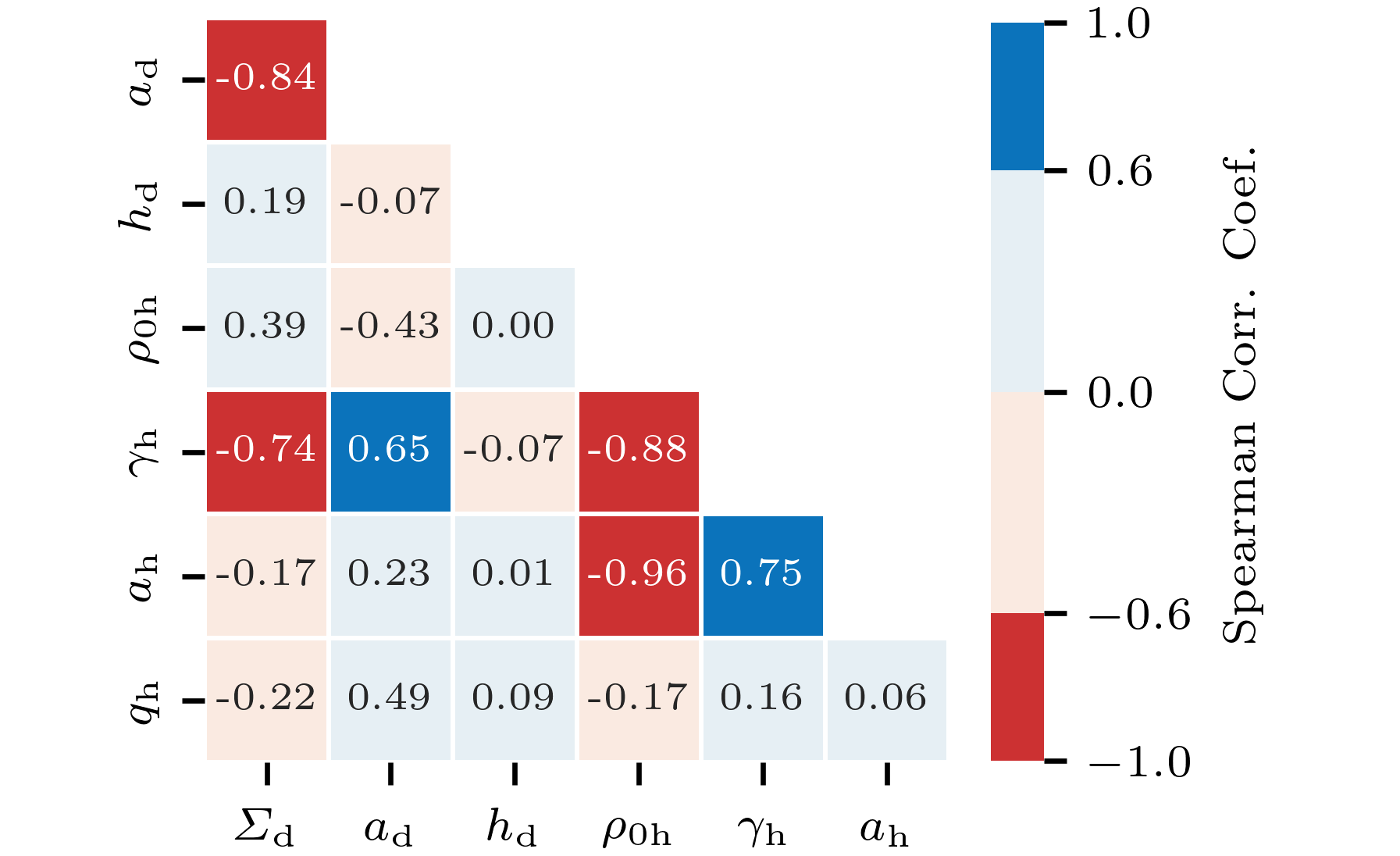}
\caption{Spearman correlation coefficient between the free parameters that characterise the disc and the dark halo. The model is constrained by the DESI star selection of the M68 stellar stream and the rotation curve of the Milky Way. The values are computed using random samples that follow the posterior distribution generated by a MCMC method.}
\label{corner_corr}
\end{figure}

Figure~\ref{corner_corr} shows the Spearman correlation coefficients between the parameters of the disc and the dark halo. Firstly, we observe that the disc scale height $h_{\rm d}$ is uncorrelated with the other parameters. Furthermore, it follows the same distribution as the prior, as it can be seen in Table~\ref{table_results}. This suggests that this parameter is not constrained by the observational data. On the other hand, there is a very strong correlation between the surface density $\varSigma_{\rm d}$, the scale length of the disc $a_{\rm d}$, and the inner slope of the dark halo $\gamma_{\rm h}$. The uncertainty in the parameter $a_{\rm d}$ is approximately half that of the prior, and the uncertainty in $\varSigma_{\rm d}$ is about 20 per cent of the median value. This means that the mass of the disc is strongly constrained. Its value is given by Eq.~\ref{mass_disc_exp} and is equal to $M_{\rm d} = 5.34 \pm 0.57 \pd{10}$~M$_{\Sun}$.

\begin{figure}
\includegraphics[width=1.0\columnwidth]{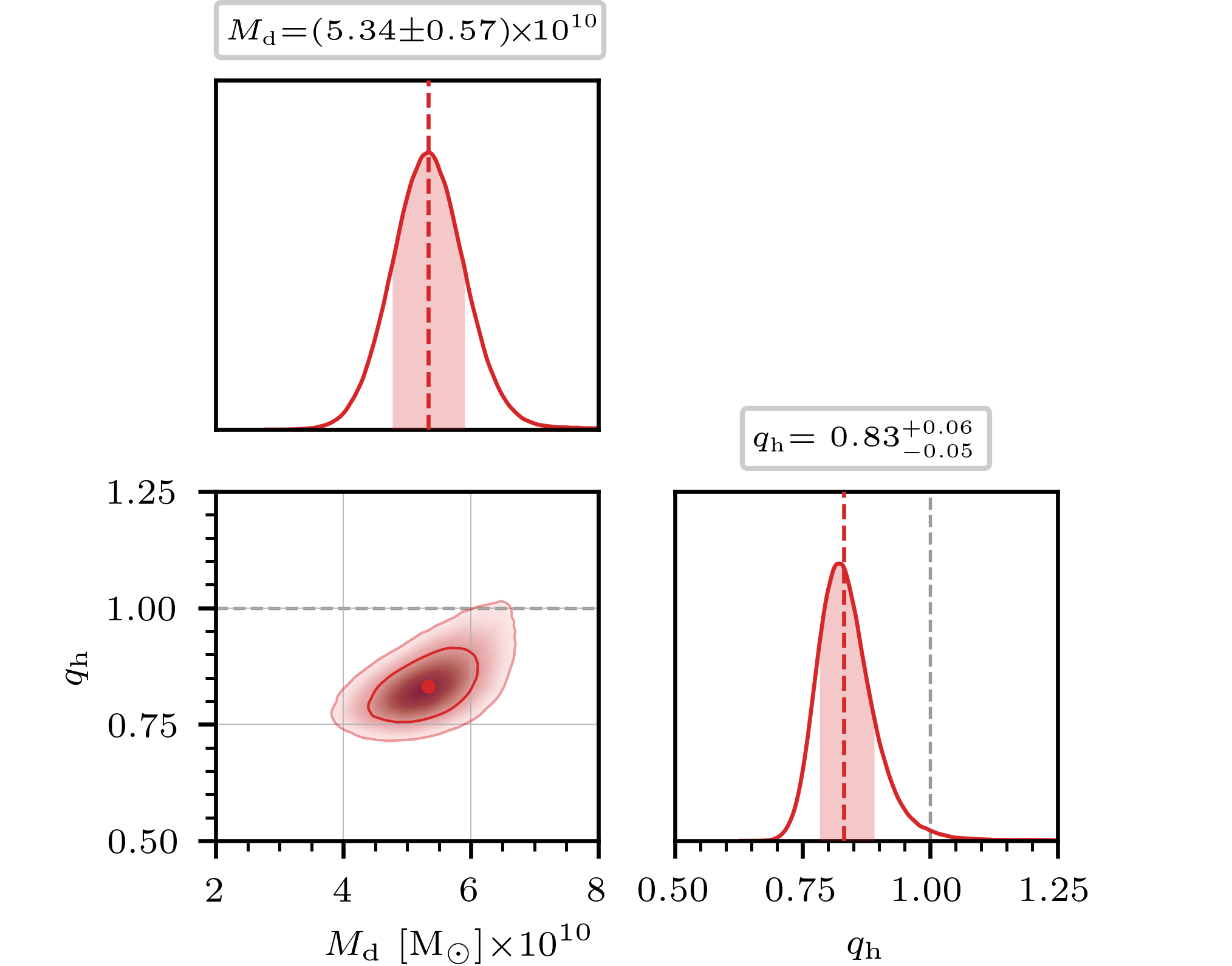}
\caption{Marginal posterior distributions of the disc mass $M_{\rm d}$ and the dark matter halo axis ratio $q_{\rm h}$. The model is constrained by the DESI star selection of the M68 stellar stream and the rotation curve of the Milky Way. These distributions are estimated using a random sample generated by a MCMC method. The grey dashed lines indicate the spherical configuration of the dark halo $q_{\rm h} = 1$. \textit{Diagonal panels:} The solid red line shows the estimated distribution using a KDE method. The vertical dashed red line marks the median, and the $1\sigma$ confidence interval is indicated by the shaded red area. The numerical values are shown in the legends.  \textit{Inferior panels:} Darker red tones indicate a higher posterior distribution. The solid red lines indicate the limits containing 68 and 95 per cent of the distribution, and the large red dot marks the median.}
\label{corner_q_Md}
\end{figure}

Figure~\ref{corner_corr} shows a weak correlation between $q_{\rm h}$ and $a_{\rm d}$, which is similar to the correlation between $q_{\rm h}$ and $M_{\rm d}$ of Spearman correlation coefficient $\rho=0.52$. The derived distribution $M_{\rm d}$ is computed by evaluating Eq.~\ref{mass_disc_exp} with the random sample generated by the MCMC method. This weak correlation implies that the strong correlation between these parameters obtained in Section~\ref{results_str} using only the stream is not present when the rotation curve of the Milky Way is included. Nevertheless, the positive correlation indicates that a massive disc requires a prolate dark halo, and that a light disc requires an oblate dark halo. This correlation is shown in the bottom-left panel of Figure~\ref{corner_q_Md}. The diagonal panels of the same figure show the marginalised posterior distributions of $M_{\rm d}$ and $q_{\rm h}$. These distributions demonstrate that, by using the DESI star selection and the Milky Way's rotation curve, we can infer the existence of an oblate halo with a value of $q_{\rm h} = 0.83^{+0.06}_{-0.05}$. This result is consistent with cosmological dark matter simulations that predict flattened dark halos aligned with the disc \citep[e.g.][]{2019MNRAS.490.4877P}.

\begin{figure}
\includegraphics[width=1.0\columnwidth]{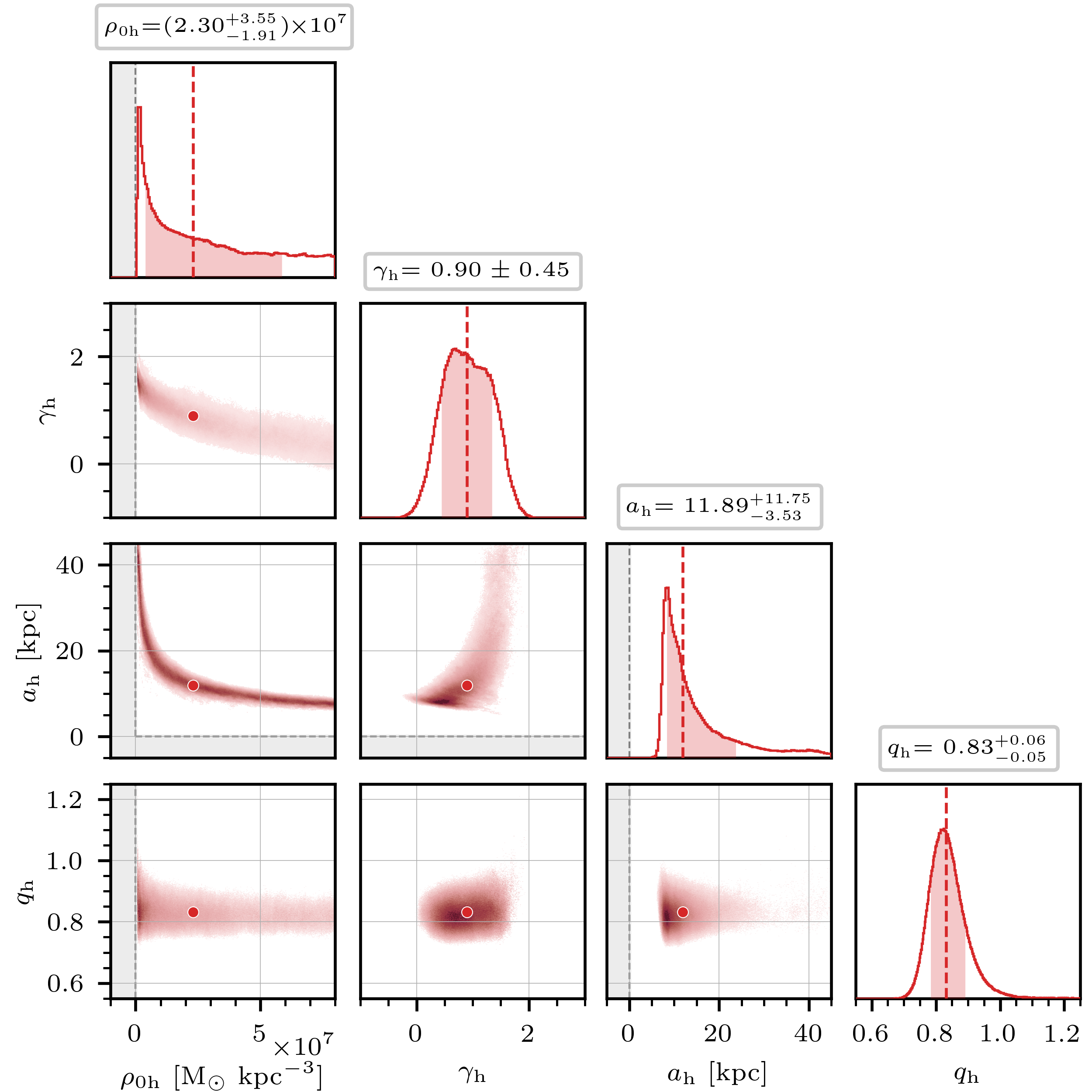}
\caption{As Figure~\ref{corner_q_Md} but for the dark halo parameters. Areas outside the valid parameter range are shaded in grey, and their boundaries are highlighted with dashed grey lines.}
\label{corner_dh}
\end{figure}

Figure~\ref{corner_corr} shows that the halo parameters $\rho_{0 \rm h}$, $\gamma_{\rm h}$, and $a_{\rm h}$ are strongly correlated. These parameters determine the radial dark matter density profile. However, they are independent of $q_{\rm h}$, which determines the shape of the dark matter halo. In Figure~\ref{corner_dh}, we show the marginalised posterior distributions of these parameters. The diagonal panels show each parameter individually, with the medians marked by vertical dashed red lines and the $1\sigma$ confidence intervals indicated by shaded red areas. The lower panels show the bivariate marginalised distributions, with the medians marked by large red dots. In this case, the dark halo inner slope $\gamma_{\rm h} \Approx 1$ is imposed by the decreasing rotation curve. We also observe that $\rho_{0 \rm h}$ and $a_{\rm h}$ cannot be constrained simultaneously, but any configuration of mass $M_{\rm h}\var{R\leqslant20\,{\rm kpc}} = 1.29\pm 0.07 \pd{11}$~M$_{\Sun}$ fits the observational data equally well. This behaviour is consistent with that obtained using the simulated model in Section~\ref{statistical}. As the distributions of $\rho_{0 \rm h}$ and $a_{\rm h}$ extend towards large values, they cannot be fully covered by the sample generated by the MCMC method. However, we assume that the areas of the distributions' tails are negligible and restrict the parameters to $\rho_{0 \rm h} \in \Range{0}{8}\pd{7}$~M$_{\Sun}$~kpc$^{-3}$ and $a_{\rm h} \in \Range{1}{50}$~kpc. This implies that their confidence intervals are underestimated, particularly for $\rho_{0 \rm h}$.

\subsection{Best-fitting configuration}\label{best_fitting}

We find the best-fitting configuration by determining the maximum value of the posterior distribution using the Nelder-Mead simplex algorithm introduced in Section~\ref{statistical}. The last column of Table~\ref{table_results} lists the results obtained using the DESI selection and the Milky Way's rotation curve as constraints. The top panel of Figure~\ref{rot_curve} shows the rotation curve for the best-fitting configuration. There is very good agreement between the best-fitting model and the observational measurements of \citet{2023ApJ...946...73Z} included in the likelihood function ($5.5 \leqslant R \leqslant 17$~kpc). For $R>17$~kpc, there is a small systematic deviation towards larger values. This can be explained by the fact that the best-fitting halo scale length is $\hat{a}_{\rm h}=14.83$~kpc and the dark halo outer slope is fixed to $\beta=3$.

\begin{figure}
\includegraphics[width=1.0\columnwidth]{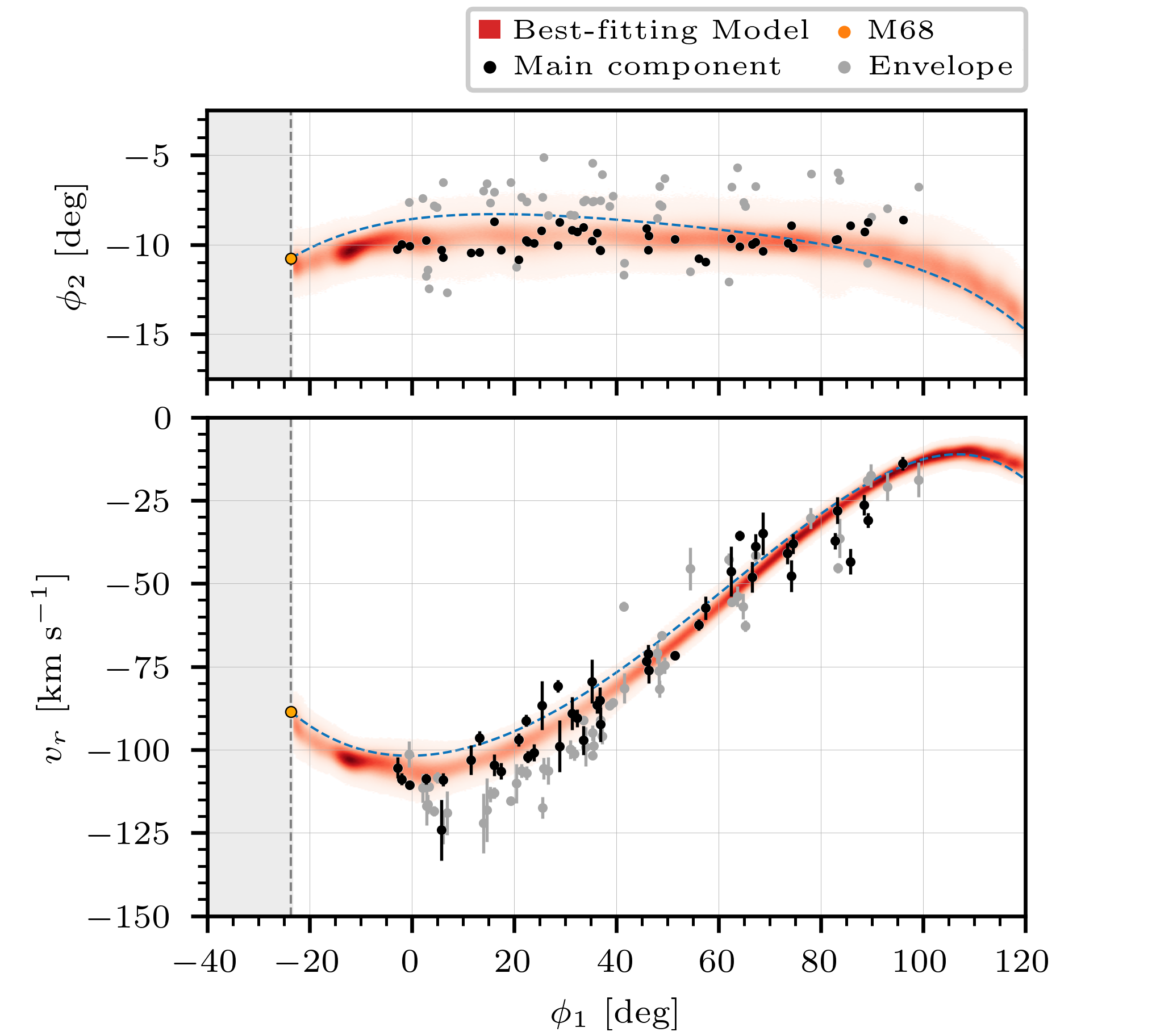}
\caption{Surface density of the best-fitting model of the leading arm of the M68 stream. Darker red tones indicate higher density. The area where the model is not depicted is shaded in grey, and its boundary is highlighted by a dashed grey line. The position of the M68 globular cluster is marked with a large orange dot, and its future orbit is shown with a dashed blue line. The $44$ stars from the DESI selection are shown as black dots, and the $52$ stars in the envelope are shown as grey dots. \textit{Top:} Sky coordinates aligned with the M68 stellar stream ($\phi_1, \phi_2$), plotted with an aspect ratio of 3.5. \textit{Bottom:} Sky coordinate $\phi_1$ against radial velocity $v_r$.}
\label{best_fitting_model}
\end{figure}

Figure~\ref{best_fitting_model} shows the best-fitting density model of the leading arm of the stream. The surface density is shown in red, with darker tones indicating higher star density. The Galactic potential determines the phase-space location of the stream track following the cluster orbit. In the same figure, the best-fitting position of M68 is marked with a large orange dot and its future orbit is marked with a dashed blue line. The width of the stream model is determined by the properties of the cluster and the age of the stream, which we consider to be independent of the Milky Way model. The overdensity at $\phi_1\Approx-10$~deg is intrinsic to the stream, and corresponds to the stars stripped during the last pericentre passage of the cluster \citepalias{2026MNRAS.545f1974P}. The surface density along the stream is determined by the chosen coordinates and phase-space projections.

The top panel of Figure~\ref{best_fitting_model} shows that the model fits the central component of the stream. These stars were selected within an area defined by horizontal lines of constant $\phi_2$ (Eq.~\ref{main_comp_limits} in Section~\ref{stream_width}). The only significant discrepancy occurs within the interval $\phi_1\in\Range{90}{100}$~deg, where the stream curves down and out of the horizontal limits. This suggests that the stars with $\phi_1\GtrSim90$~deg are more likely to be members of the envelope. The bottom panel shows the radial velocity as a function of $\phi_1$. There is a good fit along the entire stream, except for two outliers with small uncertainties that are located above the cluster orbit at $\phi_1\Approx30$~deg and $\phi_1\Approx65$~deg. Similarly, there are three outliers within the interval $\phi_1\in\Range{80}{90}$~deg. We propose that these discrepancies can be explained by stars belonging to the envelope that have been erroneously classified as members of the central component of the stream. On the other hand, unresolved binaries can introduce radial velocity offsets from the mean stream track \citep{2026ApJ..1000...33Q}. We believe that the presence of a few outliers does not affect the model's overall good fit, and that improving the star selection process is necessary to confirm the existence of these outliers.

These results differ significantly from the potential used in the \nbody\ simulation to construct the density model of the stream (Section~\ref{methodology}). For example, the fiducial potential incorporates a Miyamoto-Nagai disc and a spherical dark matter halo. However, the observational data indicate that a $\Approx20$ per cent less massive exponential disc and an oblate dark halo are required. This suggests that choosing this particular model for the fiducial potential introduces no significant bias towards the same potential used in the \nbody\ simulation. Nevertheless, the differences between the simulated streams in both potentials are small, which ensures that the assumptions used to compute the stream model and described in Section~\ref{methodology} are fulfilled. In Appendix~\ref{App6}, we present a comparison of the sky coordinates and radial velocities of the fiducial and the best-fitting potential configurations.

\subsection{Comparison to previous studies}\label{comparison_previous}

\citetalias{2023MNRAS.524.2124P} is the only study in which the M68 stream have been used to constrain the Galactic potential independently of other streams. In this study, the authors use a \textit{Gaia}-DR2 stream star selection alongside additional observational constraints such as the Milky Way's rotation curve, the vertical acceleration at the Sun's position, and the proper motion of Sgr~A$^*$. They estimated a total mass of the disc $M_{\rm d} = 7.38 \pm 0.56 \pd{10}$~M$_{\Sun}$, and a spherical-prolate dark matter halo of axis ratio $q_{\rm h} = 1.14^{+0.21}_{-0.14}$. In the present study, we obtain a disc that is about $40$ per cent lighter and an oblate halo. This discrepancy can be explained by assuming a positive correlation between $M_{\rm d}$ and $q_{\rm h}$, as obtained in Sections~\ref{results_str} and \ref{results_str_rc}. This correlation implies that a massive disc requires less dark matter in the inner region of the Galaxy or, equivalently, a prolate halo. Conversely, a less massive disc, such as that obtained in this study, requires an oblate dark matter halo.

\citet{2024ApJ...967...89I} has also used the M68 stream, in conjunction with GD-1 and streams with a known globular cluster progenitor (NGC~3201, NGC~5466, Palomar~5, and M5), to constrain the Galactic potential. This study also includes additional observational constraints such as the terminal velocity of \textsc{Hi} in the inner Galaxy and the vertical force at the Sun's position. We compare our results with their model with a dark halo outer slope $\beta_{\rm h}=3$, since we adopted the same assumption. Their model is characterised by a $M_{\rm d} = 4.2^{+0.44}_{-0.53} \pd{10}$~M$_{\Sun}$ and $q_{\rm h} = 0.75\pm0.03$. The small discrepancy between these results and ours can be explained by assuming a similar positive correlation between $M_{\rm d}$ and $q_{\rm h}$. In this case, their disc is slightly less massive, implying a slightly more flattened dark halo. Therefore, we conclude that these studies demonstrate the need for precise estimation of the disc mass in order to constrain the dark halo in the inner region of the Galaxy.

The disc and halo combination found in our study can also be compared to previous studies using the total surface density at the Sun's position integrated within $z\in\pm1.1$~kpc or surface density $\varSigma_{\Sun}$\footnote{Surface density $\varSigma_{\Sun} \equiv \int_{-\bar{z}}^{\bar{z}} \:\! \rho_{\rm tot}\var{R_{\Sun},0,z} \;\! dz$, where $\bar{z} = 1.1$~kpc and $\rho_{\rm tot}$ is the sum of the baryonic and dark matter densities.}. Our result $\varSigma_{\Sun} = 77.37 \pm 6.13$~M$_{\Sun}$ pc$^{-2}$ is larger than the mean value $\langle\varSigma_{\Sun}\rangle = 69.8\pm5.6$~M$_{\Sun}$ pc$^{-2}$ obtained in previous studies \citep[Table~1 in][]{2024ApJ...962..165H}. In general, these studies use samples of stars below $z\Approx1.4$~kpc. However, our analysis includes data at the position of the stream, which is at $z\Approx4.5$~kpc. We therefore believe that this discrepancy is due to the use of data located at different heights above the Sun. A model that can describe both the distributions of disc stars and the streams in the halo simultaneously has to be established by future research.

\section{Conclusion}\label{conclusion}

The M68 stream is one of the most significant known streams because it is long and thin, dynamically cold, and has a known progenitor. Only a part of the leading arm of this stream is visible. This section spans approximately 100 deg in the Northern Galactic Hemisphere, in an area of low star contamination. As this stream is faint, it is necessary to use proper motions to effectively separate its stars from those in the foreground. This explains why this stream was not discovered until data from the \textit{Gaia} mission became available. The most recent selection of stream stars comprises 291 likely stream members \citepalias{2026MNRAS.545f1974P}. Of these, only about 20 stars have spectroscopic data available.

In the present study, we have improved upon previous star selections by incorporating radial velocities and metallicities from DESI MWS survey. We also use photometry from the DESI Legacy Surveys, which is more precise than the available from GDR3. This enables us to reduce the amount of foreground contamination present when selecting stars that are consistent with a synthetic population of the M68 globular cluster. Using this method, we select 96 stars with a high probability of being stream members. An additional 18 stars have radial velocity measured by DESI but no photometry, and can therefore be included in the selection, albeit with a higher likelihood of being star contaminants.

This star selection reveals a much wider stream than the expected from an \nbody\ simulation. This simulation assumes that the initial internal characteristics of the cluster resemble those observed today, and that the tidal forces are caused by an axisymmetric, static potential similar to that of the Milky Way. Based on these assumptions, we define the stars that are likely to have been stripped from the cluster as forming the central component of the stream. These stars have radial velocities consistent with those of the simulated stream. Conversely, stars separated from the central component have systematically lower radial velocities, making them appear separated in $(\phi_1, v_r)$ space for $\phi_1 \in \Range{0}{50}$~deg. The cause of this separation is unknown. We hypothesise that it may be related to the progenitor cluster's complex accretion process.

We assume that the central component is a different population than the stars that are not compatible with being stripped from the cluster. Therefore, we restrict the stars to be used to constrain the Galactic potential to the 44 stars in the central component of the stream. This sample is distributed along the entire observable section, which flows almost parallel to the Galactic disc, at about $z\Approx5$~kpc, and covers a radial range of $6.82 \LessSim R \LessSim 11.66$~kpc. This implies that we can only constrain the disc and the inner dark halo. We exclude the envelope stars because most of them are located above the cluster's orbit, which is excluded for the cluster and Galactic potential models used in this research. Furthermore, including the envelope stars could introduce bias into the results of the fit, particularly in radial velocity space. The separation between the two components poses a problem for stars located at $\phi_1\GtrSim50$~deg because the two components appear mixed in the $(\phi_1, v_r)$ space. This part of the stream is more sensitive to variations in the potential because it is the farthest from the globular cluster, the phase-space position of which is measured with high precision. Therefore, we believe that reliable constraints on the Galactic potential require the correct identification of the stream stars and the correct separation of stream components at the far end of the observable section. Better-quality data will allow this to be improved upon in future work.

In order to constrain the potential of the Milky Way, we use the methodology introduced in \citetalias{2023MNRAS.524.2124P} but with several improvements included. Using a kernel density estimation method based on a sample of simulated stream stars, we model the stream in the observed ICRS coordinates. To avoid the computationally expensive task of simulating the stream for every parameter configuration, we only integrate the cluster orbit and locate a fiducial \nbody\ simulation at the corresponding relative position with respect to the integrated orbit. We then determine the best-fitting parameter values by maximising the intersection between the stream model and the observed stream stars. We have tested this methodology using a mock star selection and a potential model of the Milky Way. Due to the statistical variability of the phase-space positions of the observed stars, we conclude that we can only reliably estimate the disc mass $M_{\rm d}$ and the dark halo axis ratio $q_{\rm h}$, with respective errors of $5$ and $3$ per cent. Additionally, we find that these two parameters are strongly correlated in this model.

Including the observational uncertainties results in a loss of distance information, since the GDR3 parallaxes are imprecise for stars located approximately 5 kpc from the Sun. In addition, the observational uncertainties introduce errors in the proper motions and radial velocities. Their impact on determining the best-fitting configuration is similar to that of the previously studied statistical variability, but the magnitude of the impact is larger. We estimate that $M_{\rm d}$ and $q_{\rm h}$ can be recovered with $24$ and $12$ per cent accuracy, respectively. Additionally, the observational uncertainties introduce small systematic deviations towards lower values for both parameters. Therefore, reducing observational errors is more important than increasing the number of stars in the sample in order to obtain better constraints on the Milky Way's potential in future research.

We then use the 44 observed stars in the central component of the stream to constrain the potential of the Milky Way. We assume a static, axisymmetric potential that includes a bulge, an exponential disc, and a spheroidal dark matter halo. We also treat the positions of the Sun and the M68 globular cluster as free parameters, with priors obtained from observational measurements. While the stream model can fit the observational data, a negative dark matter halo density inner slope $\gamma_{\rm h}$ is preferred. This implies a non-physical halo, in which the rotation curve of the Milky Way increases proportionally to the Galactocentric radius. To avoid these configurations, we further constrain the halo parameters by including the observational measurements of the Milky Way's rotation curve. In this case, we can fit both the M68 stream and the rotation curve simultaneously, requiring a disc mass of $M_{\rm d} = 5.34 \pm 0.57 \pd{10}$~M$_{\Sun}$ and a dark halo axis ratio of $q_{\rm h} = 0.83^{+0.06}_{-0.05}$. Without imposing any restrictive prior, we estimate the Heliocentric distance of M68 to be $r_{\Helio}^{\SM{M68}}=10.55\pm0.09$~kpc.

These results are consistent with cosmological simulations that predict oblate dark haloes aligned with the discs of spiral galaxies. However, they are not consistent with the prolate-spherical halo obtained by \citetalias{2023MNRAS.524.2124P} using a GDR2 star selection of the M68 stream. In this case, the model requires a massive disc, which results in less dark matter at the position of the stream. This is achieved by elongating the halo perpendicular to the disc plane. In the present study, since the obtained disc is less massive, additional mass is required close to the disc. This is provided by an oblate halo. The location of the M68 stream explains this correlation, as it constrains the potential at the transition between the disc and the halo. If this correlation can be precisely determined, and if independent observational data on the disc's mass and density distribution are incorporated, the constraints on the shape of the dark halo obtained in this study can be significantly improved.


\section*{Acknowledgements}

This work is supported by NSFC (12573022, 12273021, 12595312), National Key R\&D Program of China (2023YFA1607800, 2023YFA1607801, 2023YFA1605600, 2023YFA1605601), 111 project (No. B20019), and the science research grants from the China Manned Space Project (No. CMS-CSST-2025-A04). We also acknowledge the Yangyang Development Fund. The computation of this work is done on the Gravity supercomputer at the Department of Astronomy, Shanghai Jiao Tong University.

This material is based upon work supported by the U.S. Department of Energy (DOE), Office of Science, Office of High-Energy Physics, under Contract No. DE–AC02–05CH11231, and by the National Energy Research Scientific Computing Center, a DOE Office of Science User Facility under the same contract. Additional support for DESI was provided by the U.S. National Science Foundation (NSF), Division of Astronomical Sciences under Contract No. AST-0950945 to the NSF’s National Optical-Infrared Astronomy Research Laboratory; the Science and Technology Facilities Council of the United Kingdom; the Gordon and Betty Moore Foundation; the Heising-Simons Foundation; the French Alternative Energies and Atomic Energy Commission (CEA); the Secretariat of Science, Humanities, Technology and Innovation (SECIHTI) of Mexico; the Ministry of Science, Innovation and Universities of Spain (MICIU/AEI/10.13039/501100011033), and by the DESI Member Institutions: \url{https://www.desi.lbl.gov/collaborating-institutions}.

The DESI Legacy Imaging Surveys consist of three individual and complementary projects: the Dark Energy Camera Legacy Survey (DECaLS), the Beijing-Arizona Sky Survey (BASS), and the Mayall z-band Legacy Survey (MzLS). DECaLS, BASS and MzLS together include data obtained, respectively, at the Blanco telescope, Cerro Tololo Inter-American Observatory, NSF’s NOIRLab; the Bok telescope, Steward Observatory, University of Arizona; and the Mayall telescope, Kitt Peak National Observatory, NOIRLab. NOIRLab is operated by the Association of Universities for Research in Astronomy (AURA) under a cooperative agreement with the National Science Foundation. Pipeline processing and analyses of the data were supported by NOIRLab and the Lawrence Berkeley National Laboratory. Legacy Surveys also uses data products from the Near-Earth Object Wide-field Infrared Survey Explorer (NEOWISE), a project of the Jet Propulsion Laboratory/California Institute of Technology, funded by the National Aeronautics and Space Administration. Legacy Surveys was supported by: the Director, Office of Science, Office of High Energy Physics of the U.S. Department of Energy; the National Energy Research Scientific Computing Center, a DOE Office of Science User Facility; the U.S. National Science Foundation, Division of Astronomical Sciences; the National Astronomical Observatories of China, the Chinese Academy of Sciences and the Chinese National Natural Science Foundation. LBNL is managed by the Regents of the University of California under contract to the U.S. Department of Energy. The complete acknowledgments can be found at \url{https://www.legacysurvey.org/}.

Any opinions, findings, and conclusions or recommendations expressed in this material are those of the author(s) and do not necessarily reflect the views of the U. S. National Science Foundation, the U. S. Department of Energy, or any of the listed funding agencies.

The authors are honored to be permitted to conduct scientific research on I'oligam Du'ag (Kitt Peak), a mountain with particular significance to the Tohono O’odham Nation.


\section*{Data Availability}

The Python packages \texttt{fnc}, \texttt{invi}, and \texttt{mwm} containing the codes used in this paper are available on GitHub: \url{https://github.com/cgpalau-astro}. All the plots in this paper can be reproduced using the codes and the data available on Zenodo: \url{https://zenodo.org/records/21899430}. The following files related to the M68 stellar stream are also available on Zenodo: \url{https://zenodo.org/records/19208479}:
\begin{enumerate}
 \setlength\itemsep{0.35em}
 \item \textit{N}\!\!\:-body simulation ($T=3040$~Myr)
 \item Mock \textit{Gaia}-DR3 star selection
 \item \textit{Gaia}-DR3 star selection
 \item Additional data for the \textit{Gaia}-DR3 star selection
 \item DESI star selection
\end{enumerate}


\section*{Software}

The following Tools and Python \texttt{packages} are used in this research:
\begin{center}
\begin{tabular}{lp{3.67cm}r}
\textbf{Package} & \textbf{Reference} & \textbf{Version}\\[0.15cm]

\texttt{scipy} & \citet{2020SciPy-NMeth} & \href{https://github.com/scipy/scipy}{1.16.0} \\[0.125cm]

\end{tabular}
\end{center}



\bibliographystyle{mnras}
\bibliography{bib/ref.bib}

\begin{thebibliography}{}
\makeatletter
\relax
\def\mn@urlcharsother{\let\do\@makeother \do\$\do\&\do\#\do\^\do\_\do\%\do\~}
\def\mn@doi{\begingroup\mn@urlcharsother \@ifnextchar [ {\mn@doi@}
  {\mn@doi@[]}}
\def\mn@doi@[#1]#2{\def\@tempa{#1}\ifx\@tempa\@empty \href
  {http://dx.doi.org/#2} {doi:#2}\else \href {http://dx.doi.org/#2} {#1}\fi
  \endgroup}
\def\mn@eprint#1#2{\mn@eprint@#1:#2::\@nil}
\def\mn@eprint@arXiv#1{\href {http://arxiv.org/abs/#1} {{\tt arXiv:#1}}}
\def\mn@eprint@dblp#1{\href {http://dblp.uni-trier.de/rec/bibtex/#1.xml}
  {dblp:#1}}
\def\mn@eprint@#1:#2:#3:#4\@nil{\def\@tempa {#1}\def\@tempb {#2}\def\@tempc
  {#3}\ifx \@tempc \@empty \let \@tempc \@tempb \let \@tempb \@tempa \fi \ifx
  \@tempb \@empty \def\@tempb {arXiv}\fi \@ifundefined
  {mn@eprint@\@tempb}{\@tempb:\@tempc}{\expandafter \expandafter \csname
  mn@eprint@\@tempb\endcsname \expandafter{\@tempc}}}

\bibitem[\protect\citeauthoryear{{Baumgardt} \& {Vasiliev}}{{Baumgardt} \&
  {Vasiliev}}{2021}]{2021MNRAS.505.5957B}
{Baumgardt} H.,  {Vasiliev} E.,  2021, \mn@doi [\mnras]
  {10.1093/mnras/stab1474}, \href
  {https://ui.adsabs.harvard.edu/abs/2021MNRAS.505.5957B} {505, 5957}

\bibitem[\protect\citeauthoryear{{Bennett} \& {Bovy}}{{Bennett} \&
  {Bovy}}{2019}]{2019MNRAS.482.1417B}
{Bennett} M.,  {Bovy} J.,  2019, \mn@doi [\mnras] {10.1093/mnras/sty2813},
  \href {https://ui.adsabs.harvard.edu/abs/2019MNRAS.482.1417B} {482, 1417}

\bibitem[\protect\citeauthoryear{{Bienaym{\'e}}, {Robin}, {Salomon}  \&
  {Reyl{\'e}}}{{Bienaym{\'e}} et~al.}{2024}]{2024A&A...689A.280B}
{Bienaym{\'e}} O.,  {Robin} A.~C.,  {Salomon} J.-B.,   {Reyl{\'e}} C.,  2024,
  \mn@doi [\aap] {10.1051/0004-6361/202450327}, \href
  {https://ui.adsabs.harvard.edu/abs/2024A&A...689A.280B} {689, A280}

\bibitem[\protect\citeauthoryear{{Bland-Hawthorn} \&
  {Gerhard}}{{Bland-Hawthorn} \& {Gerhard}}{2016}]{2016ARA&A..54..529B}
{Bland-Hawthorn} J.,  {Gerhard} O.,  2016, \mn@doi [\araa]
  {10.1146/annurev-astro-081915-023441}, \href
  {https://ui.adsabs.harvard.edu/abs/2016ARA&A..54..529B} {54, 529}

\bibitem[\protect\citeauthoryear{{Bonaca} \& {Price-Whelan}}{{Bonaca} \&
  {Price-Whelan}}{2025}]{2025NewAR.10001713B}
{Bonaca} A.,  {Price-Whelan} A.~M.,  2025, \mn@doi [\nar]
  {10.1016/j.newar.2024.101713}, \href
  {https://ui.adsabs.harvard.edu/abs/2025NewAR.10001713B} {100, 101713}

\bibitem[\protect\citeauthoryear{{Bonaca} et~al.,}{{Bonaca}
  et~al.}{2020}]{2020ApJ...889...70B}
{Bonaca} A.,  et~al., 2020, \mn@doi [\apj] {10.3847/1538-4357/ab5afe}, \href
  {https://ui.adsabs.harvard.edu/abs/2020ApJ...889...70B} {889, 70}

\bibitem[\protect\citeauthoryear{{Bovy}, {Bahmanyar}, {Fritz}  \&
  {Kallivayalil}}{{Bovy} et~al.}{2016}]{2016ApJ...833...31B}
{Bovy} J.,  {Bahmanyar} A.,  {Fritz} T.~K.,   {Kallivayalil} N.,  2016, \mn@doi
  [\apj] {10.3847/1538-4357/833/1/31}, \href
  {https://ui.adsabs.harvard.edu/abs/2016ApJ...833...31B} {833, 31}

\bibitem[\protect\citeauthoryear{{Bystr{\"o}m} et~al.,}{{Bystr{\"o}m}
  et~al.}{2026}]{2026arXiv260507918B}
{Bystr{\"o}m} A.,  et~al., 2026, \mn@doi [arXiv e-prints]
  {10.48550/arXiv.2605.07918}, \href
  {https://ui.adsabs.harvard.edu/abs/2026arXiv260507918B} {p. arXiv:2605.07918}

\bibitem[\protect\citeauthoryear{{Chen}, {Li}  \& {Gnedin}}{{Chen}
  et~al.}{2025}]{2025ApJ...980L..18C}
{Chen} Y.,  {Li} H.,   {Gnedin} O.~Y.,  2025, \mn@doi [\apjl]
  {10.3847/2041-8213/adaf93}, \href
  {https://ui.adsabs.harvard.edu/abs/2025ApJ...980L..18C} {980, L18}

\bibitem[\protect\citeauthoryear{{Cooper} et~al.,}{{Cooper}
  et~al.}{2023}]{2023ApJ...947...37C}
{Cooper} A.~P.,  et~al., 2023, \mn@doi [\apj] {10.3847/1538-4357/acb3c0}, \href
  {https://ui.adsabs.harvard.edu/abs/2023ApJ...947...37C} {947, 37}

\bibitem[\protect\citeauthoryear{{DESI Collaboration} et~al.,}{{DESI
  Collaboration} et~al.}{2016}]{2016arXiv161100037D}
{DESI Collaboration} et~al., 2016, \mn@doi [arXiv e-prints]
  {10.48550/arXiv.1611.00037}, \href
  {https://ui.adsabs.harvard.edu/abs/2016arXiv161100037D} {p. arXiv:1611.00037}

\bibitem[\protect\citeauthoryear{{DESI Collaboration} et~al.,}{{DESI
  Collaboration} et~al.}{2022}]{2022AJ....164..207D}
{DESI Collaboration} et~al., 2022, \mn@doi [\aj] {10.3847/1538-3881/ac882b},
  \href {https://ui.adsabs.harvard.edu/abs/2022AJ....164..207D} {164, 207}

\bibitem[\protect\citeauthoryear{{DESI Collaboration} et~al.,}{{DESI
  Collaboration} et~al.}{2025}]{2025JCAP...07..028A}
{DESI Collaboration} et~al., 2025, \mn@doi [\jcap]
  {10.1088/1475-7516/2025/07/028}, \href
  {https://ui.adsabs.harvard.edu/abs/2025JCAP...07..028A} {2025, 028}

\bibitem[\protect\citeauthoryear{{DESI Collaboration} et~al.,}{{DESI
  Collaboration} et~al.}{2026}]{2026AJ....171..285D}
{DESI Collaboration} et~al., 2026, \mn@doi [\aj] {10.3847/1538-3881/ae4c43},
  \href {https://ui.adsabs.harvard.edu/abs/2026AJ....171..285D} {171, 285}

\bibitem[\protect\citeauthoryear{{Dey} et~al.,}{{Dey}
  et~al.}{2019}]{2019AJ....157..168D}
{Dey} A.,  et~al., 2019, \mn@doi [\aj] {10.3847/1538-3881/ab089d}, \href
  {https://ui.adsabs.harvard.edu/abs/2019AJ....157..168D} {157, 168}

\bibitem[\protect\citeauthoryear{{Dillamore}, {Sanders}, {Belokurov}  \&
  {Zhang}}{{Dillamore} et~al.}{2025}]{2025MNRAS.541..214D}
{Dillamore} A.~M.,  {Sanders} J.~L.,  {Belokurov} V.,   {Zhang} H.,  2025,
  \mn@doi [\mnras] {10.1093/mnras/staf965}, \href
  {https://ui.adsabs.harvard.edu/abs/2025MNRAS.541..214D} {541, 214}

\bibitem[\protect\citeauthoryear{{Fardal}, {Huang}  \& {Weinberg}}{{Fardal}
  et~al.}{2015}]{2015MNRAS.452..301F}
{Fardal} M.~A.,  {Huang} S.,   {Weinberg} M.~D.,  2015, \mn@doi [\mnras]
  {10.1093/mnras/stv1198}, \href
  {https://ui.adsabs.harvard.edu/abs/2015MNRAS.452..301F} {452, 301}

\bibitem[\protect\citeauthoryear{{Fardal}, {van der Marel}, {Law}, {Sohn},
  {Sesar}, {Hernitschek}  \& {Rix}}{{Fardal}
  et~al.}{2019}]{2019MNRAS.483.4724F}
{Fardal} M.~A.,  {van der Marel} R.~P.,  {Law} D.~R.,  {Sohn} S.~T.,  {Sesar}
  B.,  {Hernitschek} N.,   {Rix} H.-W.,  2019, \mn@doi [\mnras]
  {10.1093/mnras/sty3428}, \href
  {https://ui.adsabs.harvard.edu/abs/2019MNRAS.483.4724F} {483, 4724}

\bibitem[\protect\citeauthoryear{{GRAVITY Collaboration} et~al.,}{{GRAVITY
  Collaboration} et~al.}{2021}]{2021A&A...647A..59G}
{GRAVITY Collaboration} et~al., 2021, \mn@doi [\aap]
  {10.1051/0004-6361/202040208}, \href
  {https://ui.adsabs.harvard.edu/abs/2021A&A...647A..59G} {647, A59}

\bibitem[\protect\citeauthoryear{{Gaia Collaboration} et~al.,}{{Gaia
  Collaboration} et~al.}{2016}]{2016A&A...595A...1G}
{Gaia Collaboration} et~al., 2016, \mn@doi [\aap]
  {10.1051/0004-6361/201629272}, \href
  {https://ui.adsabs.harvard.edu/abs/2016A&A...595A...1G} {595, A1}

\bibitem[\protect\citeauthoryear{{Gaia Collaboration} et~al.,}{{Gaia
  Collaboration} et~al.}{2023}]{2023A&A...674A...1G}
{Gaia Collaboration} et~al., 2023, \mn@doi [\aap]
  {10.1051/0004-6361/202243940}, \href
  {https://ui.adsabs.harvard.edu/abs/2023A&A...674A...1G} {674, A1}

\bibitem[\protect\citeauthoryear{{Gieles}, {Erkal}, {Antonini}, {Balbinot}  \&
  {Pe{\~n}arrubia}}{{Gieles} et~al.}{2021}]{2021NatAs...5..957G}
{Gieles} M.,  {Erkal} D.,  {Antonini} F.,  {Balbinot} E.,   {Pe{\~n}arrubia}
  J.,  2021, \mn@doi [Nature Astronomy] {10.1038/s41550-021-01392-2}, \href
  {https://ui.adsabs.harvard.edu/abs/2021NatAs...5..957G} {5, 957}

\bibitem[\protect\citeauthoryear{{Guy} et~al.,}{{Guy}
  et~al.}{2023}]{2023AJ....165..144G}
{Guy} J.,  et~al., 2023, \mn@doi [\aj] {10.3847/1538-3881/acb212}, \href
  {https://ui.adsabs.harvard.edu/abs/2023AJ....165..144G} {165, 144}

\bibitem[\protect\citeauthoryear{{Hattori}, {Valluri}  \& {Vasiliev}}{{Hattori}
  et~al.}{2021}]{2021MNRAS.508.5468H}
{Hattori} K.,  {Valluri} M.,   {Vasiliev} E.,  2021, \mn@doi [\mnras]
  {10.1093/mnras/stab2898}, \href
  {https://ui.adsabs.harvard.edu/abs/2021MNRAS.508.5468H} {508, 5468}

\bibitem[\protect\citeauthoryear{{Horta}, {Price-Whelan}, {Hogg}, {Johnston},
  {Widrow}, {Dalcanton}, {Ness}  \& {Hunt}}{{Horta}
  et~al.}{2024}]{2024ApJ...962..165H}
{Horta} D.,  {Price-Whelan} A.~M.,  {Hogg} D.~W.,  {Johnston} K.~V.,  {Widrow}
  L.,  {Dalcanton} J.~J.,  {Ness} M.~K.,   {Hunt} J. A.~S.,  2024, \mn@doi
  [\apj] {10.3847/1538-4357/ad16e8}, \href
  {https://ui.adsabs.harvard.edu/abs/2024ApJ...962..165H} {962, 165}

\bibitem[\protect\citeauthoryear{{Hunt} \& {Vasiliev}}{{Hunt} \&
  {Vasiliev}}{2025}]{2025NewAR.10001721H}
{Hunt} J. A.~S.,  {Vasiliev} E.,  2025, \mn@doi [\nar]
  {10.1016/j.newar.2024.101721}, \href
  {https://ui.adsabs.harvard.edu/abs/2025NewAR.10001721H} {100, 101721}

\bibitem[\protect\citeauthoryear{{Ibata}, {Malhan}  \& {Martin}}{{Ibata}
  et~al.}{2019}]{2019ApJ...872..152I}
{Ibata} R.~A.,  {Malhan} K.,   {Martin} N.~F.,  2019, \mn@doi [\apj]
  {10.3847/1538-4357/ab0080}, \href
  {https://ui.adsabs.harvard.edu/abs/2019ApJ...872..152I} {872, 152}

\bibitem[\protect\citeauthoryear{{Ibata} et~al.,}{{Ibata}
  et~al.}{2021}]{2021ApJ...914..123I}
{Ibata} R.,  et~al., 2021, \mn@doi [\apj] {10.3847/1538-4357/abfcc2}, \href
  {https://ui.adsabs.harvard.edu/abs/2021ApJ...914..123I} {914, 123}

\bibitem[\protect\citeauthoryear{{Ibata} et~al.,}{{Ibata}
  et~al.}{2024}]{2024ApJ...967...89I}
{Ibata} R.,  et~al., 2024, \mn@doi [\apj] {10.3847/1538-4357/ad382d}, \href
  {https://ui.adsabs.harvard.edu/abs/2024ApJ...967...89I} {967, 89}

\bibitem[\protect\citeauthoryear{{Jarvis} et~al.,}{{Jarvis}
  et~al.}{2026}]{2026arXiv260420958J}
{Jarvis} E.,  et~al., 2026, \mn@doi [arXiv e-prints]
  {10.48550/arXiv.2604.20958}, \href
  {https://ui.adsabs.harvard.edu/abs/2026arXiv260420958J} {p. arXiv:2604.20958}

\bibitem[\protect\citeauthoryear{{Johnston}, {Zhao}, {Spergel}  \&
  {Hernquist}}{{Johnston} et~al.}{1999}]{1999ApJ...512L.109J}
{Johnston} K.~V.,  {Zhao} H.,  {Spergel} D.~N.,   {Hernquist} L.,  1999,
  \mn@doi [\apjl] {10.1086/311876}, \href
  {https://ui.adsabs.harvard.edu/abs/1999ApJ...512L.109J} {512, L109}

\bibitem[\protect\citeauthoryear{{Jordi} et~al.,}{{Jordi}
  et~al.}{2010}]{2010A&A...523A..48J}
{Jordi} C.,  et~al., 2010, \mn@doi [\aap] {10.1051/0004-6361/201015441}, \href
  {https://ui.adsabs.harvard.edu/abs/2010A&A...523A..48J} {523, A48}

\bibitem[\protect\citeauthoryear{{Katz} et~al.,}{{Katz}
  et~al.}{2023}]{2023A&A...674A...5K}
{Katz} D.,  et~al., 2023, \mn@doi [\aap] {10.1051/0004-6361/202244220}, \href
  {https://ui.adsabs.harvard.edu/abs/2023A&A...674A...5K} {674, A5}

\bibitem[\protect\citeauthoryear{{Koposov} et~al.,}{{Koposov}
  et~al.}{2023}]{2023MNRAS.521.4936K}
{Koposov} S.~E.,  et~al., 2023, \mn@doi [\mnras] {10.1093/mnras/stad551}, \href
  {https://ui.adsabs.harvard.edu/abs/2023MNRAS.521.4936K} {521, 4936}

\bibitem[\protect\citeauthoryear{{Koposov} et~al.,}{{Koposov}
  et~al.}{2026}]{2026OJAp....955260K}
{Koposov} S.,  et~al., 2026, \mn@doi [The Open Journal of Astrophysics]
  {10.33232/001c.155260}, \href
  {https://ui.adsabs.harvard.edu/abs/2026OJAp....955260K} {9, 55260}

\bibitem[\protect\citeauthoryear{{Lee}}{{Lee}}{2023}]{2023ApJ...948L..16L}
{Lee} J.-W.,  2023, \mn@doi [\apjl] {10.3847/2041-8213/acd05a}, \href
  {https://ui.adsabs.harvard.edu/abs/2023ApJ...948L..16L} {948, L16}

\bibitem[\protect\citeauthoryear{{Li} et~al.,}{{Li}
  et~al.}{2025}]{2025AJ....170..171L}
{Li} S.,  et~al., 2025, \mn@doi [\aj] {10.3847/1538-3881/adf1a0}, \href
  {https://ui.adsabs.harvard.edu/abs/2025AJ....170..171L} {170, 171}

\bibitem[\protect\citeauthoryear{{Malhan} \& {Ibata}}{{Malhan} \&
  {Ibata}}{2019}]{2019MNRAS.486.2995M}
{Malhan} K.,  {Ibata} R.~A.,  2019, \mn@doi [\mnras] {10.1093/mnras/stz1035},
  \href {https://ui.adsabs.harvard.edu/abs/2019MNRAS.486.2995M} {486, 2995}

\bibitem[\protect\citeauthoryear{{Martin} et~al.,}{{Martin}
  et~al.}{2022}]{2022MNRAS.516.5331M}
{Martin} N.~F.,  et~al., 2022, \mn@doi [\mnras] {10.1093/mnras/stac2426}, \href
  {https://ui.adsabs.harvard.edu/abs/2022MNRAS.516.5331M} {516, 5331}

\bibitem[\protect\citeauthoryear{{Mateu}}{{Mateu}}{2023}]{2023MNRAS.520.5225M}
{Mateu} C.,  2023, \mn@doi [\mnras] {10.1093/mnras/stad321}, \href
  {https://ui.adsabs.harvard.edu/abs/2023MNRAS.520.5225M} {520, 5225}

\bibitem[\protect\citeauthoryear{{McMillan}}{{McMillan}}{2017}]{2017MNRAS.465...76M}
{McMillan} P.~J.,  2017, \mn@doi [\mnras] {10.1093/mnras/stw2759}, \href
  {https://ui.adsabs.harvard.edu/abs/2017MNRAS.465...76M} {465, 76}

\bibitem[\protect\citeauthoryear{{Miller} et~al.,}{{Miller}
  et~al.}{2024}]{2024AJ....168...95M}
{Miller} T.~N.,  et~al., 2024, \mn@doi [\aj] {10.3847/1538-3881/ad45fe}, \href
  {https://ui.adsabs.harvard.edu/abs/2024AJ....168...95M} {168, 95}

\bibitem[\protect\citeauthoryear{{Navarro}, {Frenk}  \& {White}}{{Navarro}
  et~al.}{1996}]{1996ApJ...462..563N}
{Navarro} J.~F.,  {Frenk} C.~S.,   {White} S. D.~M.,  1996, \mn@doi [\apj]
  {10.1086/177173}, \href
  {https://ui.adsabs.harvard.edu/abs/1996ApJ...462..563N} {462, 563}

\bibitem[\protect\citeauthoryear{{Nibauer} \& {Bonaca}}{{Nibauer} \&
  {Bonaca}}{2025}]{2025ApJ...985L..22N}
{Nibauer} J.,  {Bonaca} A.,  2025, \mn@doi [\apjl] {10.3847/2041-8213/add0a9},
  \href {https://ui.adsabs.harvard.edu/abs/2025ApJ...985L..22N} {985, L22}

\bibitem[\protect\citeauthoryear{{Nitschai}, {Cappellari}  \&
  {Neumayer}}{{Nitschai} et~al.}{2020}]{2020MNRAS.494.6001N}
{Nitschai} M.~S.,  {Cappellari} M.,   {Neumayer} N.,  2020, \mn@doi [\mnras]
  {10.1093/mnras/staa1128}, \href
  {https://ui.adsabs.harvard.edu/abs/2020MNRAS.494.6001N} {494, 6001}

\bibitem[\protect\citeauthoryear{{Palau} \& {Miralda-Escud{\'e}}}{{Palau} \&
  {Miralda-Escud{\'e}}}{2019}]{2019MNRAS.488.1535P}
{Palau} C.~G.,  {Miralda-Escud{\'e}} J.,  2019, \mn@doi [\mnras]
  {10.1093/mnras/stz1790}, \href
  {https://ui.adsabs.harvard.edu/abs/2019MNRAS.488.1535P} {488, 1535}

\bibitem[\protect\citeauthoryear{{Palau} \& {Miralda-Escud{\'e}}}{{Palau} \&
  {Miralda-Escud{\'e}}}{2023}]{2023MNRAS.524.2124P}
{Palau} C.~G.,  {Miralda-Escud{\'e}} J.,  2023, \mn@doi [\mnras]
  {10.1093/mnras/stad1930}, \href
  {https://ui.adsabs.harvard.edu/abs/2023MNRAS.524.2124P} {524, 2124}

\bibitem[\protect\citeauthoryear{{Palau}, {Wang}  \& {Han}}{{Palau}
  et~al.}{2025}]{2025MNRAS.539.2718P}
{Palau} C.~G.,  {Wang} W.,   {Han} J.,  2025, \mn@doi [\mnras]
  {10.1093/mnras/staf658}, \href
  {https://ui.adsabs.harvard.edu/abs/2025MNRAS.539.2718P} {539, 2718}

\bibitem[\protect\citeauthoryear{{Palau}, {Wang}  \& {Han}}{{Palau}
  et~al.}{2026}]{2026MNRAS.545f1974P}
{Palau} C.~G.,  {Wang} W.,   {Han} J.,  2026, \mn@doi [\mnras]
  {10.1093/mnras/staf1974}, \href
  {https://ui.adsabs.harvard.edu/abs/2026MNRAS.545f1974P} {545, staf1974}

\bibitem[\protect\citeauthoryear{{Perryman}}{{Perryman}}{2026}]{2026PhR..1150....1P}
{Perryman} M.,  2026, \mn@doi [\physrep] {10.1016/j.physrep.2025.09.004}, \href
  {https://ui.adsabs.harvard.edu/abs/2026PhR..1150....1P} {1150, 1}

\bibitem[\protect\citeauthoryear{{Poppett} et~al.,}{{Poppett}
  et~al.}{2024}]{2024AJ....168..245P}
{Poppett} C.,  et~al., 2024, \mn@doi [\aj] {10.3847/1538-3881/ad76a4}, \href
  {https://ui.adsabs.harvard.edu/abs/2024AJ....168..245P} {168, 245}

\bibitem[\protect\citeauthoryear{{Prada}, {Forero-Romero}, {Grand}, {Pakmor}
  \& {Springel}}{{Prada} et~al.}{2019}]{2019MNRAS.490.4877P}
{Prada} J.,  {Forero-Romero} J.~E.,  {Grand} R. J.~J.,  {Pakmor} R.,
  {Springel} V.,  2019, \mn@doi [\mnras] {10.1093/mnras/stz2873}, \href
  {https://ui.adsabs.harvard.edu/abs/2019MNRAS.490.4877P} {490, 4877}

\bibitem[\protect\citeauthoryear{{Qiu} et~al.,}{{Qiu}
  et~al.}{2026}]{2026ApJ..1000...33Q}
{Qiu} T.,  et~al., 2026, \mn@doi [\apj] {10.3847/1538-4357/ae422d}, \href
  {https://ui.adsabs.harvard.edu/abs/2026ApJ..1000...33Q} {1000, 33}

\bibitem[\protect\citeauthoryear{{Reino}, {Rossi}, {Sanderson}, {Sellentin},
  {Helmi}, {Koppelman}  \& {Sharma}}{{Reino}
  et~al.}{2021}]{2021MNRAS.502.4170R}
{Reino} S.,  {Rossi} E.~M.,  {Sanderson} R.~E.,  {Sellentin} E.,  {Helmi} A.,
  {Koppelman} H.~H.,   {Sharma} S.,  2021, \mn@doi [\mnras]
  {10.1093/mnras/stab304}, \href
  {https://ui.adsabs.harvard.edu/abs/2021MNRAS.502.4170R} {502, 4170}

\bibitem[\protect\citeauthoryear{{Rowell} \& {Kilic}}{{Rowell} \&
  {Kilic}}{2019}]{2019MNRAS.484.3544R}
{Rowell} N.,  {Kilic} M.,  2019, \mn@doi [\mnras] {10.1093/mnras/stz184}, \href
  {https://ui.adsabs.harvard.edu/abs/2019MNRAS.484.3544R} {484, 3544}

\bibitem[\protect\citeauthoryear{{Sanders} \& {Binney}}{{Sanders} \&
  {Binney}}{2013}]{2013MNRAS.433.1826S}
{Sanders} J.~L.,  {Binney} J.,  2013, \mn@doi [\mnras] {10.1093/mnras/stt816},
  \href {https://ui.adsabs.harvard.edu/abs/2013MNRAS.433.1826S} {433, 1826}

\bibitem[\protect\citeauthoryear{{Schlafly} et~al.,}{{Schlafly}
  et~al.}{2023}]{2023AJ....166..259S}
{Schlafly} E.~F.,  et~al., 2023, \mn@doi [\aj] {10.3847/1538-3881/ad0832},
  \href {https://ui.adsabs.harvard.edu/abs/2023AJ....166..259S} {166, 259}

\bibitem[\protect\citeauthoryear{{Schlegel}, {Finkbeiner}  \&
  {Davis}}{{Schlegel} et~al.}{1998}]{1998ApJ...500..525S}
{Schlegel} D.~J.,  {Finkbeiner} D.~P.,   {Davis} M.,  1998, \mn@doi [\apj]
  {10.1086/305772}, \href
  {https://ui.adsabs.harvard.edu/abs/1998ApJ...500..525S} {500, 525}

\bibitem[\protect\citeauthoryear{{Sch{\"o}nrich}, {Binney}  \&
  {Dehnen}}{{Sch{\"o}nrich} et~al.}{2010}]{2010MNRAS.403.1829S}
{Sch{\"o}nrich} R.,  {Binney} J.,   {Dehnen} W.,  2010, \mn@doi [\mnras]
  {10.1111/j.1365-2966.2010.16253.x}, \href
  {https://ui.adsabs.harvard.edu/abs/2010MNRAS.403.1829S} {403, 1829}

\bibitem[\protect\citeauthoryear{{Tsantaki} et~al.,}{{Tsantaki}
  et~al.}{2022}]{2022A&A...659A..95T}
{Tsantaki} M.,  et~al., 2022, \mn@doi [\aap] {10.1051/0004-6361/202141702},
  \href {https://ui.adsabs.harvard.edu/abs/2022A&A...659A..95T} {659, A95}

\bibitem[\protect\citeauthoryear{{Valluri} et~al.,}{{Valluri}
  et~al.}{2025}]{2025ApJ...980...71V}
{Valluri} M.,  et~al., 2025, \mn@doi [\apj] {10.3847/1538-4357/ada690}, \href
  {https://ui.adsabs.harvard.edu/abs/2025ApJ...980...71V} {980, 71}

\bibitem[\protect\citeauthoryear{{Vasiliev} \& {Baumgardt}}{{Vasiliev} \&
  {Baumgardt}}{2021}]{2021MNRAS.505.5978V}
{Vasiliev} E.,  {Baumgardt} H.,  2021, \mn@doi [\mnras]
  {10.1093/mnras/stab1475}, \href
  {https://ui.adsabs.harvard.edu/abs/2021MNRAS.505.5978V} {505, 5978}

\bibitem[\protect\citeauthoryear{{Vasiliev}, {Belokurov}  \&
  {Erkal}}{{Vasiliev} et~al.}{2021}]{2021MNRAS.501.2279V}
{Vasiliev} E.,  {Belokurov} V.,   {Erkal} D.,  2021, \mn@doi [\mnras]
  {10.1093/mnras/staa3673}, \href
  {https://ui.adsabs.harvard.edu/abs/2021MNRAS.501.2279V} {501, 2279}

\bibitem[\protect\citeauthoryear{Virtanen et~al.,}{Virtanen
  et~al.}{2020}]{2020SciPy-NMeth}
Virtanen P.,  et~al., 2020, \mn@doi [Nature Methods]
  {10.1038/s41592-019-0686-2}, \href {https://rdcu.be/b08Wh} {17, 261}

\bibitem[\protect\citeauthoryear{{Wan} et~al.,}{{Wan}
  et~al.}{2023}]{2023MNRAS.519..192W}
{Wan} Z.,  et~al., 2023, \mn@doi [\mnras] {10.1093/mnras/stac3566}, \href
  {https://ui.adsabs.harvard.edu/abs/2023MNRAS.519..192W} {519, 192}

\bibitem[\protect\citeauthoryear{{Wang}, {Iwasawa}, {Nitadori}  \&
  {Makino}}{{Wang} et~al.}{2020}]{2020MNRAS.497..536W}
{Wang} L.,  {Iwasawa} M.,  {Nitadori} K.,   {Makino} J.,  2020, \mn@doi
  [\mnras] {10.1093/mnras/staa1915}, \href
  {https://ui.adsabs.harvard.edu/abs/2020MNRAS.497..536W} {497, 536}

\bibitem[\protect\citeauthoryear{{Wegg}, {Gerhard}  \& {Bieth}}{{Wegg}
  et~al.}{2019}]{2019MNRAS.485.3296W}
{Wegg} C.,  {Gerhard} O.,   {Bieth} M.,  2019, \mn@doi [\mnras]
  {10.1093/mnras/stz572}, \href
  {https://ui.adsabs.harvard.edu/abs/2019MNRAS.485.3296W} {485, 3296}

\bibitem[\protect\citeauthoryear{{Yanny} et~al.,}{{Yanny}
  et~al.}{2009}]{2009AJ....137.4377Y}
{Yanny} B.,  et~al., 2009, \mn@doi [\aj] {10.1088/0004-6256/137/5/4377}, \href
  {https://ui.adsabs.harvard.edu/abs/2009AJ....137.4377Y} {137, 4377}

\bibitem[\protect\citeauthoryear{{Yavetz}, {Johnston}, {Pearson},
  {Price-Whelan}  \& {Weinberg}}{{Yavetz} et~al.}{2021}]{2021MNRAS.501.1791Y}
{Yavetz} T.~D.,  {Johnston} K.~V.,  {Pearson} S.,  {Price-Whelan} A.~M.,
  {Weinberg} M.~D.,  2021, \mn@doi [\mnras] {10.1093/mnras/staa3687}, \href
  {https://ui.adsabs.harvard.edu/abs/2021MNRAS.501.1791Y} {501, 1791}

\bibitem[\protect\citeauthoryear{{Yavetz}, {Johnston}, {Pearson},
  {Price-Whelan}  \& {Hamilton}}{{Yavetz} et~al.}{2023}]{2023ApJ...954..215Y}
{Yavetz} T.~D.,  {Johnston} K.~V.,  {Pearson} S.,  {Price-Whelan} A.~M.,
  {Hamilton} C.,  2023, \mn@doi [\apj] {10.3847/1538-4357/ace7b9}, \href
  {https://ui.adsabs.harvard.edu/abs/2023ApJ...954..215Y} {954, 215}

\bibitem[\protect\citeauthoryear{{Zhao}, {Zhao}, {Chu}, {Jing}  \&
  {Deng}}{{Zhao} et~al.}{2012}]{2012RAA....12..723Z}
{Zhao} G.,  {Zhao} Y.-H.,  {Chu} Y.-Q.,  {Jing} Y.-P.,   {Deng} L.-C.,  2012,
  \mn@doi [Research in Astronomy and Astrophysics]
  {10.1088/1674-4527/12/7/002}, \href
  {https://ui.adsabs.harvard.edu/abs/2012RAA....12..723Z} {12, 723}

\bibitem[\protect\citeauthoryear{{Zhou}, {Li}, {Huang}  \& {Zhang}}{{Zhou}
  et~al.}{2023}]{2023ApJ...946...73Z}
{Zhou} Y.,  {Li} X.,  {Huang} Y.,   {Zhang} H.,  2023, \mn@doi [\apj]
  {10.3847/1538-4357/acadd9}, \href
  {https://ui.adsabs.harvard.edu/abs/2023ApJ...946...73Z} {946, 73}

\bibitem[\protect\citeauthoryear{{Zhou} et~al.,}{{Zhou}
  et~al.}{2025}]{2025OJAp....8E..83Z}
{Zhou} R.,  et~al., 2025, \mn@doi [The Open Journal of Astrophysics]
  {10.33232/001c.141680}, \href
  {https://ui.adsabs.harvard.edu/abs/2025OJAp....8E..83Z} {8, 83}

\makeatother
\end{thebibliography}



\clearpage
\appendix

\section{Metallicity M92}\label{App0}

\begin{figure}
\includegraphics[width=1.0\columnwidth]{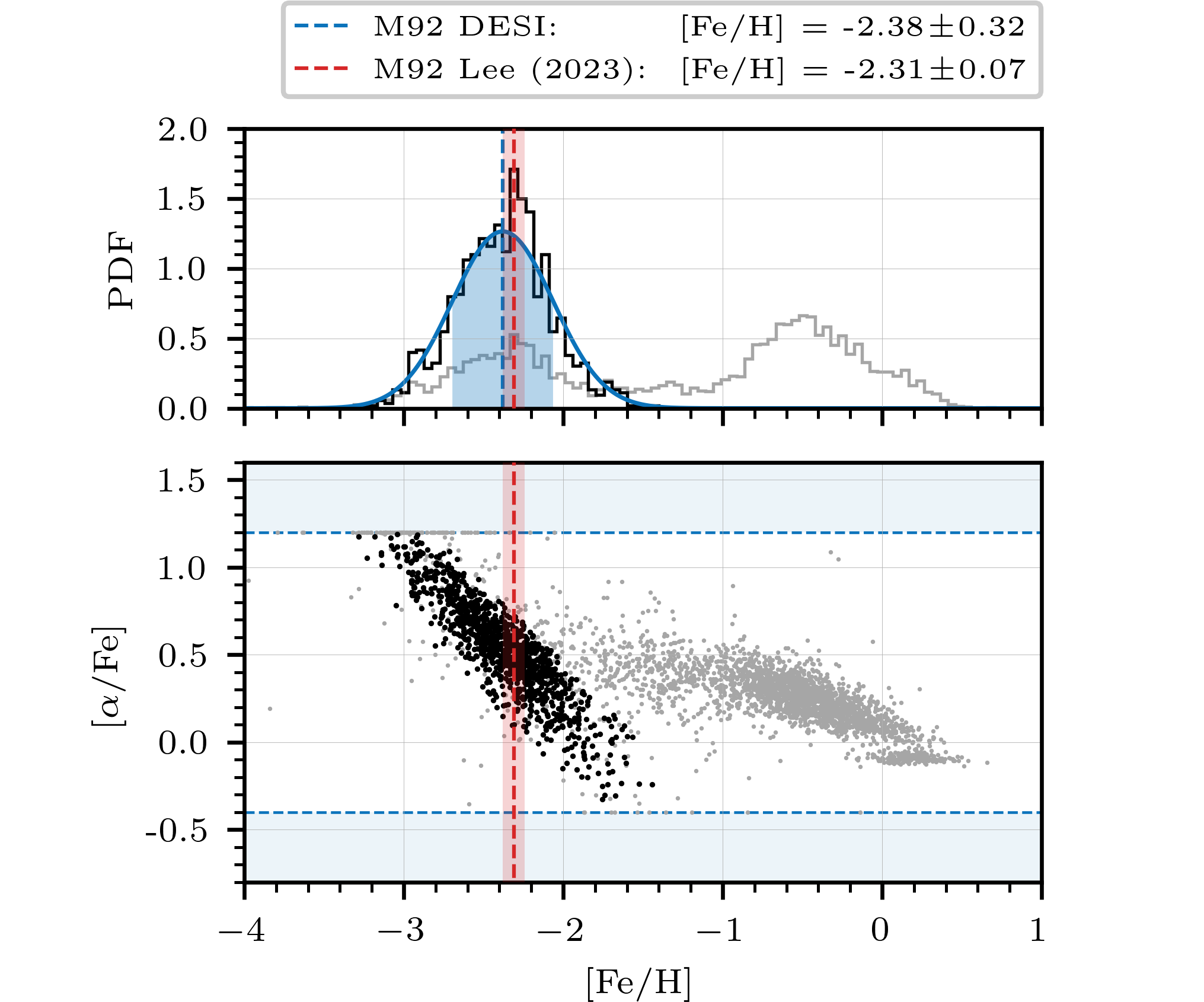}
\caption{\textit{Top:} Distributions of iron abundance $\FeH$. The grey histogram corresponds to the $3854$ stars that pass the cuts (1-5), and the black histogram corresponds to the $1074$ stars in the M92 cluster. The blue line shows the best-fitting Gaussian distribution to the cluster stars. Its mean is marked with a dashed blue line, and its standard deviation with a shaded blue area. The dashed red line indicates the metallicity of M92 as measured by \citet{2023ApJ...948L..16L}, and the shaded red area indicates the error bars. \textit{Bottom:} Iron abundance $\FeH$ against alpha-elements abundance $\AlphaFe$. The stars that pass the cuts (1-5) are shown as grey dots, and the stars in the M92 globular cluster are shown as black dots. The areas outside the DESI observational limits are shaded in blue, and their boundaries are highlighted with dashed blue lines.}
\label{metallicity_M92}
\end{figure}

We present the alpha-elements abundance $\AlphaFe$ and iron abundance $\FeH$ of the globular cluster M92, as measured by the DESI MWS survey. We analyse the stars that pass the quality cuts (1-3), as defined in Section~\ref{selection}, and that are located within a square area on the sky defined by the following conditions:
\begin{enumerate}
\setlength\itemsep{0.5em}
    \item[(4)] $|\alpha - 259.218| < 0.5$~deg
    \item[(5)] $|\delta - 43.136| < 0.5$~deg,
\end{enumerate}
where the coordinates of the cluster centre are obtained from \citet{2021MNRAS.505.5978V}. The top panel of Figure~\ref{metallicity_M92} shows the distribution of $\FeH$ of the $3854$ that pass the above cuts as a grey histogram, and the bottom panel shows the abundances of each star as grey dots.

The stars that belong to M92 are defined by the following additional conditions:
\begin{enumerate}
\setlength\itemsep{0.5em}
    \item[(6)] $-40 < v_r < -100$~km s$^{-1}$
    \item[(7)] $-0.39 < \AlphaFe < 1.19$
    \item[(8)] $-1 \leqslant \dfrac{\AlphaFe -a -b\,\FeH}{\epsilon} \leqslant 1$,
\end{enumerate}
where $a\simeq-1.682$, $b\simeq-0.9$, and $\epsilon=-0.336$. Figure~\ref{metallicity_M92} shows the distribution of $1074$ stars passing the cuts (1-8) as a black histogram. The best-fitting Gaussian distribution is shown in blue in the top panel. Its mean and standard deviation constitute the DESI measurement of $\FeH = -2.38\pm0.32$. We compare this result to $\FeH = -2.31\pm0.07$ measured by \citet{2023ApJ...948L..16L}, which is shown in red in the same figure. The iron abundance of M92 estimated by DESI is significantly wider, indicating that the DESI observational uncertainties are underestimated. The bottom panel shows as black dots the $\AlphaFe$ and $\FeH$ abundances of the stars in M92. They appear negatively correlated, which we consider to be non-physical and caused by correlated observational errors.


\section{Supplement stream stars}\label{App1}

\begin{figure}
\includegraphics[width=1.0\columnwidth]{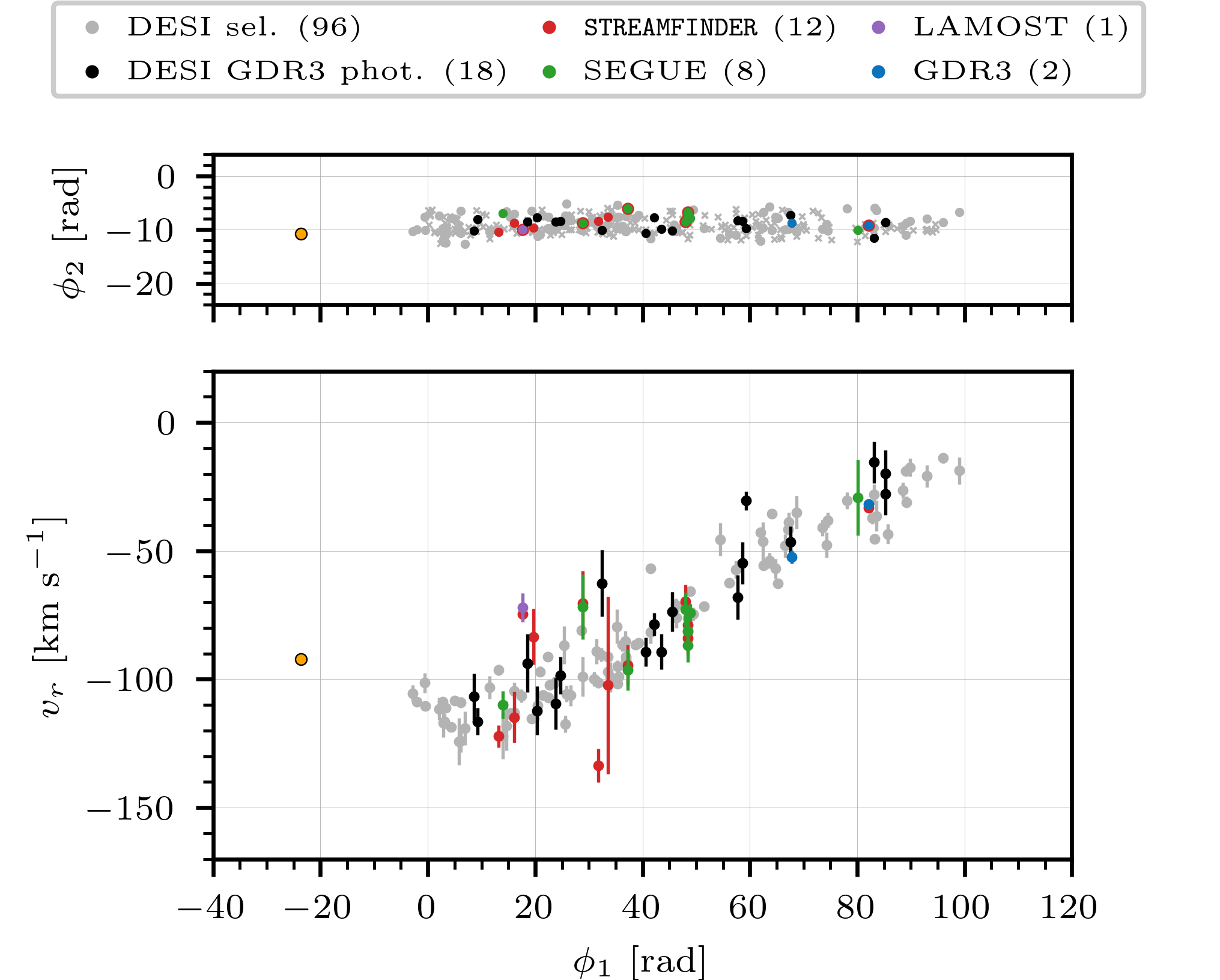}
\caption{Results of cross-matching a sample of M68 stream stars with several spectroscopic surveys. The sample comprises $96$ stars from the DESI selection (Section~\ref{final_sel}) and $225$ stars from the GDR3 selection (\S4 in \citetalias{2026MNRAS.545f1974P}) that lack complete DESI photometry or spectroscopy. Each point with error bars corresponds to a star and is colour-coded according to its survey of origin. The legend indicates the number of stars matched for each survey. The stars from the DESI selection are shown as grey dots. The location of the M68 globular cluster is marked with a large orange dot. \textit{Top:} Sky coordinates aligned with the stream ($\phi_1, \phi_2$). The grey crosses indicate the position of the GDR3 stars for which no DESI data is available. \textit{Bottom:} Radial velocity $v_r$ against the sky coordinate $\phi_1$.}
\label{comparison}
\end{figure}

A total of $528$ stars that pass the cuts (1-11) have no available DESI photometry, but have measured DESI radial velocity and metallicity. For these stars, we apply the CMD selection in cut (12) using \textit{Gaia} $\BPRP$ colour index and $M_G$ absolute magnitude. We correct for dust reddening by following the App.~B in \citetalias{2025MNRAS.539.2718P} and select stars that are compatible with a synthetic stellar population of M68 as described in \S4 of \citetalias{2026MNRAS.545f1974P}. A total of $107$ stars pass the cuts (1-16). Of these, $18$ are newly selected stars that are not included in the DESI selection. These stars are shown as black dots in Figure~\ref{comparison}. They are located along the entire stream and are consistent with the DESI star selection.

\section{Radial velocities from other surveys}\label{App2}

We investigate whether the radial velocity measurements of stars in the M68 stream are included in other spectroscopic surveys. We compile a sample of stream stars by combining the final DESI star selection of $96$ stars (Section~\ref{final_sel}) with the $225$ stars from the GDR3 selection (\S4 in \citetalias{2026MNRAS.545f1974P}) that lack complete DESI photometry or spectroscopy. We then cross-match this sample with the catalogues listed below:

\begin{enumerate}
 \item The Survey of Surveys (SoS) DR1 compilation \citep{2022A&A...659A..95T} includes measurements from RAVE, LAMOST, GES, GALAH, APOGEE, and SEGUE spectroscopic surveys. We find a total of $8$ matches from the SEGUE \citep{2009AJ....137.4377Y}, and $1$ from LAMOST \citep{2012RAA....12..723Z}.

 \item The \texttt{STREAMFINDER} survey \citep{2021ApJ...914..123I} includes $12$ matches with stars from the Fjörm stream\footnote{The M68 stream (Fjörm) is labelled 22 in the catalogue.}. The star located at ($\phi_1,v_r$)=($31.75$~deg, $-133.68$~km~s$^{-1}$) is considered to be foreground contamination.

 \item The GDR3 catalogue includes $2$ matches.
\end{enumerate}

Figure~\ref{comparison} shows the matched stars, colour-coded as indicated in the legend. The top panel shows the $\phi$ coordinates of the stream stars. The grey dots correspond to the final DESI selection, and the grey crosses correspond to the GDR3 stars for which no DESI data is available. The bottom panel shows the radial velocity $v_r$ against the coordinate $\phi_1$. The stars from the final DESI selection are shown in grey for reference. The error bars indicate the measurement uncertainty. In both panels, the M68 cluster is shown as a large orange dot. Overall, there is good agreement between these measurements and those from the DESI survey.


\section{Smoothing parameter or bandwidth}\label{App3}

The smoothing parameter, also known as bandwidth, is determined by the weights used to compute the covariance matrices (Eq.~C4 in \citetalias{2023MNRAS.524.2124P}). In this paper, we adopt the following definition:
\begin{equation}
c_{nm} = \exp\big[ \:\!\! -(l_{nm}/\epsilon)^\alpha \,\big],
\end{equation}
where the parameters $\epsilon\simeq0.63$~kpc and $\alpha\simeq7.218$ have been optimised to produce the best-fit of the stream model to the \nbody\ simulation computed with the fiducial potential.


\section{Gaussian inner product}\label{App4}

As shown in Eq.~C6 of \citetalias{2023MNRAS.524.2124P}, the convolution of a star with the stream model is defined as the inner product of the PDF representing the star and its observational uncertainties, and the PDF modelling the stream. In general, the inner product of two functions $f\var{x}$ and $g\var{x}$, where $x\in\mathbb{R}^d$ and $d\in\mathbb{Z}^{+}$, is defined as:
\begin{equation}
\langle f\var{x}, g\var{x}\rangle \equiv \int_{-\infty}^{+\infty} f\var{x}\,g\var{x}\,dx.
\end{equation}
If the functions are the multivariate Gaussian distributions $N\var{x\:\!|\:\!a,A}$ and $N\var{x\:\!|\:\!b,B}$, where $a$ and $b$ are the means, and $A$ and $B$ are the covariance matrices, then:
\begin{equation}\label{Nab}
\langle N\var{x\:\!|\:\!a,A}, N\var{x\:\!|\:\!b,B}\rangle = N\var{0\:\!|\:\!b-a,A+B}.
\end{equation}
Consequently, if $J$ is the number of observed stars and $N$ is the number of stars used to construct the stream density model, evaluating the likelihood function requires $J\!\times\! N$ evaluations of Eq.~\ref{Nab}. Therefore, computing the inverse of $A+B$, as well as computing the covariance matrices of the stream model, are the most computationally expensive steps of this algorithm. Evaluating the likelihood function can be accelerated by neglecting the correlations of the observational uncertainties of the stream stars, and by modelling the stream with uncorrelated Gaussians. In this case, $A+B$ is a diagonal matrix that can be inverted easily. Further development is required to determine the impact of these assumptions on the results obtained when using them to speed up the evaluation of the posterior distribution.


\section{Distance estimates}\label{App5}

We use the \texttt{SpecDis} database, which provides Heliocentric distances $r_{\Helio}$ for $\Approx 10.7$ million stars in the internal Year-3 data of the DESI MWS. This catalogue is an improvement on the publicly available version\footnote{\url{https://data.desi.lbl.gov/public/dr1/vac/dr1/mws-specdis/}}, which only includes distances for stars in the DESI Data Release 1 \citep{2025AJ....170..171L}. The distances are estimated from the spectra of the stars and GDR3 parallaxes using a feed-forward multilayer perceptron neural network. The results have been validated by comparing them with independent distance measurements of stars in globular clusters, dwarf galaxies, the Sagittarius stream, and blue horizontal branch stars.

\begin{figure}
\includegraphics[width=1.0\columnwidth]{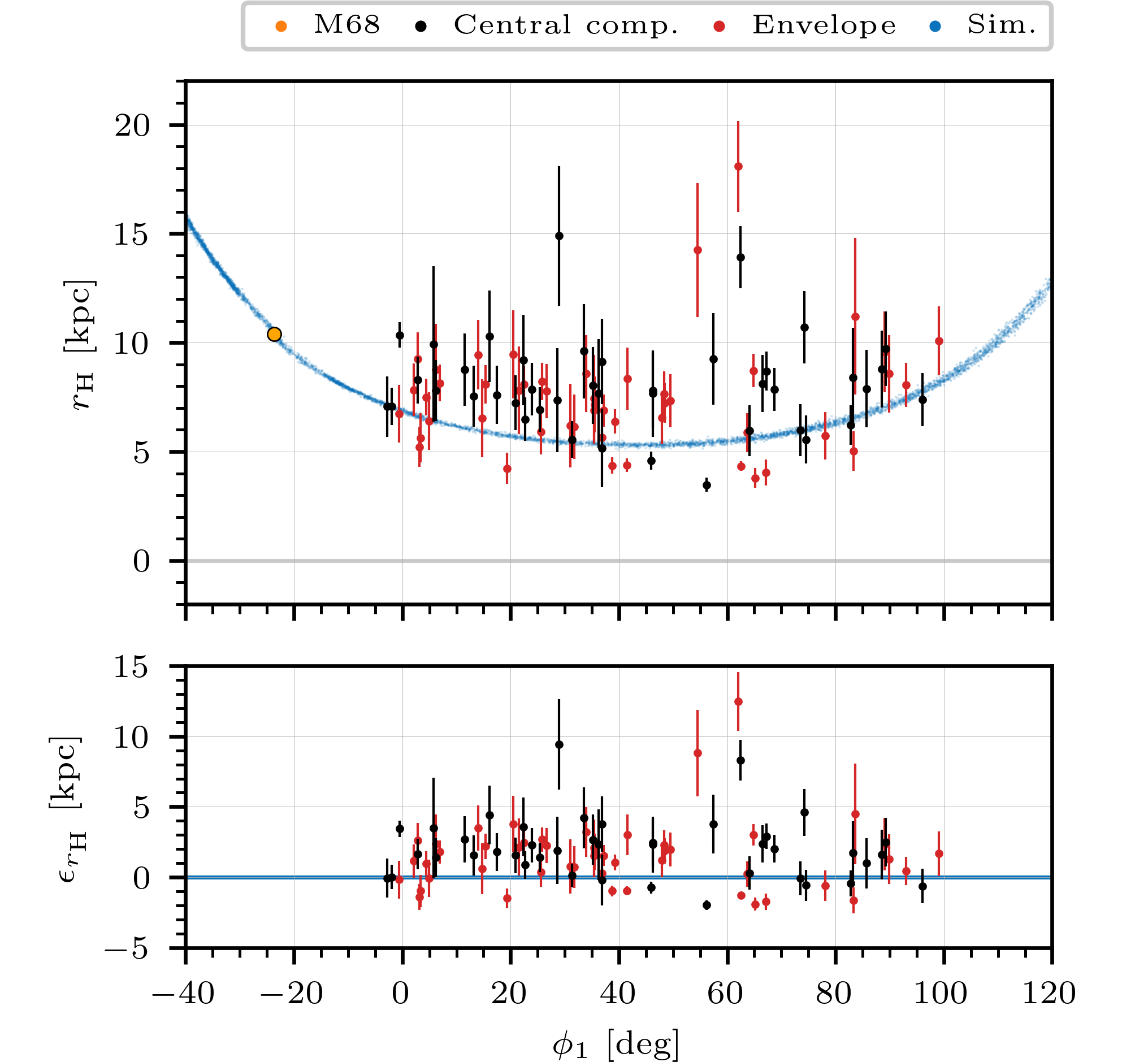}
\caption{\textit{Top:} Heliocentric distance $r_{\Helio}$ with error bars against the sky coordinate aligned with the M68 stream $\phi_1$ of $91$ stars in the DESI selection (Section~\ref{final_sel}). The distances have been estimated from the spectra of the stars and are available in the \texttt{SpecDis} catalogue. The stars in the central component of the stream are shown in black and those in the envelope in red. The position of the cluster is marked by a large orange dot, and the \nbody\ simulation of the stream is shown by the small blue dots. The value $r_{\Helio}=0$ is highlighted in grey. \textit{Bottom:} Difference between the \texttt{SpecDis} distance and the distance estimated from the cluster orbit $\epsilon_{r_{\Heliosmall}}$ against the sky coordinate $\phi_1$. The value $\epsilon_{r_{\Heliosmall}}=0$ is highlighted in blue.}
\label{distance}
\end{figure}

A total of $91$ stars in the DESI selection have an available distance estimate. In the top panel of Figure~\ref{distance}, we plot the mean values with error bars against the sky coordinate $\phi_1$. We also show the position of the globular cluster as a large orange dot, and the \nbody\ simulation of the stream in blue. In the bottom panel, we plot the difference between the \texttt{SpecDis} distance and the distance estimated from the cluster orbit $\epsilon_{r_{\Heliosmall}}$ (\S5.3 in \citetalias{2025MNRAS.539.2718P}) against $\phi_1$. The estimated mean distances of the stars align well with the simulation, following the arc described by the stream in the space $(\phi_1, r_{\Helio})$, within $r_{\Helio} \in \Range{5}{7}$~kpc. This implies that the error is independent of $\phi_1$. The stream stars present a slight systematic deviation of $\epsilon_{r_{\Heliosmall}} \simeq 1.75\pm2.82$~kpc, especially within the range $\phi_1\in\Range{0}{50}$~deg, with respect to the \nbody\ simulation. In addition, there are four outliers with $r_{\Helio} \GtrSim 13$~kpc that could be background stars. These stars are not eliminated from the DESI selection because we consider their mean distance to be potentially biased or their uncertainties to be underestimated. There is no significant difference between the central component of the stream and the envelope stars. Even if a small systematic separation exists, large observational uncertainties prevent us from distinguishing between the two components. In general, we consider the estimates obtained from the cluster orbit to be more precise. This is because the cluster distance is estimated with high precision \citep[$D=10.4\pm0.1$~kpc from][]{2021MNRAS.505.5957B} and errors caused by an incorrect potential only become significant for stream stars that are far from the cluster, for $\phi_1 \GtrSim 80$~deg.

\section{Comparison between fiducial and best-fitting model}\label{App6}

\begin{figure}
\includegraphics[width=1.0\columnwidth]{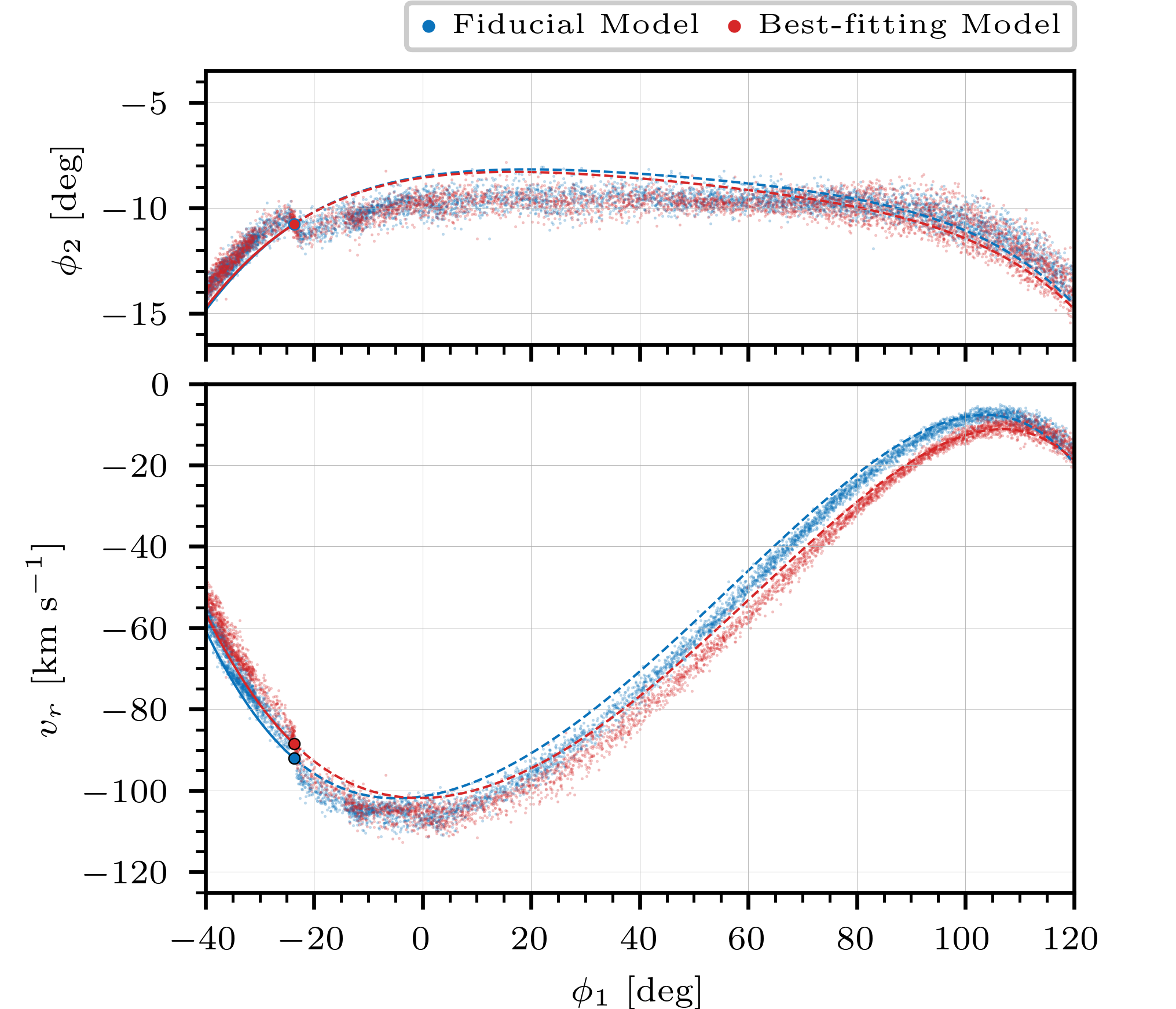}
\caption{Comparison between the \nbody\ simulations of the M68 stellar stream obtained using the fiducial model (blue, Section~\ref{methodology}) and the best-fitting model (red, Section~\ref{best_fitting}). For each model, the stream stars are shown as small dots, the current position of the cluster is marked with a large dot, and its past and future orbits are shown as solid and dashed lines, respectively. \textit{Top:} Sky coordinates aligned with the stream ($\phi_1, \phi_2$). \textit{Bottom:} Radial velocity $v_r$ against the sky coordinate $\phi_1$.}
\label{comparison_fid_bf}
\end{figure}

Figure~\ref{comparison_fid_bf} shows the sky coordinates and the radial velocities of the \nbody\ simulation used to determine the model of the stream, computed in the fiducial potential (blue, Section~\ref{methodology}). This simulation is compared with that obtained using the best-fitting configuration (red, Section~\ref{best_fitting}). The distribution in sky coordinates is almost identical for both models. The cluster's orbit differs only slightly as it moves further away from the cluster along the leading arm, since the cluster's position is measured with high precision. The differences are larger in radial velocity space, indicating that the model is comparatively more constrained by the radial velocities measured by DESI than by the sky coordinates. In both cases, the models have approximately the same internal velocity dispersion, separation from the cluster orbit and overall density variations along the stream. This ensures that the assumption that the fiducial model do not differ significantly from those used to determine the posterior distribution, particularly the best-fitting configuration, is valid.

\clearpage
\section*{Affiliations}

$^{1}$ Department of Astronomy, Shanghai Jiao Tong University, 800 Dongchuan Road, Shanghai 200240, China\\
$^{2}$ Shanghai Key Laboratory for Particle Physics and Cosmology, 800 Dongchuan Road, Shanghai 200240, China\\
$^{3}$ Lawrence Berkeley National Laboratory, 1 Cyclotron Road, Berkeley, CA 94720, USA\\
$^{4}$ Department of Physics, Boston University, 590 Commonwealth Avenue, Boston, MA 02215 USA\\
$^{5}$ Institute for Astronomy, University of Edinburgh, Royal Observatory, Blackford Hill, Edinburgh EH9 3HJ, UK\\
$^{6}$ Dipartimento di Fisica ``Aldo Pontremoli'', Universit\`a degli Studi di Milano, Via Celoria 16, I-20133 Milano, Italy\\
$^{7}$ INAF-Osservatorio Astronomico di Brera, Via Brera 28, 20122 Milano, Italy\\
$^{8}$ Department of Physics \& Astronomy, University College London, Gower Street, London, WC1E 6BT, UK\\
$^{9}$ Departamento de Astrof\'{\i}sica, Universidad de La Laguna (ULL), E-38206, La Laguna, Tenerife, Spain\\
$^{10}$ Instituto de Astrof\'{\i}sica de Canarias, V\'{\i}a L\'{a}ctea, s/n, E-38205 La Laguna, Tenerife, Spain\\
$^{11}$ Institut d'Estudis Espacials de Catalunya (IEEC), Esteve Terradas 1, Edifici RDIT, Campus PMT-UPC, 08860 Castelldefels, Spain\\
$^{12}$ Institute of Space Sciences, ICE-CSIC, Campus UAB, Carrer de Can Magrans s/n, 08913 Bellaterra, Barcelona, Spain\\
$^{13}$ Instituto de F\'{\i}sica, Universidad Nacional Aut\'{o}noma de M\'{e}xico,  Circuito de la Investigaci\'{o}n Cient\'{\i}fica, Ciudad Universitaria, Cd. de M\'{e}xico  C.~P.~04510,  M\'{e}xico\\
$^{14}$ Departamento de F\'isica, Universidad de los Andes, Cra. 1 No. 18A-10, Edificio Ip, CP 111711, Bogot\'a, Colombia\\
$^{15}$ Observatorio Astron\'omico, Universidad de los Andes, Cra. 1 No. 18A-10, Edificio H, CP 111711 Bogot\'a, Colombia\\
$^{16}$ Institute of Cosmology and Gravitation, University of Portsmouth, Dennis Sciama Building, Portsmouth, PO1 3FX, UK\\
$^{17}$ University of Virginia, Department of Astronomy, Charlottesville, VA 22904, USA\\
$^{18}$ Fermi National Accelerator Laboratory, PO Box 500, Batavia, IL 60510, USA\\
$^{19}$ Department of Physics and Astronomy, University of California, Irvine, 92697, USA\\
$^{20}$ Sorbonne Universit\'{e}, CNRS/IN2P3, Laboratoire de Physique Nucl\'{e}aire et de Hautes Energies (LPNHE), FR-75005 Paris, France\\
$^{21}$ Department of Astronomy and Astrophysics, UCO/Lick Observatory, University of California, 1156 High Street, Santa Cruz, CA 95064, USA\\
$^{22}$ Department of Astronomy and Astrophysics, University of California, Santa Cruz, 1156 High Street, Santa Cruz, CA 95065, USA\\
$^{23}$ Department of Astronomy \& Astrophysics, University of Toronto, Toronto, ON M5S 3H4, Canada\\
$^{24}$ NSF NOIRLab, 950 N. Cherry Ave., Tucson, AZ 85719, USA\\
$^{25}$ Instituci\'{o} Catalana de Recerca i Estudis Avan\c{c}ats, Passeig de Llu\'{\i}s Companys, 23, 08010 Barcelona, Spain\\
$^{26}$ Institut de F\'{i}sica d’Altes Energies (IFAE), The Barcelona Institute of Science and Technology, Edifici Cn, Campus UAB, 08193, Bellaterra (Barcelona), Spain\\
$^{27}$ Instituto de Ciencias F\'{\i}sicas, Universidad Nacional Aut\'onoma de M\'exico, Av. Universidad s/n, Cuernavaca, Morelos, C.~P.~62210, M\'exico\\
$^{28}$ Instituto de Estudios Astrof\'isicos, Facultad de Ingenier\'ia y Ciencias, Universidad Diego Portales, Av. Ej\'ercito Libertador 441, Santiago, Chile\\
$^{29}$ Steward Observatory, University of Arizona, 933 N. Cherry Avenue, Tucson, AZ 85721, USA\\
$^{30}$ IRFU, CEA, Universit\'{e} Paris-Saclay, F-91191 Gif-sur-Yvette, France\\
$^{31}$ Department of Physics and Astronomy, University of Waterloo, 200 University Ave W, Waterloo, ON N2L 3G1, Canada\\
$^{32}$ Perimeter Institute for Theoretical Physics, 31 Caroline St. North, Waterloo, ON N2L 2Y5, Canada\\
$^{33}$ Waterloo Centre for Astrophysics, University of Waterloo, 200 University Ave W, Waterloo, ON N2L 3G1, Canada\\
$^{34}$ Instituto de Astrof\'{i}sica de Andaluc\'{i}a (CSIC), Glorieta de la Astronom\'{i}a, s/n, E-18008 Granada, Spain\\
$^{35}$ Departament de F\'isica, EEBE, Universitat Polit\`ecnica de Catalunya, c/Eduard Maristany 10, 08930 Barcelona, Spain\\
$^{36}$ Universit\'{e} Clermont-Auvergne, CNRS, LPCA, 63000 Clermont-Ferrand, France\\
$^{37}$ Department of Physics and Astronomy, Sejong University, 209 Neungdong-ro, Gwangjin-gu, Seoul 05006, Republic of Korea\\
$^{38}$ Queensland University of Technology,  School of Chemistry \& Physics, George St, Brisbane 4001, Australia\\
$^{39}$ CIEMAT, Avenida Complutense 40, E-28040 Madrid, Spain\\
$^{40}$ Max Planck Institute for Extraterrestrial Physics, Gie\ss enbachstra\ss e 1, 85748 Garching, Germany\\
$^{41}$ Department of Physics, University of Michigan, 450 Church Street, Ann Arbor, MI 48109, USA\\
$^{42}$ University of Michigan, 500 S. State Street, Ann Arbor, MI 48109, USA\\
$^{43}$ National Astronomical Observatories, Chinese Academy of Sciences, A20 Datun Road, Chaoyang District, Beijing, 100101, P.~R.~China\\


\bsp	
\label{lastpage}

\end{document}